\documentclass[10pt,a4paper]{article}
\usepackage{amsmath}
\usepackage{amsfonts}
\usepackage{amssymb} 
\usepackage[scr=boondoxupr,cal=euler]{mathalpha}
\usepackage{lmodern}
\usepackage{graphicx} 
\usepackage{caption}
\usepackage[left=2cm,right=2cm,top=2cm,bottom=2cm]{geometry}
\usepackage{tensor}
\usepackage{braket}
\usepackage[nottoc]{tocbibind} % Puts references in toc 
\usepackage{bbm} % provides \mathbbm{1} 
\usepackage{slashed} % provides \slashed{#}
\usepackage{cite}
\usepackage{booktabs}
\usepackage{xcolor} 

\usepackage{color} % not to have squares on links
\usepackage{hyperref} % internal links from toc to document 
\hypersetup{
    colorlinks=true, % make the links colored
    linkcolor= black, % color TOC links in blue
    urlcolor= black, % color URLs in red
    citecolor = black,
    linktoc=all, % 'all' will create links for everything in the TOC
    pdftitle = {2606.01167}
}

\allowdisplaybreaks  % allow to break the content of align between pages

\numberwithin{equation}{section} % numbering X.Y

\DeclareMathOperator{\tr}{tr}

\newcommand{\DD}[1]{\mathcal{D}#1\ }
\newcommand{\dd}[1]{d#1\ }
\newcommand{\eg}{e.g.}

\newcommand{\ads}{ { \mathrm{AdS} } }

\newcommand{\cd}{ { \mathcal{\nabla} } }

\newcommand{\CP}{ \mathbb{CP} }
\newcommand{\intl}{\int\limits  }
\newcommand{\vol}{\mathrm{vol}}
\newcommand{\VOL}{\mathrm{V}}

    \newcommand{\psib}{\overline{\psi}}
    
    \newcommand{\phib}{\overline{\phi}}
    
    \newcommand{\Psib}{\overline{\Psi}}

    \newcommand{\chib}{{\overline{\chi}}}

    \newcommand{\zetab}{{\overline{\zeta}}}

    \newcommand{\mD}{{\mathcal{D}}}

    \newcommand{\cD}{ \mathcal{D} }

        \newcommand{\D}{\mathcal{D}}

        \newcommand{\nablasl}{\slashed{\nabla}}
       
\begin{document}

	\begin{center}
\vspace*{2cm}

{\Large\bf
One-loop divergences for KK theories on \(\ads\times S\) spaces;
\\ \vspace{0.2cm}
a reanalysis of \(\ads_4 \times S^7 \)/ABJM precision holography
}
 
\vspace{1cm}

{
Federico Arrighi 
and
Lorenzo Casarin\footnote[1]{Current affiliation: Max-Planck-Institut für Kolloid- und Grenzflächenforschung, 
Am M\"u{}hlenberg 2, 14476 Potsdam, Germany}
}

\vspace{0.5cm}
 
{\em
\vspace{0.5cm} 
Institut f\"ur Theoretische Physik  \\ 
Leibniz Universit\"at Hannover \\
Appelstra\ss{}e 2, 30167 Hannover, Germany   }
\\[1cm]
 \url{federico.arrighi@itp.uni-hannover.de}, 
 \url{lorenzo.casarin@itp.uni-hannover.de} 

\end{center}
\vspace{0.9cm}
%
%%%%%%%%%%%%%%%%%%%%%%%%%%%%%%%%%%%%%%%
\begin{abstract}
\noindent  
We provide a systematic framework for computing the logarithmically divergent part of one-loop partition functions on product spaces $\ads_{d_A} \times S^{d_S}$ of arbitrary dimension. By expanding the higher-dimensional kinetic operators in spherical harmonics, we reduce the ($d_A+d_S$)-dimensional spectral problem to an infinite tower of $d_A$-dimensional determinants, which are then represented via spectral $\zeta$-function methods. We isolate the logarithmic divergences arising from the interplay between the individual AdS determinants and the infinite Kaluza--Klein sum, carefully accounting for the contributions of zero modes on the sphere that produce additional AdS determinants. We test this framework on different fields and apply it to the complete multiplet of 11-dimensional supergravity on $\ads_4 \times S^7$. We recover in a 4d language the result of  
% 	arXiv:1210.6057,   %<-- use this in the arxiv abstract
\cite{Bhattacharyya:2012ye},  % <--remove from arxiv abstract
namely that the only non-vanishing logarithmic divergence originates entirely from the 2-form AdS mode in the ghost sector, reproducing the well-known $\frac{1}{4}\log N$ correction to the ABJM free energy predicted by supersymmetric localization.

\end{abstract}
%%%%%%%%%%%%%%%%%%%%%%%%%%%%%%%%%%%%%%%
 
\newpage
\tableofcontents

\setcounter{footnote}{0}

\newpage
%%%%%%%%%%%%%%%%%%%%%%%%%%%%%%%%
\section{Introduction}
%%%%%%%%%%%%%%%%%%%%% 

The AdS/CFT conjecture~\cite{Maldacena:1997re, Gubser:1998bc, Witten:1998qj} is a cornerstone of modern theoretical physics, and providing precision tests of this correspondence is an active area of research. On the field-theory side, advances in non-perturbative techniques, most notably supersymmetric localization~\cite{Pestun:2007rz,Pestun:2016zxk}, have made it possible to compute observables to a precision well beyond the leading large-\(N\) behavior. On the gravity side, matching these results requires going beyond the classical supergravity approximation and including the one-loop determinants of the fluctuating fields around the dual background. This program of precision holography has been pursued with considerable success in recent years; see e.g.~\cite{Bhattacharyya:2012ye,Bobev:2023dwx,Gautason:2023igo,Gautason:2021vfc,Beccaria:2023ujc,Tseytlin:2024euk,Beccaria:2024gkq,Buchbinder:2014nia} and references therein.

A canonical example, and the one we focus on in this work, is the 
\(\ads_4 / \text{CFT}_3\) duality between M-theory on
\(\ads_4\times S ^7 \) 
and the ABJM theory~\cite{Aharony:2008ug}. The ABJM free energy on the round three-sphere admits a large-\(N\) expansion~\cite{Marino:2011eh,Fuji:2011km}, see also \cite{Bobev:2022wem,Bobev:2023lkx,Bobev:2025ltz } for recent developments. The leading $\mathcal{O}(N^{3/2})$ and subleading $\mathcal{O}(N^{1/2})$ power-law contributions have been successfully matched with bulk supergravity calculations in the past~\cite{Drukker:2010nc, Jafferis:2011zi, Benini:2015eyy, Hosseini:2016tor,Azzurli:2017kxo, Zaffaroni:2019dhb,Bobev:2020egg,Bobev:2021oku,Bobev:2021wir,Bobev:2022nxr}, and the logarithmic term was matched in \cite{Bhattacharyya:2012ye}.
Logarithmic corrections in AdS\(_4\)/CFT\(_3\) have also been of interest   from the viewpoint of black-hole entropy and Sen's quantum entropy function~\cite{Sen:2008yk, Banerjee:2010qc,Banerjee:2011jp,Sen:2012dw,Sen:2012kpz,Sen:2012cj}, as well as from AdS\(_4\) gauged supergravity and its extensions~\cite{Liu:2016dau,Liu:2017vll,Liu:2017vbl,Gang:2019uay,Benini:2019dyp,PandoZayas:2020iqr,Hristov:2021zai,David:2021eoq,Karan:2022dfy,Karan:2024gwf,Jeon:2026rjy}.

We are interested in the universal logarithmic term \(\tfrac{1}{4}\log N\). 
Reproducing the result from the bulk has been a non-trivial test of the duality. In the 11D supergravity setting this was achieved in~\cite{Bhattacharyya:2012ye}, where the entire contribution was shown to originate from the zero mode of the 2-form ghost associated with the quantization of the 3-form potential. More recently, the same quantity has been studied from a purely four-dimensional perspective as a sum over the infinite tower of Kaluza--Klein modes of gauged \(\mathcal{N}=8\)
supergravity~\cite{Bobev:2023dwx}, with intriguing results.

On general grounds, up to discrete zero-mode contributions, logarithmic divergences on a \(D\)-dimensional space are controlled by the Seeley--DeWitt (SdW) coefficient \(b_D\), which vanishes for odd \(D\) and is only fully available in the literature for \(D\leq 6\) \cite{Vassilevich:2003xt,Gilkey:1975iq,Barvinsky:1985an}.  A successful strategy, pursued in several precision-holography setups with product geometry  \(\ads \times S\)~\cite{Buchbinder:2014nia, Giombi:2023vzu, Beccaria:2023ujc,Beccaria:2024gkq}, is to expand the higher-dimensional operators in harmonics on the compact factor, express the  resulting tower of AdS determinants by spectral \(\zeta\)-function methods \cite{Vassilevich:2003xt,Hawking:1976ja,Camporesi:1991nw,Camporesi:1993mz,Camporesi:1994ga,Camporesi:1995fb,Camporesi199457} and appropriately resumming the result. This effectively trades \(D\)-dimensional heat-kernel data for AdS\(_{d_A}\) spectral data combined with information on the spectrum of \(S\), both of which are available.
The resummation of the harmonic expansion is, however, subtle. The logarithmic divergence of the \(D\)-dimensional theory does not equal the naive sum of logarithmic contributions of the lower-dimensional determinants: on dimensional grounds, additional terms must arise from the interplay between the spectral \(\zeta\)-regularization of the individual AdS determinants and the sum over KK levels \(l\). Extracting these terms consistently, and understanding which prescriptions are compatible with the higher-dimensional answer, is the central technical problem addressed in this paper.

We provide a systematic framework for computing the logarithmically divergent part of one-loop partition functions on product spaces \(\ads_{d_A}\times S^{d_S}\) of arbitrary dimension, starting from the spectral representation of the AdS determinants and the known spectrum of the sphere. The prescription applies to fields of arbitrary spin and form degree, in both even- and odd-dimensional AdS factors, and reduces the \(D\)-dimensional log divergence to an explicit computation in \(d_A\) dimensions supplemented by a  specific large-\(l\) analysis of the KK sum. A key ingredient is the careful treatment of  zero modes of the sphere Laplacian that retain a non-trivial dependence on the AdS coordinates and behave as additional dynamical fields on the AdS factor. We refer to these as `AdS modes'; they arise   from  the transverse/longitudinal decompositions used to isolate irreducible representations with definite spectral data, and their proper inclusion is crucial for recovering the expected divergences.

We test the framework in several settings of increasing complexity  and   apply it to the full bosonic and fermionic content of 11D supergravity on \(\ads_4\times S^7\). In this case the individual KK contributions conspire so that the field-theoretic log divergences from different determinants cancel in both the bosonic and fermionic sectors, and the net \(\tfrac14 \log N\) coefficient originates entirely from a single zero mode of the 2-form AdS ghost, in full analogy with~\cite{Bhattacharyya:2012ye}.

Our calculation is conceptually analogous to~\cite{Bhattacharyya:2012ye}, whose result we recover, but the framework we propose generalizes the low-dimensional treatments of \(\ads \times S\) partition functions of~\cite{Buchbinder:2014nia,Giombi:2023vzu,Beccaria:2023ujc,Beccaria:2024gkq}. Our analysis is technically similar to the four-dimensional KK approach of~\cite{Bobev:2023dwx}; both approaches prescribe how to combine the AdS spectral representation with the harmonic sum, but they do so in inequivalent orders, see~\cite{Bobev:2023dwx,Adhikari:2026rfb} for related considerations and further details.

\subsubsection*{Structure of the paper} 

In section~\ref{sec::genHK} we discuss the technical novelty of this paper, namely the prescriptions for resumming the KK contributions to the determinant. In particular, 
in section~\ref{sec::HK} we review standard results on determinant representations of the logarithmic divergences for  one-loop determinants and in section~\ref{sec::KKsum} we present our prescription to recover the divergence on \(\ads \times S\) after harmonic expansion on the sphere factor on even-dimensional AdS. Section~\ref{sec::op4} provides further considerations on the treatment of fourth-order operators appearing in KK reductions. 
Finally, section~\ref{sec::path-integral-general} reviews the decomposition of fields into irreducible (transverse-traceless) representations on the sphere and the treatment of  the corresponding zero modes, which lead to determinant factors living in AdS alone (`AdS modes').

In section~\ref{sec:11d-sugra}, we apply our formalism to  11D supergravity on  $\ads_4 \times S^7$. 
Specifically, in section~\ref{sec:11d-sugra-partf} we construct  the partition function and introduce the gauge fixing terms required to gauge-fix the bosonic symmetries. Then,
in section~\ref{sec:quadratic_exp} we expand the 11D supergravity lagrangian to quadratic order in the fluctuations, using appropriate field decompositions   to isolate non-diagonal interactions.
The bosonic path integral is then evaluated in section~\ref{subsec:bosonic-1loop}, and the corresponding logarithmic divergences are discussed in section~\ref{sec:bosonic_resuts}. 
Section~\ref{sec::fermions} is devoted to the fermionic sector, where the analysis proceeds in close analogy to the bosonic case. 

We then gather and summarize our final results in section~\ref{sec::results}, before concluding in section~\ref{sec:conclusions} with some final comments and directions for future extensions.

Several appendices are included to supplement the main text with technical details and generalizations that extend beyond the specific $\ads_4 \times S^7$ framework. 
Appendix~\ref{appA} summarizes our conventions and notation and gives detailed formulae that we did not include in the main text.
In appendix~\ref{app:partf} we explicitly compute the partition functions for vectors, symmetric rank-two tensors, as well as two- and three-forms on $\ads_{d_A} \times S^{d_S}$.
In appendix~\ref{appendix:divergences} we evaluate the one-loop divergences of several cases from appendix~\ref{app:partf}, performing  explicit checks of our technique by comparing with the standard SdW coefficient results wherever available. This appendix also collects the explicit divergence results for the operators appearing in the 11D supergravity analysis of the main text as well as in closely related cases.
Finally, we discuss the case of odd-dimensional $\ads$  in appendix~\ref{app::adsodd}; we present the general strategy  to  calculate the corresponding divergences after harmonic expansion on the sphere factor, testing it on several examples and once again finding perfect agreement with SdW coefficient expectations.

\section{Functional determinants and heat kernel across dimensions}\label{sec::genHK}

\subsection{General remarks on determinant representations}\label{sec::HK}

The calculation of one-loop effects reduces to the quadratic expansion around classical solutions, resulting in the functional determinant of the differential operator of the quadratic fluctuation. In terms of the free energy \(\Gamma = -\log Z = S_\text{cl} + \Gamma_{(1)} + \ldots \)   we have
    \begin{equation}\label{ffd}
    \begin{aligned}
    \Gamma_{(1)} &= 
    \sum_{\text{fields}} (-)^F 
        \big(\tfrac{1}{2}\log \det{}' \Delta  + n_0    \beta_0\log \Lambda L \big)
    \\
    & = 
    \sum_{\text{fields}} (-)^F \Big[ 
                -B_D(\Delta)  + n_0 \hat    \beta_0   \Big]  \log \Lambda L 
    + \text{finite},
    \end{aligned}
    \end{equation} 
where \(F\) is fermion number, \(n_0\) is the number of zero modes of the operator \(\Delta\), \( \beta_0\) and \(\hat \beta_0\) are field-specific pure numbers, and \(B_D(\Delta) \) is the integral of the Seeley--DeWitt coefficient. In \eqref{ffd},   $\det'$ indicates the determinant of the operator after removal of the zero modes, which are separately accounted for by the term \(\sim n_0 \beta_0\).  
The sum runs over all fields (including ghosts) and we suppressed explicit dependence  of \(F, \Delta,n_0, \beta_0,\hat \beta_0\) on the fields. In \eqref{ffd}  we assume that there is no boundary term and we suppressed power law divergences that can be cancelled via local counterterms. At this stage \(L\) is an arbitrary regularization scale and \(\Lambda\) is a UV cutoff.
\(B_D(\Delta) \)   has the form
\begin{equation}\label{kag}   
B_D(\Delta)  = \frac{1}{(4\pi)^{\frac12 D} }  \int \dd{^D x \sqrt{g}   } b_D(\Delta) \,.
\end{equation} 
where \(b_D\) is the Seeley--DeWitt coefficient appearing in the heat kernel expansion \cite{Vassilevich:2003xt,Gilkey:1975iq}.
We also  have \(\hat \beta_ 0 = \beta_0 + m\), \(m \in \mathbb Z\) field-dependent, because \(B_D(\Delta)\) contains both zero and non-zero modes  \cite{Hawking:1976ja,Fradkin:1983mq,Tseytlin:2013jya,Bhattacharyya:2012ye}.

Alternatively, the determinant can be expressed in terms of the zeta function of the differential operator \cite{RAY1971145,Hawking:1976ja}
\begin{equation}\label{dsj}
\zeta_\Delta (z) =  \int  d \mu (v ) \frac{1}{\lambda(v)^z} 
\,,
\qquad
\log \det \Delta{}'
=- \zeta_\Delta'(0) -  2 \zeta_\Delta(0) \log \Lambda L \,,
\end{equation}
which goes over the spectrum of the operator \(\Delta\) and   becomes a discrete sum for compact spaces. With this analytic regularization one has
\begin{equation}\label{ffd2}
\begin{aligned}
    \Gamma_{(1)} & = 
        \sum_{\text{fields}} (-)^F 
            \big(\tfrac{1}{2}\log \det{}' \Delta  + n_0    \beta_0\log \Lambda L \big)
        \\ 
       & = 
    \sum_{\text{fields}} (-)^F \big(
                -\zeta_\Delta (0) + n_0 \beta_0 \big)  \log \Lambda L 
    - \sum_{\text{fields}}  \frac{(-)^F} 2\zeta_\Delta'(0).
\end{aligned}
\end{equation}
The spectral function provides therefore an explicit representation of finite and divergent parts of the effective action, but requires knowing the details of the spectrum of all operators involved. This is usually available on very special geometries, such as \(\ads\).

In \(\ads_{d_A}\),
for the particular cases of interest for this paper, namely second order operators acting on spin-\(s\) fields (transverse-traceless symmetric rank-\(s\) tensors) and transverse  \(p\)-forms, the spectral function \(\zeta(z)\) takes the form (cf.\ \eqref{baa},\eqref{cqc})
\begin{equation}\label{dfs}
\zeta_\Delta (z) 
	= 
	(L_A)^{{d_A} + 2z}
	c_{{d_A}}
	\intl_0 ^{+\infty} \! d v\,
	\frac{ \mu_{d_A}(v)}{( v^2 + \mathrm{b}^2  )^z},
\end{equation}
where  \(c_{d_A}\) and \(\mathrm b^2\) are \(s\)- (resp.\ \(p\)-) and \(d_{\! A}\)-dependent  constants.  The same is true for the spectral measure \(\mu_{d_{\!A}}(v)\), which has the schematic form  
 \begin{equation}\label{bae2}
\begin{aligned}
 \mu _{d_A}(v ) & =P _{d_A}(v)  \qquad \text{\(d_A\) odd}\,,
 \\ 
 \mu _{d_A}(v ) & =  P _{d_A}(v) \ 
   \tanh(\pi v )   \qquad \text{\(d_A\) even}\,,
\end{aligned} 
 \end{equation}
where \(P _{d_A}(v)\) is a polynomial in \(v\) with dependence on the field representation.

For   product spaces  \(\ads_{d_A} \times S^{d_S}\) of interest in this paper, no analogous closed-form spectral representation is available. However, one can  formally expand the higher-dimensional operator in spherical harmonics on \(S^{d_S}\), thereby reducing it to an infinite tower of \(d_A\)-dimensional operators, each labelled by a KK level \(l\) and carrying an \(l\)-dependent mass. Formally,
\begin{equation}\label{fjs}
\log \det \Delta = \sum_{l \geq 0} m_l \log \det \Delta(l)\,,
\end{equation}
where \(m_l\) is the multiplicity of the \(l\)-th sphere mode, which suggests identifying
\begin{equation}\label{fps}
\zeta_\Delta(z) = \sum_{l\geq 0} m_l\, \zeta_{\Delta(l)}(z)\,.
\end{equation}
Each \(\zeta_{\Delta(l)}\) is of the form~\eqref{dfs} and can in principle be handled by the spectral methods reviewed above. The question, then, is how to reproduce the higher-dimensional result from  the infinite KK sum combined with the analytic continuation \(z\to 0\).

\subsection{Regularization and resummation of the KK spectrum} \label{sec::KKsum}

Using the decomposition~\eqref{fjs},\eqref{fps}, we write (\(D=d_A + d_S\))
\begin{equation}\label{saf}
\Gamma_D = \Gamma_{D,\text{fin}} + \hat \Gamma_D \log \Lambda L 
\,,\qquad
 \hat \Gamma_D \log \Lambda L
= \Big[\sum_{l\geq 0} m_l\, \Gamma_l\Big]_{\log{\Lambda L}}
\,, \qquad
\Gamma_l = \Gamma_{l, \text{fin}} + \hat \Gamma_l \log \Lambda L \,,
\end{equation}
which makes manifest that the \(D\)-dimensional divergence has two potentially distinct origins: the divergences \(\hat\Gamma_l\) already present at each KK level (which vanish whenever \(d_A\) is odd), and divergences induced by the infinite sum over \(l\) of the   finite terms \(\Gamma_{l,\text{fin}}\).  We shall argue that these are relevant for the cases \(d_A\) even and \(d_A\) odd respectively.

For even \(d_A\), the determinants in AdS space are generically divergent and
\begin{equation}\label{sbf} 
\hat \Gamma _D 
= \sum_ {l\geq 0} m_l\, \hat \Gamma_{l } 
= - \sum_ {l\geq 0} m_l\,  \zeta_{\Delta(l)}(z)
\,, \qquad
z \to 0\,.
\end{equation}
The direct evaluation of \( \zeta_{\Delta(l)}(0)\) with subsequent \(\zeta\)-function regularization of the sum in \(l\) cannot yield correct results, as it is clear on dimensional grounds. Indeed, for an operator \(\Delta=  - \cd^2 + \kappa\) with \(\kappa\) constant, one expects powers of \(\kappa\) from \(0\) to \(( d_A + d_S )/2\), but, using this prescription, only powers up to \(d_A/2\) come from \(\zeta_{\Delta(l)}(0)\). Furthermore, the \(\zeta\)-function regularization does not introduce new dimensionful parameters. Indeed, it requires an ad-hoc regularization, as  discussed  in~\cite{Bobev:2023dwx}. 
Note that, in writing~\eqref{sbf} we have implicitly assumed that, for even \(d_A\), no additional divergence arises from the sum over \(l\) of the finite parts; we have verified this in all examples we considered, but we do not have a general proof.

The problem at hand is therefore to find a prescription for the regularization  of the single \(l\)-dependent terms in \eqref{fjs}. Such a regularization should allow one to recover the terms of appropriate dimensionality.
Particular examples of such a procedure have been produced in low-dimensional cases in
\cite{Buchbinder:2014nia,Giombi:2023vzu,Beccaria:2023ujc}. Here we show that their method can be generalized to spaces of larger dimensionality and can be extended to fields living in  other representations relevant for precision holography.

For concreteness we refer to the case of a (transverse-traceless) spin-\(s\) field, but the same discussion applies almost identically to a transverse \(p\)-form, see \eqref{baa},\eqref{cqc}. Assuming an operator  \(\Delta = -  \cd_A^2 -\cd^2_S + \kappa \) (note that the radii enter implicitly via \eqref{paa},\eqref{pab} etc.), expanding \(-\cd^2_S\) in spherical harmonics we have 
\begin{equation}\label{dfs2}
\zeta_\Delta (z) 
	= 
	(L_A)^{{d_A}+2z} c_{{d_A}} \sum_{l \geq 0} m^{s,{d_S}}_l
	\intl_0 ^{+\infty} \! d v\,
	\frac{ \mu_{d_A}(v)}{( v^2 + b^2_l )^z}
	\,, 
	\qquad 
	b^2_l  = \left(\frac{{d_A}-1}{2}\right)^2 + s + L_A^2\left ( \kappa + \frac{\lambda^{s,{d_S}}_l}{L_S^2} \right),
\end{equation}
where \(\lambda^{s,{d_S}}_l\) and \(m^{s,{d_S}}_l\) are the eigenvalues and multiplicity of the Laplacian on the sphere \(S^{d_S}\). 
They have generically the form 
\begin{equation} \label{redef_l-to-X}
    \lambda^{s,{d_S}}_l = \hat a \, l^2 + \hat b \, l + \hat c 
    \equiv \hat a \left(l + b \right)^2 + \hat c - b^2
    \,,\qquad\qquad
    m^{s,{d_S}}_l
     = \sum_{k = 0}^{{d_S}-1} a^{s,d_S}_{k} (l + b   )^k
    \,, \qquad
     b = \frac {\hat b} {2\hat a} >0\,,
\end{equation}
where  the multiplicities are polynomials of order ${d_S}-1$ in $l$ (or equivalently of \(l+b\), which will be  convenient later). Note that \(b\) is always an integer or half-integer.

Here we focus on the case \(d_A\) \emph{even}. In close analogy to~\cite{Buchbinder:2014nia}, we split  \( \tanh(\pi v) = 1- 2 (e^{2\pi v}+1)^{-1}\) and consider 
\begin{align}\label{dsf}
\zeta_1(z) & =
	(L_A)^{{d_A}+2z}  c'_{{d_A}}
    \sum_{l\geq 0} m_{l}^{s,{d_S}}
	\intl_0 ^{+\infty} \! d v \, v 
	\frac{v ^2+\rho^2}{( v ^2 + b_l^2)^z} 
\prod_{j=\frac12}^{ \frac{{d_A}-5}2} 
	 {(v ^2 + j^2)} 
\\*\nonumber &
	 \equiv  
	    (L_A)^{{d_A}+2z}  \sum_{l\geq 0} m_{l}^{s,{d_S}}
	 	\intl_0 ^{+\infty} \! d v  \
	 	{( v ^2 + b_l^2)^{-z}}
	 \sum_{j=0}^{ \frac{{d_A}-2 }2}  q^{(1)}_{j} v ^{ 2j+1 },
\\
\zeta_2(z) & = 
    - 2 (L_A)^{{d_A}+2z}  c'_{{d_A}}
    \sum_{l\geq 0} m_{l}^{s,{d_S}}
	\intl_0 ^{+\infty} \! d v \, 
\frac{v }{1+e^{2\pi v }}
	\frac{v ^2+\rho^2}{( v ^2 + b_l^2)^z} 
\prod_{j=\frac12}^{ \frac{{d_A}-5}2}
	 {(v ^2 + j^2)} 
\\*\nonumber &	 	 \equiv  
	 	    (L_A)^{{d_A}+2z}  \sum_{l\geq 0} m_{l}^{s,{d_S}}
	 	 	\intl_0 ^{+\infty} \! d v  \
	 	 	\frac{{( v ^2 + b_l^2)^{-z}}}{1+e^{2\pi v }}
	 	 \sum_{j=0}^{ \frac{{d_A}-2 }2}  q^{(2)}_{j} v ^{ 2j+1 },
	\\ 
	\zeta_\Delta (z) 
	& =
	\zeta_1 (z)+ \zeta_2(z)
	\,,\qquad
	c'_{d_A} = \frac{\pi }{(2^{{d_A}-2}\Gamma\left[\frac{d_A}{2}\right])^2} c_{{d_A}}
	\,,\qquad
	\rho =   s + \frac {{d_A}-3}2 \, ,
	\end{align}
where \(q^{(1)}_j,q^{(2)}_j\) are \(l\)-independent coefficients.

In \(\zeta_1\), the integral in \(v\) can be done term by term using
\begin{equation} \label{fpd}
    \intl_0^\infty dv  \, \frac{v ^p}{(v ^2 + A)^z} 
    = A^{\frac{1}{2} p+\frac{1}{2} - z} 
    \frac{\Gamma  (\frac{p+1}{2} ) \Gamma  (z -\frac{1}{2}p-\frac{1}{2} )}{2 \Gamma (z)},
\end{equation}   
and the general result is  
\begin{equation}\label{fdn}
\zeta_1(z) = 	(L_A)^{{d_A}+2z}      \sum_{l\geq 0} m_{l}^{s,{d_S}}
\sum_{j=0}^{ \frac{{d_A}-2 }2}  q^{(1)}_{j} \,
	 	b_l^{2 (j+1  - z)} 
	 	    \frac{\Gamma  ( j+1 ) \Gamma  (z -j -1)}{2 \Gamma (z)}.
\end{equation}   
We now consider  the \(l\)-dependence.
Writing \(b_l ^2= a ( l + b   )^2 + c\)  (so that e.g.\  \(b = \hat b /2\hat a\) in \eqref{redef_l-to-X}),
\begin{equation}\label{sum-b_l^r}
    (b_l^2 )^r  
    = a^r (l + b)^{2 r} \Big[ \frac{c}{a (l + b)^2}+1 \Big]^r
    = a^r (l + b)^{2 r} \sum_{n\geq 0}    
    	\frac{ (-)^n  c ^n \Gamma (n-r)  }{a^n \Gamma (n+1) \Gamma (-r)}
    	(l + b)^{-2n}\,,
\end{equation}  
replacing in \(\zeta_1\) and  inserting the generic expression for the multiplicity  \eqref{redef_l-to-X}
\begin{equation}
\begin{aligned}
\zeta_1(z) = 	 (L_A)^{{d_A}+2z}     \sum_{n\geq 0}    \sum_{l\geq 0}\sum_{k = 0}^{{d_S}-1} 
\sum_{j=0}^{ \frac{{d_A}-2 }2}  (-)^n q^{(1)}_{j} a^{ j - z-n+1 }
				&	 a_{k} c ^n \, 
	\frac{ 
				 \Gamma  ( j+1 ) 
	}{  
					2 
					 \Gamma (n+1) 
	 }  
 \\
 &  {}\times 
	 \frac{   		\Gamma (n +  z - j - 1  )  	 	}{    \Gamma (z)        }     (l + b)^{2 (j - z +1) -2n  + k } 
     \,.
\end{aligned}
\end{equation}    
We regularize  the sum over $l$ via Hurwitz \(\zeta\)-function  \(\zeta_\text{H}\) with \(\sum_{l \geq 0} (l + b )^p = \zeta_{\text H} (-p,b)\)
    \begin{equation}\label{fnh}
        \begin{split}
            \zeta_1(z) = 	 (L_A)^{{d_A}+2z}     \sum_{n\geq 0}  \sum_{k = 0}^{{d_S}-1} 
            \sum_{j=0}^{ \frac{{d_A}-2 }2} (-)^n q^{(1)}_{j}a^{ j - z-n+1 }  \, 
                & a_{k}\, c ^n
                \frac{  
                                \Gamma  ( j+1 ) 
                }{ 
                                2 
                                \Gamma (n+1)
                }  
            \\
                &\times 
                \frac{
                                \Gamma (n +  z - j - 1  ) 
                }{
                                \Gamma (z) 
                }
                \zeta_{\text H}(2 z + 2n -2j - k -2 ,b)
                \,.
        \end{split}
    \end{equation} 
We now wish to understand the behavior of this expression as $z \to 0$ before performing the sum over $n$. Given the overall factor $\Gamma(z)^{-1} = z + \mathcal{O}(z^2)$, we can obtain a non-zero finite result only when this simple zero is compensated by a simple pole from the numerator. There are two possible origins for such a pole:
\begin{itemize}
    \item[(i)] \(n  - j -1 = -m\) for $m = 0, 1, 2, \dots$, arising from the \(\Gamma\) function,
    \item[(ii)] \(2n - 2 j    - k -2= 1\), arising from the pole of  \(\zeta_{\text H}\).
\end{itemize}
These two conditions cannot be satisfied simultaneously, as this would imply  $k = 2(1-m) - 3 = -2m - 1 \le -1$.  
This guarantees the absence of double poles, allowing us to obtain a well-defined result for the sum over $n$. Furthermore, the conditions (i) and (ii) above
are satisfied for finitely many values of the indices, hence we obtain a well-defined result for   the sum over \(n\), which, explicitly,   goes from \(0\) to \( \left\lfloor\frac12({d_S}+{d_A})\right\rfloor \). 

Note that the dimensionful parameter \(\kappa\), which plays the role of a squared mass, sits linearly inside \(c\) (recall \(b^2_l = a ( l + b   )^2 + c\) after \eqref{fdn}). For even \({d_S}\), the result in \eqref{fnh} therefore fits with the expectation that,
on general grounds from the full \(\ads_{d_A} \times S^{d_S}\) perspective, the divergent part of the effective action should be a polynomial of degree at most \(({d_S}+{d_A})/2\) in \(\kappa\) (with dimensionful coefficients given by powers of \(L_A\), \(L_S\)).
For odd \({d_S}\), hence odd  total dimension \({d_A}+{d_S}\), one expects the divergence to vanish, however this requires cancellations among different terms, as we shall comment later on in appendix~\ref{appendix:divergences}.

For \(\zeta_2\), we expand the \(z\)-dependent factor as in \eqref{sum-b_l^r}, up to the  shift  $c \to   c + v^2$, to obtain
\begin{equation}\label{fdj}
\zeta_2(z) =  (L_A)^{{d_A}+2z}  \sum_{n\geq 0}\sum_{l\geq 0} \sum_{k=0}^{{d_S}-1} \sum_{j=0}^{ \frac{{d_A}-2 }2} 
			( -)^n a_k^{s,d_S} a^{-z-n}    
				 	   q^{(2)}_{j}  
	 	 	\intl_0 ^{+\infty} \!\! d v  \
	 	 	\frac{  v ^{ 2j+1 }  (v^2+c )^n  }{1+e^{2\pi v }}
    	\frac{ \Gamma (n+z)  }{\Gamma (n+1) \Gamma (z)}
    	(l + b)^{k-2n - 2z }\,.
\end{equation}
In this case, in contrast to \(\zeta_1(z)\), we have a \(z\)-independent exponential damping of  \(v\) in the integrand, so we can commute the discrete sums and the limit $z \to 0$ inside the integral.
We again perform  Hurwitz \(\zeta\)-function regularization for the sum over \(l\) obtaining
\begin{equation}\label{fdj2}
\zeta_2(z) =  
	 	 	(L_A)^{{d_A}+2z}  \intl_0 ^{+\infty} \! \!d v  
	 	 	 \sum_{n\geq 0}\sum_{k=0}^{{d_S}-1} \sum_{j=0}^{ \frac{{d_A}-2 }2} 
			( -)^n a_k^{s,d_S} a^{-z-n}    
				 	   q^{(2)}_{j}  
	 	 	\frac{  v ^{ 2j+1 }  (v^2+c )^n  }{1+e^{2\pi v }}
    	\frac{ \Gamma (n+z)  }{\Gamma (n+1) \Gamma (z)}
    	\zeta_{\text H}(2n + 2z - k,b)\,.
\end{equation}
As before, to obtain a non-zero result as $z \to 0$, we must compensate the zero of $\Gamma(z)^{-1}$ with a pole. This can arise either from $n=0$ (where $\Gamma(n+z)/\Gamma(z) = 1$) or from the pole of $\zeta_{\text{H}}$ when $2n - k = 1$, in analogy with conditions (i) and (ii) above. 
%The latter implies that $k$ must be odd also in this sector,     \(k = 2n-1 \).  RELEVANT BECAUSE?
This limits the non-vanishing contributions to $n$ ranging from $0$ to $\left\lfloor \frac{1}{2}d_S\right\rfloor$. The remaining integrals over $v$ can then be evaluated using the standard identity
\begin{equation}\label{fsn}
\intl_0^{+\infty}\! \!dv  \frac{v^p}{e^{2\pi v} +1 } =
\left(1-2^{-p}\right) (2 \pi )^{-p-1} \zeta (p+1) \Gamma (p+1)\,.
\end{equation}

We conclude this discussion with a comment on the comparison between our procedure and the approach of \cite{Bobev:2023dwx}. In that case, the sum over \(l\) is effectively postponed and the \(z\to 0\) limit is taken for fixed \(l\). As a result, only the poles of type (i) are present in \(\zeta_1\) and only the \(n=0\) term is present in \(\zeta_2\). Dimensionally, indeed, this restricts the divergence to a polynomial of degree \({d_A}\) in the mass  (correctly with the \(\ads_{d_A}\) expectation). It may be therefore possible to formally understand the particular regularization advocated in~\cite{Bobev:2023dwx} with the additional contributions from \eqref{fnh},\eqref{fdj2}, which might hint to a physical justification of that procedure.

The situation is  different for odd \(d_A\),  where \(\hat \Gamma_{l } =0 \) in \eqref{saf}, but divergences in \(D\) dimensions  arise from the infinite sum of the finite contributions \( \Gamma_{l , \text{fin} }  \),
\begin{equation}\label{svf} 
  \Gamma _D 
= -\sum_ {l\geq 0} m_l\,  \zeta_{\Delta(l)}'(z) 
\,, \qquad
z \to 0\,.
\end{equation}
The integrals in \eqref{dfs2} can be done immediately, and the result can be expanded in a large-\(l\) power series. The subsequent sum over \(l\) may be divergent if a term \(l^{-1}\) is present;  regularizing \(\sum 1/l \sim \log l_{\text {max}} \sim \log \Lambda L\) reproduces the divergence.   This is discussed in detail in appendix~\ref{app::adsodd}.

\subsection{Fourth-order operators factorizing on \texorpdfstring{\(S^{d_S}\)}{S}}\label{sec::op4}
 KK reductions often involve higher-order operators that do not properly factorize to local quadratic operators on generic \(\ads \times X\) backgrounds but do so for the specific case of \(\ads \times S\), namely have the structure 
\begin{equation}\label{opf}
\Delta  =\nabla^4 - 2\kappa \nabla^2   + A^2\nabla_S^2 - A^2   C^{\mathcal  R}_{d_S} + \kappa^2
=\Delta_{+}\Delta_{-}
\,,\qquad
\Delta_{\pm} = -\nabla^2 + \kappa \pm  A \sqrt{-\nabla_S^2 +  C^{\mathcal  R}_{d_S}  }\,,
\end{equation}
where \(A\) is a constant and  \(C^{\mathcal  R}_{d_S} \) is the particular value that, upon substitution of the sphere eigenvalues, allows one to simplify the square root. This latter constant depends on the sphere parameters and field type (denoted by its representation \(\mathcal R\)).
Explicitly, for spin \(s=r\)  transverse-traceless  fields   or transverse \((p=r)\)-forms   it is
\begin{equation}\label{feb}
C^{r{\perp}}_{d_S}  = \frac{1}{4\,L_S^2}\left(d_S^2-2d_S+4r+1\right) \,,
\end{equation}
cf.\ \eqref{eq:eigenvaluesSphericalHarmonicsforLaplacian},\eqref{eq:eigenvaluesSphericalHarmonicsPformSymmetric}.

Using the specific eigenvalues of the Laplacian on \(S^{d_S}\), the factors become
\begin{equation}
\Delta_\pm  
= -\nabla_A^2  + \kappa +  \hat a \, l^2 + \hat b_\pm \, l  + \hat c_\pm
= -\nabla_A^2  + \kappa +  \hat a (l + b_\pm)^2 + c_\pm  
\,,\qquad
b_\pm = \frac { \hat b_\pm} { 2 \hat a},
\end{equation}
where \(\hat b_\pm \), \(\hat c_\pm \), \(b_\pm\), \(c_\pm\) are  different constants. Compared to the discussion around \eqref{redef_l-to-X}, the operators \(\Delta_\pm\) formally contain modified  eigenvalues of the sphere Laplacian, which however remain quadratic in \(l\). 
The procedure discussed in section~\ref{sec::KKsum} to evaluate the divergence of the determinant can then be formally followed, but there is an important caveat: \(\hat b_-\) can be  a negative integer, and this happens in relevant cases, such as the one presented in this paper. This leads to sums of the form 
\(\sum_{l \geq 0} (l - n )^k\), which contain divergent contributions for negative \(k\).
We regularize them via 
\begin{equation}\label{fjf}
\sum_{l \geq 0} (l + \beta )^k = \zeta_{\text H} (-k,\beta)
\qquad
\forall \beta.
\end{equation}
It is important here to pay attention to the correct branch cut choice in the analytic continuation of the zeta function \eqref{fjf}. In particular, \(\zeta_\mathrm H(-1,-1) = -\frac{13}{12}\) etc., which is implemented in Mathematica by the function \texttt{HurwitzZeta}. 
This works in the case we discuss in this paper and we tested it on some other example given in appendix~\ref{subsec:zRegScalar}.

Finally, note that on \(\ads_{d_A} \times S^{d_S}\)
the particular operator \(\Delta\) appearing in \eqref{opf} is also special because it can be cast in a fully \((D = d_A + d_S)\)-covariant form. As discussed in  \cite{Beccaria:2024gkq} for the \(\ads_3\times S^3\) case, since the two factors are Einstein spaces, one can introduce the projector on the second one 
\begin{equation}\label{edn}
P_{MN} =  a \left(g_{MN} + b R_{MN}\right), \qquad a = \left(1 + \frac{L_A^2}{L_S^2}\frac{d_S-1}{d_A-1}\right)^{-1} \, ,\quad b = \frac{L_A^2}{d_A-1}\, ,
\end{equation}
which satisfies 
\begin{equation}
    P_{\mu\nu} = 0\, ,\qquad P_{mn} = g_{mn}\, ,
\end{equation}
so that 
\begin{equation}\label{opg}
\begin{aligned}
\Delta 
 & =  
\nabla^4 + V _{MN} \nabla^M\nabla^N - A^2   C^{\mathcal  R}_{d_S} + \kappa^2
\,,
\qquad  V _{MN} =  A^2 P _{MN}  - 2 \kappa g_{MN}.
\end{aligned}
\end{equation}
Written in this form, \(\Delta\) is thus amenable to the heat-kernel treatment for higher-order operators developed e.g.\ in~\cite{Gusynin:1988zt,Casarin:2019aqw,Casarin:2021fgd,Casarin:2023ifl,Barvinsky:1985an,Fradkin:1981iu,Avramidi:2000bm}.  
As a result, we can leverage the standard facts that there should be no logarithmic divergence in odd \(D\) and that  in even \(D\) the dependence  of the logarithmic divergence on the various parameters must obey definite polynomiality and dimensionality constraints.
We emphasize that these considerations are \emph{not} valid for the single factors \(\Delta_\pm\). In several examples we considered, the logarithmic divergences of their determinants are indeed more  complicated but simplify when combining them together to obtain the result for the fourth-order operator \(\Delta\).

\subsection{Irreducible  decompositions and zero modes on the sphere}
\label{sec::path-integral-general}

To apply spectral techniques one usually   decomposes the fields into their transverse, longitudinal, and trace components,  as   the spectral functions are defined only for fields transforming in irreducible representations. However, this is not inconsequential, as we discuss in this section.

The splitting is a field redefinition with differential operators, which introduces non-trivial Jacobians. In the example of a vector field (cf.~\eqref{det1}),
\begin{equation}\label{dns} 
A_M = A^\perp_M + \cd_M \sigma
\,, \qquad \cd^M A^\perp_M =0
\,, \qquad
J = \Big[  \int\! \DD \sigma e^{-\int \sigma[-\cd^2]\sigma }\Big]^{-1}
=  \det_0 [-\cd^2] ^{\frac12 }\,.
\end{equation}
The Jacobian \(J\) is computed via the standard procedure (cf.\ e.g.\ \cite{Casarin:2024qdn} for complementary considerations)
    \begin{equation}\label{fdsf}
    1  {}= \int \DD A e^{-\int A^M A_M}
    = J  \int \DD A^\perp e^{-\int A_\perp ^M A^\perp_M}
    \int \DD \sigma e^{-\int \sigma(-\cd^2)\sigma }
    = J\,  \det\limits_0 [-\cd^2] ^{\frac12 }\,.
    \end{equation}
Consider the partition function
\begin{equation}\label{dnt}  
\begin{aligned}
Z = e^{-\Gamma}& = \int \! \DD A e^{-\int A^M \Delta A_M}  = \det_1 [\Delta] ^ {-\frac12}
\,,
\\
[\Gamma]_{\log \Lambda L}  &= \frac 12 \big[\log \det_1 \Delta\big]_{\log \Lambda L}  
= 
   - B_D(\Delta)   \log \Lambda L \,,
\end{aligned}
\end{equation}
where \(\Delta\) is a self-adjoint differential operator.
We can alternatively compute this partition function via the change of variable \eqref{dns}
\begin{equation}\label{djw}
Z
= J \int \! \DD {A^\perp} \DD \sigma  
			e^{-\int A^\perp_M \Delta A_\perp ^M} 
			e^{-\int \sigma  \hat\Delta (-\cd^2)  \sigma}
= 
\det_{1 \perp} [\Delta] ^{- \frac 12 }
\det_{0 }  [\hat\Delta] ^{- \frac 12 }  ,
\end{equation}
where the Jacobian cancels a contribution from the integral in \(\sigma\) (this is typical). 
One expects the relation 
\begin{equation}\label{www}
    B_D(\Delta) \log \Lambda L = -\frac{1}{2}[\log \det_{1\perp} \Delta + \log \det_0 \hat \Delta] _{\log \Lambda L} = \left(\zeta_\Delta(0) + \zeta_{\hat \Delta}(0)\, \right) \log \Lambda L .
\end{equation}

However, as explained \eg\ in~\cite{Christensen:1979iy,Fradkin:1983mq,Tseytlin:2013fca}, the decomposition in \eqref{dns} entails   zero modes, which correct the expected relation to 
\begin{equation}\label{dnv}
 B_D(\Delta) = \zeta_\Delta(0) + \zeta_{\hat \Delta}(0) - k\,, 
\end{equation}
where \(k\) is the number of zero modes on \(S^D\) introduced by the decomposition \eqref{dns}.
These modes are absent in \(\ads\) (hyperbolic space) because they are non-normalizable.

However, when working with product spaces and zero modes, there are further consequences.
In this work, we are interested  in \(\ads_{d_A} \times S^{d_S}\) and we split the indices in the two factors,  \(A_M \to (A_\mu , A_m) \), retaining only \(d_A\)- and \(d_S\)-dimensional covariance separately.  We introduce longitudinal and transverse components with respect to the single factors  (cf.~\eqref{det2})
\begin{equation}\label{pdg}
\begin{aligned}
A_\mu &= A^\perp_\mu + \cd_\mu \sigma
\,, \qquad& 
 \cd^\mu A^\perp_\mu & =0
\,,\qquad &
J_A &{}= \Big[  \int\! \DD \sigma e^{-\int \sigma[-\cd_A^2]\sigma }\Big]^{-1}
=  \det_{0,0} [-\cd_A^2] ^{\frac12 }
\,,
\\
A_m &= A^\perp_m + \cd_m \rho
\,, \qquad &
\cd^m A^\perp_m& =0
\,,\qquad &
J_S &= \Big[  \int\! \DD \rho e^{-\int \rho[-\cd_S^2]\rho }\Big]^{-1}
= \det_{0,0} [-\cd_S^2]^{\frac12 }\,.
\end{aligned}
\end{equation}
We now consider the analogous partition function from \eqref{dnt}, where
\begin{equation}\label{dsp}
\begin{aligned}
A^M \Delta_1 A_M &{}= 
A^\mu \Delta_{1,0} A_\mu + 
A^m \Delta_{0,1} A_m  
\\ &{}= 
A^\mu_\perp \Delta_{1,0} A_\mu^\perp + 
A^m_\perp \Delta_{0,1} A_m^\perp +
\sigma \hat \Delta_{0,0}  [-\cd^2_A] \sigma + 
\rho \tilde \Delta_{0,0} [-\cd^2_S] \rho \,,
\end{aligned}
\end{equation}
with some other differential operators,
and we emphasized in the subscript the \((\ads , S)\) index structure of the fields on which the operator acts on. On the one hand, the evaluation \eqref{dnt} is still valid; on the other hand, proceeding as in \eqref{djw} with \eqref{pdg} gives
\begin{equation}\label{dms}
Z = J_A \, J_S \,
\det_{1 \perp,0} [\Delta_{1,0}]^{- \frac 12 } \,
\det_{0,1 \perp} [\Delta_{0,1}]^{- \frac 12 }
\int \!   \DD \sigma \DD \rho   
			e^{-\int \sigma  \hat\Delta_{0,0} [-\cd^2_A]  \sigma}
			e^{-\int \rho  \tilde\Delta_{0,0} [-\cd^2_S]  \rho}\,.
\end{equation}
This requires some discussion due to the fact that \(-\cd^2_S\) appearing on the \(\rho\) action induces zero modes. 
In contrast to the pure sphere case where the zero modes are constants and amounts to the shift \eqref{dnv}, here we have fields with \(\ads\) and \(S\) dependence. Zero modes of  \(-\cd^2_S\) that still retain an arbitrary dependence on \( \ads\) coordinates are therefore annihilated by the kinetic operator.  In the present case \eqref{pdg}-\eqref{dms} they are of the form  \(\rho (x_A,y_S) = f(x_A)\), which does not change \(A_m\); this is effectively a gauge ambiguity in the definition of \(\rho\) in \eqref{pdg}. We refer to them as ``\(\ads\) modes''.

We need to account for the presence of these \(\ads\) modes that are absent in the original expression of the functional integral. These modes live in the lower-dimensional \(\ads\) space only, and modify the evaluation of the path integral as
\begin{equation}\label{dfw}
\int \!   \DD \rho    
			e^{-\int \rho  [\tilde\Delta_{0,0}] [-\cd^2_S]  \rho}
=  \frac{
			     \det\limits_{0,\emptyset} [\tilde\Delta_{0,0}]   ^{\frac12}  
  }{   
  	     \det\limits_{0,0} [\tilde\Delta_{0,0}]   ^{\frac12}   	
		     \det\limits_{0,0}  [-\cd^2_S]  ^{\frac12}   
  } \, ,
\end{equation}
where the notation \({0,\emptyset}\) emphasizes that this determinant is acting on fields which depend only on \(x_A\) while the \(y_S\) dependence has been dropped. Furthermore, in the present example, the \(\ads\) modes satisfy \(-\cd^2_S f(x_A) =0 \). Thus, if \( \tilde\Delta_{0,0} = -\cd^2 + \kappa = -\cd^2_A - \cd^2_S+ \kappa \), the operator in the numerator of \eqref{dfw} is \( \tilde\Delta_{0,0} f(x_A)  = (-\cd^2_A  + \kappa )f(x_A)  \).

The integral in \(\sigma\) can be  done immediately, as there is no normalizable zero mode from \(\ads\). Completing the evaluation of \eqref{dms} we thus have  (as usual, \(J_A,J_S\) cancel out)
\begin{equation}\label{rsfe}
Z =  
\frac 1
{  \det\limits_{1 } [\Delta_{1}  ]^{\frac 12 }}
=
\frac{
				 \det\limits_{0,\emptyset} [\tilde\Delta_{0,0}]^{\frac12}  
  }{   
    \det\limits_{1 \perp,0} [\Delta_{1,0}] ^{\frac 12 } 
    \det\limits_{0,1 \perp} [\Delta_{0,1}] ^{ \frac 12 } 
	 \det\limits_{0,0}  [\hat\Delta_{0,0}]^{\frac12}   	 
	 \det\limits_{0,0} [\tilde\Delta_{0,0}]^{\frac12}   	 
  } \,.
\end{equation}

Jacobian determinants as in \eqref{dns} are evaluated via functional integrals and these considerations apply to them as well. Consider for example the following decomposition, relevant later on, 
\begin{equation}\label{asd}
 h_{\mu m}  = h^{\perp}_{\mu m} + \cd_ m q^\perp_\mu + \cd_\mu q_m^\perp + \cd_m \cd_\mu \, q.
\end{equation} 
The scalar \(q\) induces a nontrivial \(\ads\) mode in the Jacobian determinant as well. Indeed,
the Jacobian factor related to the scalar \(q\) contains the operator \([-\cd^2_A][-\cd^2_S]\), so that the correct evaluation is
\begin{equation}\label{fdo}
J_q = \Big[  \int\! \DD q e^{-\int q[-\cd^2_A][-\cd^2_S]q }\Big]^{-1}
=\frac{ \det\limits_{0,0}[-\cd^2_A][-\cd^2_S]^\frac12}{\det\limits_{0,\emptyset}[-\cd^2_A]^{\frac12} } .
\end{equation}

Note that, although in this section we focused on scalar \(\ads\) modes for ease of exposition, they can live in an arbitrary representation, and the contribution has to be weighted with the respective multiplicity.

Another important case is that of a field \(\varphi\) defined on \(\ads \times S\) but always appearing under the action of \(-\cd^2_S\). A non-singular field redefinition of the type \(\varphi = F\phi \) naively brings a Jacobian factor \(\det _ {0,0} (F \equiv (F\phi)') \) (we restrict to linear \(F\) for simplicity).  
    However,  considering \(-\cd^2_S\) as part of the metric in field space, we proceed in a manner similar to \eqref{fdsf}, 
    \begin{equation}\label{fdk}
  \begin{aligned}
    \det\limits_{0,0}[-\cd^2_S] ^{-\frac12} 
    & = \int \!\DD{\varphi} e^{-\int \varphi[-\cd^2_S] \varphi } 
    = J  \int \!\DD{\phi}  e^{-\int \phi F^2[-\cd^2_S] \phi } \\
    &
    = J    \frac{    \det\limits_{0,\emptyset}F   }
            {  \det\limits_{0,0}F  \,
            \det\limits_{0,0}[-\cd^2_S]^{\frac12} }
    \qquad \implies \qquad
    J  = \frac{ \det\limits_{0,0}F  } {    \det\limits_{0,\emptyset}F   }
  \end{aligned}
    \end{equation}
    correcting the naive expectation with an \(\ads\) mode term.
    An example of this setting is  \(S = \int \varphi \Delta [-\cd^2_S] \varphi\) where \(\Delta\) is a regular operator (i.e.\ without zero modes); only properly accounting for  \eqref{fdk} does the result for \(F = \Delta^{-1}\) agree with the direct evaluation of the path integral on \(\varphi\) via \eqref{dfw}.

    We tested formulae of the type \eqref{rsfe},\eqref{fdo} in several cases, and the appropriate inclusion of the \(\ads\)-mode contribution is always key in recovering  expected results, see appendix \ref{appendix:divergences}. 
    Note the particular case \(d_A\) even and \(d_S\) odd, where from the odd total dimension  \(d_A + d_S\) one expects no divergence, but the \(\ads\) modes living in the even dimensional subspace bring generically nonzero contributions. Indeed, we observe that fields with transverse traceless indices on \(S\) generically give nonzero divergences when evaluated following the procedure outlined in section~\ref{sec::KKsum}.

\section{11D supergravity on \texorpdfstring{\(\ads_4 \times S^7\)}{AdS4 x S7}} \label{sec:11d-sugra}
\subsection{Structure of the 1-loop partition function}\label{sec:11d-sugra-partf}
We now apply the framework developed in  the previous section to 11D supergravity on $\ads_4\times S^7$ \cite{Sezgin:1983ik,Biran:1983iy, Casher:1984ym,Castellani:1984vv,Duff:1986hr}. We address the bosonic and fermionic sectors in turn. In both cases the local KK contribution to the logarithmic divergence is found to vanish, as expected from the odd total dimensionality. The full one-loop result therefore reduces to the normalizable zero-mode contribution of the 2-form ghost, which we recall below. 

We start from the 11D supergravity Lagrangian \cite{Cremmer:1978km,Freedman:2012zz}
\begin{align}
\label{uaa}
e^{-1} L _{\text{B}} &=\frac{1}{4} R - \frac 1 {48}  F_{MNRS} F^{MNRS}+ \frac{2}{12^4}   \varepsilon^{M_1\dots M_{11} } F_{M_1 \dots M_4} F_{M_5 \dots M_8 } A_{M_9 \dots M_{11}},   
\\
\label{eq:Lgravitino1}
e^{-1}L _{\text{F}}& = -\frac{1}{2} \Psib_M \Gamma^{MNP}\nabla_N \Psi_P + \frac{1}{96}\left[\Psib_M\Gamma^{MNXYWZ}\Psi_N 
+ 12 \Psib^X \Gamma^{YW}\Psi^Z\right] F_{XYWZ}  + 
\text{\(\Psi^4\) terms},
\end{align}
and consider the Freund--Rubin $\ads_4 (\frac12L) \times S^7(L)$ solution \cite{Freund:1980xh}, 
\begin{equation}\label{baa2}
ds^2_{\text{11D}} =    L^2 ds^2_{\ads_4} + L^2   ds^2_{S^7} \,, \qquad
F_4 = dC_3 =  -  3 L^3 \vol_{\ads_4} \,,
\end{equation}
where   \(ds^2_{\ads_4}\) and \( ds^2_{S^7}\) are the metric of \(\ads_4\) and \(S^7\) with radius \(L_A = \frac12\) and \(L_S =1\) respectively.
We  split the 11D components \(x^M = (x^\mu, x^m)\) so that e.g.\
$\nabla^2 = \nabla^M\nabla_M = 
\cd^\mu \cd_\mu + \cd^m \cd_m \equiv \nabla_A^2 + \nabla_S^2$, and 
\begin{equation}\label{aaa}
   F_{\mu\nu\rho\sigma} = 3 \epsilon_{\mu\nu\rho\sigma}.
\end{equation} 
We note that, since we are interested in logarithmic divergences of the one-loop effective action, specific choices and discussions of boundary conditions and  boundary terms are not relevant.

The gauge-fixed partition function is
\begin{equation}\label{pqd}
Z_{(1)} = \int e^{-S^\star} Z_{\text{gf}}
=
	Z_{\text B}\, 
	Z_{\text F}\, 
	Z_{\text{B,gh}} \,
	Z_{\text{F,gh}} \, 
	Z_{\text {0m}}
   \,,
   \qquad\qquad
   S^\star = \int \!  \big\{ L_\text{B} + L_{\text F} + L_{\text {gf}} \big\} \, ,
\end{equation}
 where $L_\text{gf}$ is the gauge-fixing lagrangian and \(Z_{\text{gf}} \) is the  Faddeev--Popov-like term.
 We have three  symmetries that require fixing: diffeomorphism invariance for the metric, gauge invariance of the three-form fields, and local supersymmetry for the gravitino.

For the bosonic sector, we implement the following gauge conditions for the gauge 3-form and the metric 
\begin{equation}
    f_{MN} = \nabla^P A_{MNP} = 0\,, \qquad f_M = -\nabla^N h_{MN} + \frac{1}{2}\nabla_M h\indices{^N_N} =  0\, ,
\end{equation}
and the partition function for the ghost sector is given by~\cite{Thierry-Mieg:1980ihu, Siegel:1980jj, Copeland:1984qk}
\begin{align}
    Z_{\text{gf}}^{\text{grav}} &= \det \Delta_\text{FP}^\text{grav}\, , 
    \qquad [\Delta_\text{FP}^{\text{grav}}]_{MN} =  -g_{MN}\nabla^2  - R_{MN} \, ,\label{eq:gravityFPZ}\\
    Z_{\text{gf}}^{\text{gauge}} &= \frac{\left(\det \Delta_{2}\right)\left(\det \Delta_{0}\right)^2}{\left(\det \Delta_{1}\right)^\frac{3}{2}}\, ,\label{eq:gaugeFPZ}
\end{align}  
  $\Delta_r$ being the Hodge--de Rham operator acting on $r$-forms given in appendix \ref{appA}.
The gauge fixing of the gravitino will be discussed in section \ref{sec::fermions}.

Let us recall at this point the important result of~\cite{Bhattacharyya:2012ye} and connect the present discussion with the general formula \eqref{ffd}.
Among all the fields appearing in the quantization of 11D supergravity on \(\ads_4 \times S^7\), the only one possessing normalizable zero modes is the anticommuting 2-form ghost \(C_{MN}\) introduced above for the gauge-fixing of the 3-form potential. Using the regularized volume \(\VOL_{\ads_4} = \tfrac{4\pi^2}{3} L_A^4\), the number of such zero modes evaluates to \(n_0 = 1\), and \(\hat \beta_0 = \frac32\). The corresponding contribution to the free energy  reads
\begin{equation}\label{logL-final}
- \log Z_{0 \text m} = n_0 \, \hat \beta_{0 }\,  \log \Lambda L = \tfrac{3}{2}\log \Lambda L 
\,,
\end{equation}
which, upon using the AdS/CFT dictionary \(L^6 \propto N\), reproduces precisely the \(\tfrac{1}{4}\log N\) coefficient of the ABJM free energy~\cite{Fuji:2011km,Marino:2011eh}.
From the 11D perspective, \eqref{logL-final} is the entire contribution to the divergence of \(\Gamma_{(1)}\), as the SdW coefficients vanish in odd dimensions.
Recovering this result from the four-dimensional perspective occupies the rest of this section.

\subsection{Quadratic expansion of the bosonic sector} \label{sec:quadratic_exp}
We start by computing the quadratic expansion. We expand
\begin{equation}\label{dfl}
g_{MN} \to g_{MN} + h_{MN}\,,
\qquad
C_{MNR} \to C_{MNR}  + A_{MNR},
\end{equation}
where \(g,C\) now refer to the background values
and focus on the term quadratic in the fluctuations \(h,A\).

To organize the quadratic expansion, let us decompose it in the following form,
\begin{equation}\label{kfn}
L_{\text B}^{(2)} =
   L_{h^2}
   	+  L_{h^2 F}
	+ L_{A^2} 
	+   L_{hA }
+ \text{gauge terms}\,,
\end{equation}
where the first term comes from the expansion of the Einstein--Hilbert term, the second one  comes from the \(F^2\) term and is quadratic in the metric fluctuation, the third one is quadratic in the fluctuation of the 3-form, and the last one is a mixing between the 3-form and the metric, also coming from \(F^2\).

The expansion of the Einstein--Hilbert term is standard and reads
\begin{equation}\label{cnd}
\begin{aligned}
e^{-1}  L_{h^2}&=
 e^{-1}  L_{h^2}^\star 
    + \tfrac18 f^M f_M
 \,,
 \qquad\qquad
 f_N = - \cd^M   h_{MN} + \tfrac 12 \cd_N h   \,,
 \\[0.5em]
 e^{-1}  L_{h^2}^\star  &=
\tfrac{1}{16} h^{M N} \nabla ^2h_{MN} 
   - \tfrac{1}{32} h^M_M \nabla ^2h^N_N 
    + \tfrac{1}{8} h^{M N} h^{RS} R_{MRNS} 
    \\
    & \qquad  
+ \tfrac{1}{8} h_{MR} h^{MN} R\indices{_N^R} 
 - \tfrac{1}{8} h^M_M  h^{NR} R_{NR} 
 - \tfrac{1}{16} h_{M N} h^{M N} R 
 + \tfrac{1}{32} h^M_M h^N_N R \,,
\end{aligned}
\end{equation} 
where the \( f_M f^M\) term is cancelled by the gauge-fixing contributions.
Specializing to the \(\ads_4 \times S^7\) background
\begin{equation}\label{dih}
 \begin{aligned}
 e^{-1}  L_{h^2}^\star 
  &  
  = 
   \tfrac{1}{16} h^{ \mu \nu } \nabla ^2h_{ \mu \nu }  
  + \tfrac{1}{16} h^{ m n } \nabla ^2h_{ m n } 
  + \tfrac{1}{8}   h^{ \mu n } \nabla ^2h_{ \mu n} 
   - \tfrac{1}{32}  h^r_r   \nabla ^2  h^n_n  
 - \tfrac{1}{32} h^\nu _\nu  \nabla ^2 h^\sigma _\sigma 
 - \tfrac{1}{16}    h^\nu _\nu   \nabla ^2  h^n_n  
 \\
  & \qquad
 + \tfrac{13 }{16}  h^\mu_\mu h^\nu_\nu
 -\tfrac{5 }{8  } h^{\mu \nu} h_{\mu\nu}
  + h^{mn} h_{mn}
   + \tfrac{3 }{8  }  h^\mu_\mu h^m_m
  - \tfrac{13 }{16} h^n_n h^m_m.
 \end{aligned}
\end{equation}
The other term quadratic in \(h\) comes from the Maxwell term of the 3-form and is
\begin{equation}\label{bnd}
\begin{aligned}
e^{-1}  L_{h^2F} &= 
\tfrac{1}{192} F_{ ABCD } F\indices{^A^{BCD} } h_{M  N } h^{M  N } 
- \tfrac{1}{12} F_{ P ABC } F\indices{ _N ^{ABC} } h_{M    P } h^{M  N } 
 - \tfrac{1}{384} F_{ ABCD } F^{ ABCD } h^{M   M } h^{ N    N } \\
& \qquad\qquad+ \tfrac{1}{24} F_{ N  ABC } F\indices{ _P^{ABC} } h^{M   M } h^{ N  P } 
 - \tfrac{1}{8} F_{M  P    AB } F\indices{ _N   _S  ^A^B} h^{M  N } h^{ P   S}
 \\ 
 & =
\frac{9 }{16}  h^m_m  h^n_n
- \frac{9 }{8}    h^m_m h^\mu_\mu
-\frac{9}{8}    h^{mn} h_{mn}
+\frac{9 }{4}   h^{\mu m} h_{\mu m}
+\frac{9  }{16}   h^\mu_\mu  h^\nu_\nu
+\frac{9 }{8}   h^{\mu \nu} h_{\mu \nu}.
\end{aligned}
\end{equation}

For the three form we consider the kinetic term \(F^2\), which after gauge fixing reads
\begin{equation}\label{dvq}
    \begin{split}
        e^{-1} L^\star_{A^2} &= 
        \tfrac{1}{12} A^{MN  P } \nabla ^2A_{MN  P }
        + \tfrac{1}{2} A_{M}{}^{RS} A^{MN  P } R_{ N R P S} 
        - \tfrac{1}{4} A_{M}{}^{ P R} A_{ N  P R} R^{MN }  
        \\
        &\qquad
        - \tfrac{1}{864} A^{MNR} \varepsilon_{M NR ABCK PQST} 
        F^{PQST}   \nabla ^{K}A^ {ABC}
        \\& 
        =
        -\frac1 {12} A^{\mu \nu \rho }   [-\cd^2  -12 ] A_{\mu \nu \rho }
        -\frac1 {12} A^{mnr }  [-\cd^2 + 12 ] A_{ mnr }
        + \frac 1 {12} \varepsilon_{mnrabcs} A^{mnr} \cd^s A^{abc}
        \\
        &\qquad
        -\frac14  A^{\mu \nu m }  [-\cd^2  -10 ] A_{\mu \nu m }
        -\frac14  A^{\mu m n }  [-\cd^2 -2  ] A_{\mu m n } .
    \end{split}
\end{equation}

Finally, the \(hA\) mixing term is:
\begin{equation}\label{bni}
\begin{aligned}
e^{-1}  L_{hA} & = 
\tfrac{1}{6} F_{NPKR} h^{MN} \nabla _{M}A^{PKR} 
 - \tfrac{1}{2} F_{NPKR} h^{MN} \nabla ^{R}A_{MPK} 
 + \tfrac{1}{12} F_{NPKR} h^{M}_{M} \nabla ^{R}A^{NPK}
 \\
 & = 
 \tfrac{1}{2}  h^{\mu \nu }  \varepsilon_{ \nu \alpha  \rho \sigma } \nabla _{\mu }   
 		A^{\alpha\rho\sigma} 
- \tfrac{3}{2}  h^{\mu \nu } \varepsilon_{\nu \rho \sigma \alpha }  \nabla ^{\alpha }
  			 	A_{\mu \rho \sigma } 
+   \tfrac{1}{2}    h^{m \nu } \varepsilon_{\nu \alpha \rho \sigma }  \nabla _{m}
				A^{\alpha\rho\sigma} 
				\\
				&  \qquad
- \tfrac{3}{2} h^{m\nu } 	\varepsilon_{\nu \rho \sigma \alpha }  \nabla ^{\alpha }
			 A_{m\rho \sigma } 
+ \tfrac{1}{4} ( h^{\nu}_\nu + h^m _m )\varepsilon_{ \alpha \rho \sigma \mu}  \nabla ^{\mu }
				A^{\alpha\rho\sigma} .
\end{aligned}
\end{equation}

It is convenient to present the results obtained so far dualizing the \(\ads\) three-form \(A_{\mu \nu \rho }\) and    decomposing \(h_{\mu\nu}\) and \(h_{mn}\) into symmetric traceless and scalar parts, 
\begin{equation}\label{car} 
A^{\alpha \beta \gamma } = \frac16 \varepsilon^{\alpha\beta\gamma \mu} \theta_\mu\,,\qquad
h_{\mu \nu } = k_{\mu \nu } + \frac14 g_{\mu \nu } U  
 \,,\qquad  h_{m n } = k_{mn } + \frac17 g_{m n } V 
 \,, \qquad
 k^\mu_\mu = 0 = k^m_m\,,
\end{equation}  
These decompositions are algebraic and do not bring any Jacobian factor to the partition function.
We arrive at a quadratic term that is only partially diagonal and can be represented as
\begin{equation}\label{fsg}
\begin{aligned}
L_{\text B}^{[2]\star} &{}=
   L_{\text{diag}}
   	+  L_{\text{mix}}  
   	\,,\qquad Q A_{mnr}= \frac 16 \varepsilon_{mnrabcs}\cd^s A^{abc},
\\
   e^{-1} L_{\text{diag}} &{}= - \tfrac{1}{16} k^{ \mu \nu } [- \nabla ^2 -8  ]k_{ \mu \nu }  - \tfrac{1}{16} k^{ m n } [- \nabla ^2 +2 ] k_{ mn}   \\ 
            &\qquad{}  -\tfrac1{4}  A^{\mu m n }[ -\cd^2 -2   ]  A_{\mu m n }  -\tfrac1{12} A^{mnr } [ -\cd^2 + 12  -6 Q  ]  A_{ mnr }
 \\
     e^{-1} L_{\text{mix}} &{}=
      \tfrac1{72} \theta^\mu   [ - \cd^2 -12  ]  \theta_\mu 
      -\tfrac1{4}  A^{\mu \nu m }[ - \cd^2 - 10    ] 	A_{\mu \nu m}   
      - \tfrac{1}{8} h_{ \mu n}[-\cd^2 - 18]  h ^{\mu n} ,
    \\&\quad {}
      + \tfrac{1}{64} U [-\nabla^2 + 96 ] U + \tfrac{5}{224}  V [-\nabla^2-12 ]  V
      +\tfrac{1}{16}U [- \nabla ^2 - 12  ]V
      \\&\quad{}
      + \tfrac{1}{2}   h^{m \nu }  \nabla _{m}\theta_\nu+ \tfrac{1}{4}  U \nabla ^{\mu } \theta_\mu - \tfrac{1}{4} V \nabla ^{\mu } \theta_\mu - \tfrac{3}{2} h^{\nu }_m 	\varepsilon_{\nu \rho \sigma \alpha }  \nabla ^{\alpha } A^{\rho \sigma m}  .
\end{aligned}
\end{equation}

Fields appearing in \(L_{\text{diag}} \) can be straightforwardly integrated.
On the other hand,
\(L_{\text{mix}} \) requires more work to be brought to a useful form.
We  proceed to isolate the interaction between the different degrees of freedom.  To this aim, we split into longitudinal + transverse part with respect to the single \(\ads\) and \(S\) factors,
\begin{align}
\label{dksh}
 h_{\mu m}  &= h^{\perp}_{\mu m} + \cd_ m q^\perp_\mu + \cd_\mu q_m^\perp + \cd_m \cd_\mu \, q, \quad 
\\
&J = \det_{1\perp,0}[-\cd^2_S]^\frac12\det_{0,1\perp}[-\cd^2_A]^\frac12\det_{0,0}[-\cd^2_A][-\cd^2_S]^\frac12\det_{0,\emptyset}[-\cd^2_A]^{-\frac12},
\\
  \theta_\mu  &= \theta_\mu^\perp + \cd_\mu \vartheta,   
\qquad\qquad
 J = \det_{0,0}[-\cd^2_A]^\frac12.
\end{align}
We stress that    $\perp$ means transverse in all indices, i.e. $\nabla^\mu \theta_\mu^\perp =0$ but also \(\cd^m h^\perp_{\mu m} = 0 =\cd^\mu h^\perp_{\mu m} \). 
For the components \( A_{\rho \sigma m}  \) which have \(\ads\) 2-form indices we also dualize the transverse 2-form components,
\begin{align}\label{dsv} 
\hat{B}^{\perp}_{\mu n} & = \epsilon_{\mu\nu\rho\sigma}\nabla^\nu A_{\perp  } ^{\rho \sigma n}, &
{A}^{\perp }_{ \mu \nu n} & = \frac{1}{2}   \epsilon_{\mu\nu\rho\sigma} \frac{1}{-\nabla_A^2 - 16}\nabla^\sigma \hat{B}_{\perp}^{ \rho n}, 
\\
    \hat{b} ^{\perp}_\mu &= \epsilon_{\mu\nu\rho\sigma}\nabla^\nu a_\perp^{  \rho \sigma }, & 
  {a}^\perp_{ \mu \nu } &= \frac{1}{2}   \epsilon_{\mu\nu\rho\sigma} \frac{1}{-\nabla_A^2 - 16}\nabla^\sigma \hat{b}^\rho_{\perp},  
\end{align}
so that
\begin{align}\label{dst} 
    A_{\rho \sigma m} & =  
     \frac{1}{2} \frac{  \epsilon_{\rho\sigma\alpha\beta} }{-\nabla_A^2 - 16}\nabla^\beta \hat{B}_{\perp} ^{\alpha m} 
     + 
     				 \frac{1}{2}   \frac{\epsilon_{\rho\sigma\alpha\beta} }{-\nabla_A^2 - 16}\nabla^\beta \cd_m  \hat{b}^\alpha_{\perp}  
     + \cd_{[\rho} a_{\sigma] m }^{\perp} 
     + \cd_{m}\cd_{[\rho}  a_{\sigma]}^\perp,  \quad 
     \\
     &  {J} = \frac{ 
     	\det\limits_{1\perp,0}[-\cd^2_S]^\frac12
     	\det\limits_{1\perp,1\perp}[-\cd^2_A-12 ]^\frac12
     	\det\limits_{1\perp,0}[-\cd^2_A-12 ][-\cd^2_S]^\frac12
      }
     {
     	\det\limits_{1 \perp, 1 \perp}  [-\nabla_A^2 - 12 ]^{\frac{1}{2}} 
     	\det\limits_{1 \perp,0}  [-\nabla_A^2 - 12 ]^{\frac{1}{2}} 
      	\det\limits_{1\perp,\emptyset}[-\cd^2_A-12 ]^{\frac12} 
     }  
       \, .
\end{align} 
As a result,
\begin{equation} \label{dso}
\begin{aligned}
e^{-1}L_{ \text{B,mix}} & { }=
          - \tfrac18   a_\perp^{ \mu n   }  [ - \cd^2  -6   ][-\cd_A^2-12]  a_{\mu n}^{\perp}
          - \tfrac18   a_\perp^{ \mu  }  [ - \cd^2  -12   ][-\cd^2_S][-\cd_A^2-12]  a_{\mu}^\perp     
          \\& \qquad
              - \tfrac{1}{8}    q_{m  } ^\perp   [ -  \cd^2 - 6  ][- \cd^2_A] q^{m}_\perp 
+\tfrac1{72}      \theta^\mu_\perp   [ - \cd^2 -12 ]  \theta_\mu  ^\perp
+ \tfrac1{72}    \vartheta [ -  \cd^2 ]  [ -\cd^2_A]  \vartheta 
\\
& \qquad 
          +\tfrac{1}{8}\hat b_\perp^{ \mu }\frac{-\nabla^2 - 12}{-\nabla^2_A-12} [-\nabla^2_S]\hat b_{\mu }^\perp  	
+ \tfrac{1}{8}\hat B_\perp ^{\mu m} \frac{-\cd^2 - 6}{-\cd_A^2 - 12}\hat B^{\perp}_{\mu m} 
\\
& \qquad   
+ \tfrac{1}{64} U[ - \nabla ^2  + 96  ]   U  
+ \tfrac{5}{224}  V [ -\nabla ^2 -12   ] V  
\\
& \qquad
- \tfrac{1}{8}    h_{ \mu n  }^{\perp}  [ -  \cd^2 - 18  ]h _\perp^{ \mu n  }
- \tfrac{1}{8}    q_{ \mu  } ^\perp   [ -  \cd^2 - 24  ][- \cd^2_S] q_{\perp}^{\mu} 
- \tfrac{1}{8}    q     [ -  \cd^2 - 12  ][- \cd^2_S] [- \cd^2_A ]  q
\\&\qquad
- \tfrac{1}{2}   q^\nu_\perp  \nabla ^2_S \theta_\nu^\perp
+ \tfrac{1}{2}     q  \nabla ^2_S \nabla^2_A \vartheta
+ \tfrac{1}{4}  U \nabla^2_A \vartheta  
- \tfrac{1}{4}  V \nabla^2_A \vartheta  
+ \tfrac{1}{16}    U   [- \nabla ^2  -12 ] V  
\\
& \qquad
- \tfrac{3}{2}h^{\mu m}_{\perp} \hat B^{\perp}_{ \mu m}
- \tfrac{3}{2}q_\perp^\mu [-\nabla_S^2]\hat b_{\mu}^\perp .
\end{aligned}
\end{equation}

\subsection{Bosonic contribution to the one-loop effective action}
\label{subsec:bosonic-1loop} 
We have several bosonic contributions to the partition function (cf.~\eqref{det2} for notation). Let us start with the diagonal piece in \eqref{fsg} that can be immediately integrated and gives
\begin{equation}\label{dsn}
Z_{\text {B,diag}} =  
	\det_{(2),0} [-\nabla^2 - 8   ]^{-\frac{1}{2}}  
	\det_{0,(2)} [-\nabla^2 + 2   ]^{-\frac{1}{2}}
	\det_{1,2} [-\nabla^2 - 2   ]^{-\frac{1}{2}} 
	\det_{0,3} [-\nabla^2 +12   -  6 Q ]^{-\frac{1}{2}}\,.
\end{equation}

The rest of the lagrangian, resulting in \eqref{dso}, has been manipulated by replacing fields with their constrained (transverse) forms. This brings Jacobians in intermediate stages of the calculation, that however systematically cancel, cf.\ discussion leading to \eqref{rsfe}. Now we proceed to integrate all the fields in \eqref{dso}; for ease of exposition, we will implicitly cancel all Jacobian determinants with the associated contributions from the integration.

To understand the contribution of \(L_{ \text{B,mix}} \), we need to diagonalize the quadratic term.  We note a group of fields that are not actually coupled (\(a_{\mu n}, a _\mu,  q_m\), which we label (0)), yielding
\begin{equation}\label{cds}
Z_{\text{B,mix}(0) } =   
\frac{\det\limits_{1\perp,\emptyset} [-\nabla_A^2 - 12   ]^{\frac{1}{2}}}
{
	\det\limits_{0, 1\perp}\left[-\nabla^2 -6\right]^{\frac{1}{2}} 
	\det\limits_{1\perp,1\perp}\left[-\nabla^2 - 6  \right]^{\frac{1}{2}} 
	\det\limits_{1\perp,0}\left[-\nabla^2 - 12  \right]^{\frac{1}{2}}
}
\,.
\end{equation}
As said, in writing \eqref{cds} we already suppressed contributions that cancel with Jacobians associated with the introduction of \(a_{\mu n}, a _\mu,  q_m\) from \eqref{dksh} and \eqref{dst}.
Besides this effectively diagonal term, a proper mixing takes place in the following three groups:
(1)  AdS vectors with sphere indices $h^{\perp}_{\mu m},\hat B^{\perp}_{\mu m}$;
(2)  AdS vectors $\theta^\perp_\mu, q^\perp_\mu, \hat b^\perp_{\mu  }$;
(3) scalars $\vartheta, U, V, q$.
We now consider these three families separately.

\paragraph{(1) AdS vectors with sphere indices.} 

The relevant part of the lagrangian is
\begin{equation}
\begin{split}
e^{-1} L_{ \text{B,mix}}|_{(1)} &=       -\frac{1}{8}h_{\mu n}^{\perp} \left[-\nabla^2 - 18   \right]h_\perp^{\mu n} - \frac{3}{2}h_{\mu m}^{\perp} \hat B_\perp^{  \mu m}+ \frac{1}{8}\hat B_\perp^{  \mu m} \frac{-\cd^2 - 6}{-\cd_A^2 - 12}\hat B^{\perp}_{\mu m} \, .
\end{split}
\end{equation}
The mixing can be eliminated via the field redefinition (with unit Jacobian)
\begin{equation}
\begin{split}
h^{ \perp}_{\mu n} &= h'^{\perp}_{\mu n} - \frac{6}{-\cd^2-18}B_{\mu n}^\perp,
\end{split}
\end{equation} 
which yields
\begin{equation}\label{daa}
\begin{split}
e^{-1} L_{ \text{B,mix}}|_{(1)}  & = -\frac{1}{8}h'^{\mu n}_{\perp} [-\nabla^2 - 18  ] h'^{\perp}_{\mu n} 
+ \frac{1}{8}\hat B_\perp^{\mu m} \, \frac{\nabla^4 - 12 \nabla^2 + 36 \nabla_S^2 - 324 }{ [-\cd^2-18 ] [-\cd_A^2-12 ]} \,\hat B^{\perp}_{\mu m} .
\end{split} 
\end{equation} 
The partition function can be now expressed in terms of standard determinants; after the cancellations of the Jacobian term and the simplification of the two-derivative terms, only the four-derivative operator survives and the result is surprisingly simple
\begin{equation}\label{dsn2}
Z_{ \text{B,mix(1)}}  =
\det_{1\perp,1\perp }\left[
\nabla^4 - 12 \nabla^2 + 36 \nabla_S^2 - 324 \right]^{-\frac{1}{2}}\,.
\end{equation}

\paragraph{(2) AdS vectors.}

The starting point is 
\begin{equation}\label{bqtheta1}
\begin{split}
e^{-1} L_{ \text{B,mix}}|_{(2)}  = &\frac{1}{8}\hat b_{\perp }^{ \mu }\frac{-\nabla^2 - 12}{-\nabla^2_A-12}[-\nabla^2_S]\hat b_{\mu }^\perp -\frac{1}{8}q_\mu^\perp[-\nabla^2-24  ][-\nabla^2_S]q_\perp^\mu + \frac{1}{72}\theta_\perp^\mu[-\nabla^2-12  ]\theta_\mu^\perp \\
&- \frac{3}{2}q_\perp^\mu [-\nabla_S^2]\hat b_{\mu}^\perp+\frac{1}{2} q_\perp^\nu [-\nabla_S^2]\theta_\nu^\perp.
\end{split}
\end{equation}
We have three different fields. 
Define 
\begin{equation}
\hat b_{\mu}^{\perp} = \hat b_\mu'^\perp + 6 \frac{-\nabla^2_A-12}{-\nabla^2 - 12} q_\mu^\perp\,,
\qquad\qquad
\theta^{\perp}_\mu = \theta_\mu'^\perp - 18  \frac{1}{\nabla^2-12}(-\nabla_S^2)q_\mu^\perp.
\end{equation}
In the end 
\begin{equation}\label{LbqthetaFinal1}
\begin{split}
e^{-1} L_{ \text{B,mix}}|_{(2)}= &  \frac{1}{72}\theta'^{\mu}_\perp(-\nabla^2-12  )\theta'^{\perp}_\mu
+\frac{1}{8}\hat b'^{  \mu\nu}_\perp
\frac{-\nabla^2 - 12}{-\nabla^2_A-12}[-\nabla^2_S]\hat b'^{\perp}  _{\mu\nu}
-\frac{1}{8}q_\mu^\perp\left[-\cd^2 + 12\right]\left[-\cd_S^2\right]q^{\perp\,\mu},
\end{split}
\end{equation}
remarkably, the operator on \(q_\mu\) is local. The result for the partition function is 
\begin{equation}\label{ZbqthetaFinal1} 
Z_{ \text{B,mix}(2)} 
= \frac{\det\limits_{1\perp,\emptyset} [-\cd_A^2 + 12 ]^{\frac{1}{2}}
        \det\limits_{1\perp,\emptyset} [-\cd_A^2 - 12 ]^{\frac{1}{2}}}
        {\det\limits_{1\perp,0} [-\cd^2 + 12 ]^{\frac12}   \, 
 \det\limits_{1\perp,0} [-\cd^2 - 12 ]^{\frac12}  } .
\end{equation}

\paragraph{(3) scalars.}

The relevant piece of the lagrangian is
\begin{equation}
\begin{split}
e^{-1} L_{ \text{B,mix}}|_{(3) } &= \tfrac{1}{64} U [-\nabla^2 + 96 ] U + \tfrac{5}{224}  V [-\nabla^2-12 ]  V  - \tfrac{1}{8} q [-\nabla^2-12 ][-\nabla_S^2][-\nabla^2_A] q + \frac{1}{72}\vartheta [-\nabla^2][-\nabla_A^2]\vartheta\\
&\qquad	+ \tfrac{1}{16}U[-\nabla^2 -12 ]V + \tfrac{1}{4}\vphantom{m}U\nabla^2_A \vartheta -\tfrac{1}{4}  \vphantom{m} V   \nabla ^2_A \vartheta +\tfrac{1}{2} \vphantom{m}  \vartheta  \nabla^2_S \nabla^2_A q\, ,
\end{split}
\end{equation}
which has the structure 
\begin{equation} 
L_{(3)}=  \vec \xi \cdot \mathrm M \vec \xi + q \,  [-\nabla^2_S] \Delta_q \, q 
+ 2 \vec{ \mathrm b }\cdot \vec\xi  [-\nabla^2_S]q
\,,
\qquad\qquad
\vec \xi= (U,V,\vartheta).
\label{eq:mixingScalars} 
\end{equation}
with non-singular \(\mathrm M\).
The full kinetic operator on \((\vec\xi,q)\) is given by
\begin{equation}
\mathcal M =  
\begin{pmatrix} 
 \mathrm M &  (-\nabla^2_S) \vec { \mathrm b } \\  
(-\nabla^2_S)  \vec { \mathrm b } ^{\text t} &  (\nabla^2_S)  \Delta_q 
\end{pmatrix} 
\,,\qquad
\Delta_q = -\frac{1}{8}  \left[-\nabla^2-12  \right][-\nabla^2_A]
\,,
\qquad
\vec {\mathrm b}^{\text t} =\left( 0,0,\tfrac{1}{4}  [-\nabla^2_A]\right)\,,
\end{equation} 
and the naive result for the path integral \(\sim [\det \mathcal M]^{-\frac12}\) needs to be properly corrected to account for AdS-modes, which in this case only concern the \(q\) field. Integrating the \(\vec \xi\) fields first, we immediately get
\begin{equation}\label{dsd2}
\begin{aligned}
Z_\text{scal} &= \int \! \DD {\vec {\xi} } \DD {q }
 e^{- \int L_{(3)}}
 =   {  \det_{0,0} \mathrm M^{-\frac12} }
 \int \!   \DD {q } 
 e^{-\int q\,   \tilde \Delta_q  [-\nabla^2 _S]  \, q }
 \,,
 \qquad\quad
 \tilde \Delta_q  = \Delta_q - \vec { \mathrm b }^\text t \, \mathrm M^{-1} \,  \vec { \mathrm b } [-\cd_S^2] \,.
\end{aligned}
\end{equation}
The integration of \(q\) picks up the AdS-mode contribution; using that \( \tilde \Delta_q  \equiv   \Delta_q \) on \(\ads\) modes we finally have
\begin{equation}\label{dsd3}
\begin{aligned}
Z_\text{scal}  &  =
   \det_{0,0} \mathrm M^{-\frac12} \
   \det_{0,0}  [ \tilde\Delta_q  (-\nabla^2 _S) ] ^{\frac12} \
   \det_{0,\emptyset}      \Delta_q   ^{\frac12}   
   =
  \det_{0,0} \mathcal M ^{- \frac12} \
  \det_{0,\emptyset}    \Delta_q     ^{\frac12}   	\,,
\end{aligned}
\end{equation}
where  we recognized \(\det \mathcal M = \det \mathrm M\det [  \Delta_q [-\cd^2_S] - \cd_S^4 \vec { \mathrm b }^\text t   \, \mathrm M^{-1} \, \vec { \mathrm b }  ]\). 
This discussion implies that, once again dropping the Jacobian terms, in our case we have
\begin{equation}\label{dsd} 
Z_{ \text{B,mix} (3)} 
=
\frac{
    \det\limits_{0,\emptyset} [-\nabla_A^2 - 12   ]^{\frac{1}{2}} 
 }{
\det\limits_{0,0} [-\cd^2 + 24 ]^{\frac{1}{2}}
\det\limits_{0,0} [-\cd^2 -12 ]^{\frac{1}{2}} 
\det\limits_{0,0} [ 
					\nabla^4 - 72 \nabla^2 + 144 \nabla_S^2 
			 ]^{\frac{1}{2}} 
				} \,.
\end{equation}

\subsection{Result for the bosonic sector and logarithmic divergences} \label{sec:bosonic_resuts}

The bosonic contribution to the partition function is finally
\begin{equation}\label{jer}
Z_\text{B}
= 
Z_{\text {B,diag}} \ 
Z_{ \text{B,mix} (0)}  \ 
Z_{ \text{B,mix} (1)} \ 
Z_{ \text{B,mix} (2)} \ 
Z_{ \text{B,mix} (3)} 
\,,\qquad 
\Gamma_\text{B} = - \log Z_\text{B}\,,
\end{equation}
where each contribution was explicitly given above in \eqref{dsn}, \eqref{cds}, \eqref{dsn2}, \eqref{ZbqthetaFinal1}, \eqref{dsd}.
We now consider the UV logarithmic divergences.
\begin{equation}\label{zsp} 
\begin{aligned}
\Gamma_\text{B} & { }=  
- \log Z_{\text {B,diag}}
+ \Gamma^{S^{7}}_{\text{scal}} 
-
\frac 12
 \log \frac{
	\det\limits _{0,\emptyset} [-\nabla_A^2 - 12   ] 
}{
	\det\limits _{0, 1\perp} [-\nabla^2 -6 ]  
}
-
\frac 12
\log \frac{
	\det\limits _{1\perp,\emptyset} [-\nabla_A^2 - 12   ] 
}{
	\det\limits _{1\perp,1\perp} [-\nabla^2 - 6    ] 
} 
\\ & \qquad { }
-
\frac 12
\log \frac{
	\det\limits _{1\perp,\emptyset} [-\cd_A^2 + 12 ] 
        \det\limits _{1\perp,\emptyset} [-\cd_A^2 - 12 ] 
 }{
	 \det\limits _{1\perp,1\perp } [\nabla^4 - 12 \nabla^2 + 36 \nabla_S^2 - 324 ] 
} 
\\ & \qquad { }
+ \frac 12
\log 
	\det\limits_{0,0} [ 
					\nabla^4 - 72 \nabla^2 + 144 \nabla_S^2
			 ] \,,
\end{aligned}
\end{equation}
where we separated out the contribution of the fields that are  scalars on \(S^7\) with standard two-derivative operators,
\begin{equation}\label{dtv}
\begin{aligned}
\Gamma^{S^{7}}_{\text{scal}} 
&{}= 
\log 
	\det _{1\perp,0} [-\cd^2 - 12   ] 
	+ \frac12 
	\log 
      \det\limits_{1\perp,0} [-\cd^2 + 12 ]    
\\
& {}	\qquad 
+
 \frac12 
	 \log 
	\det\limits_{0,0} [-\cd^2 + 24 ] 
	+ \frac12 
	\log 
	\det\limits_{0,0} [-\cd^2 -12 ] 
	\,.
\end{aligned}
\end{equation}

Each term in the algebraic sums \eqref{zsp},\eqref{dtv} is finite. As a result, we find  a vanishing logarithmic divergence for the bosonic contribution to the effective action \(\Gamma_\text B\).  In some cases, the determinants are finite by themselves, while in other cases cancellations between different contributions take place.

We now discuss in more detail how these cancellations take place. 
Many tedious technical details that do not really depend on the example under discussion have been worked out in appendices~\ref{app:partf} and \ref{appendix:divergences}; here we mostly focus on nontrivial aspects specific to this case.

To start, \(Z_{\text {B,diag}}\) \eqref{dsn} is expressed in terms of determinants on unconstrained fields, namely fields that do not obey differential constraints. As a consequence 
 of the discussion in \eqref{atm}, it is UV finite.
    Similarly, following \eqref{doq},  determinants of two-derivative operators acting on fields that are scalars on \(S^7 \) are finite by themselves.

The next term in the first line of  \eqref{zsp} combines $h^\perp_{\mu n}$ and an AdS-mode from
$q^\perp_m$, which comes from the decomposition of $h_{\mu n}$. 
As we discuss in appendix \ref{app::AdSevenSodd}, the  log-divergent contribution to $\log Z$ coming from the $(0,1 {\perp})$ component \(\det_{0, 1{\perp}} [-\nabla^2 +\kappa ] \), cancels with the \(\ads\)-mode coming from the scalar $(0,0)$ component \(\det_{0, \emptyset} [-\nabla^2_A+ \kappa - 6 ] \). 
Analogously, the last term in the first line of  \eqref{zsp}  combines  $A^\perp_{\mu\nu n} \simeq B^\perp_{\rho n} $ with an AdS-mode from  $a_\mu^\perp$, which come from the decomposition of $A_{\mu\nu n}$. Again, for a  $(2,1)$ form $A_{\mu \nu m}$, the divergent contribution to $\log Z$ of \(\det_{1\perp, 1\perp} [-\nabla^2 + \kappa ]^{-\frac{1}{2}}\) from  $a_{\mu m}^\perp$   cancels with  \(\det_{1\perp, \emptyset} [-\nabla^2_A +\kappa -6 ]^{\frac{1}{2}}\) of the \(\ads\)-mode  from the vector $(1{\perp},0)$ component $a_\mu^\perp$. 
These  sectors of the partition function behave effectively like  unconstrained fields in 11D, which do not give   divergent contributions.

We are left with the last two lines of \eqref{zsp}, which exhibit fourth-order operators due to the mixing. These are exactly of the type discussed in section~\ref{sec::op4}, as they have the right combination of coefficients to factorize on the sphere \(S^7\) and can be computed as explained there. 
Let us first consider the scalar operator 
\begin{equation} 
\begin{aligned}
    \nabla^4 - 72 \nabla^2 + 144 \nabla_S^2\Big |_{(0,0)}
     = & {}  \left[-\nabla^2 + 36   +  12 \sqrt{-\nabla_S^2 + 9  }\right]\left[-\nabla^2 + 36   -  12 \sqrt{-\nabla_S^2 + 9  }\right] \\
        = & {}\left[-\nabla^2_A+l^2   -6l \right] \left[-\nabla^2_A+l^2   +18 l   +72   \right].
\end{aligned}
\end{equation}
From the 11D perspective we do not expect any divergence, and indeed this is verified by the result \eqref{alla}: the contributions of the two factors cancel each other.
On the other hand, for the vector operator
\begin{equation}
\begin{aligned}
   \nabla^4 - 12 \nabla^2 + 36 \nabla_S^2 - 324 \Big |_{(1{\perp},1{\perp})}
    = & {}\left[-\cd_A^2 - \cd_S^2 + 6 + 6\sqrt{-\cd_S^2 + 10}\right]\left[-\cd_A^2 - \cd_S^2 + 6 - 6\sqrt{-\cd_S^2 + 10}\right] \\
      = & {}\left[-\cd_A^2 +l^2+14 l+36\right]\left[-\cd_A^2 +l^2+2 l-12\right],
\end{aligned}
\end{equation}
we have the result \eqref{eq:logDetsqrt}, which gives 
\begin{equation}\label{abc}
\log \det_{ 1\perp,1\perp }  [\nabla^4 - 12 \nabla^2 + 36 \nabla_S^2 - 324   ]
: \quad
\zeta(0) = \frac{17}{15} \,.
\end{equation}
Notably, for the \(\ads\)-mode determinants of the operators \(-\cd^2 \pm 12 \), we have from \eqref{eq:log-div-results}
\begin{equation}\label{nts2}
\log \det_{ 0,\emptyset }  [ -\cd^2 \pm 12 ]: \quad
\zeta(0) =   \frac{17}{30} \pm \frac{3}{4} \,.
\end{equation}
As a result,   the total combination given by the sum of  the two terms in \eqref{nts2} cancels exactly \eqref{abc}.
This final cancellation between the divergences of \(\det_{1\perp, 1\perp}\left[\nabla^4 - 12 \nabla^2 + 36 \nabla_S^2 - 324 \right]^{-\frac{1}{2}}\) and those arising from the  \(\ads\)-modes \(\det_{1\perp,\emptyset}\left[-\nabla_A^2 - 12  \right]^{\frac{1}{2}}  \det_{1\perp,\emptyset}\left[-\nabla_A^2 + 12  \right]^{\frac{1}{2}}\)
is not automatic in the formalism and it provides an important consistency check of the technique.

As discussed in \cite{Bobev:2023dwx}, this result does not depend on the duality frame of the scalar fields, and the overall result is that the dynamic bosonic contribution to the partition function is finite.

\subsection{Fermionic contributions}\label{sec::fermions}

Starting with the gravitino action \eqref{eq:Lgravitino1}, we proceed to the construction of the partition function
\begin{equation}\label{sdk}
\begin{gathered}
Z_\text F = \int \!\DD{\psi}  \,  e^{-S_\psi }
\,,\qquad\qquad
S_\psi  = \int L_{\psi_\mu}^\star
\,,\qquad\qquad
L_{\psi_\mu}^\star = L_{\psi_\mu} - \frac{9}{8}i e\psib_M \Gamma^M H \Gamma^N \psi_N\, ,
\\
Z_{\text{F,gh}} = 
\det_{\frac{1}{2}} [i\nablasl ]^{-1}\, \det_{\frac{1}{2}}[H ]^{-\frac{1}{2}}
\,,
\qquad \qquad 
H   
    = i \nablasl + \frac{i}{16} F_{MNRS}\Gamma^{MNRS}
\,.
\end{gathered}
\end{equation}
The specific form of gauge-averaging functional \(H \) is chosen to simplify some terms later on.
We also  redefine the gravitino field as \cite{Endo:1985km}
\begin{equation} 
\begin{aligned}
        \label{eq:redefPsitoPhi} 
        \phi_M & =  \psi_M - \frac{1}{2}\Gamma_M \Gamma^N\psi_N\,,
        \qquad&
        \phib_M & =   \psib_M - \frac{1}{2}\psib_N\Gamma^N \Gamma_M\,,
        \\
        \psi_M & = \phi_M - \frac{1}{D-2}\Gamma_M\Gamma^N \phi_N\,,
        & 
        \psib_M & = \phib_M - \frac{1}{D-2}\phib_N\Gamma^N\Gamma_M \,,
\end{aligned}
\end{equation}
this redefinition is algebraic and does not bring any Jacobian.

Substituting these redefinitions in the quadratic lagrangian and
specializing to the background solution \eqref{baa2}-\eqref{aaa} with the 4+7-covariant notation explained in \eqref{not:gamma}, we obtain  
\begin{gather}  \label{eq:fermion-quadratic}
    L^{(2) \star }_{\text{F}} =
     -\frac{1}{2}\phib_\mu \slashed{\mathbb{D}}^{\mu\nu} \phi_\nu -\frac{1}{2}\phib_m \slashed{\mathbb{D}}^{mn} \phi_n
     \,,
     \qquad
     H    
         =i\nablasl - \frac{9}{2}\gamma_* \,,
\\ 
    \label{eq:fermion-quadratic-operators}
    \slashed{\mathbb{D}}^{\mu\nu} = g^{\mu\nu} \nablasl  -\frac{3}{2}i\left(g^{\mu\nu}\gamma_*  + \frac{4}{3}\gamma^\mu \gamma_* \gamma^\nu\right), \qquad 
    \slashed{\mathbb{D}}^{mn} = g^{mn} \nablasl  -\frac{3}{2}i\left(-g^{mn}\gamma_*  + \frac{2}{3}\gamma^m \gamma_* \gamma^n\right) .
\end{gather}
The quadratic operators are already diagonal in the \(\ads\) and \(S\) components of the modified gravitino field. We further isolate the spin-\(\frac32\) and spin-\(\frac12\) components, we decompose into transverse traceless, trace and longitudinal as
\begin{align}
    \phi_\mu &= \phi_\mu^\perp + \frac{1}{4}\gamma_\mu \chi_A + \nabla_\mu\zeta_A - \frac{1}{4}\gamma_\mu \gamma_\nu \nabla^\nu \zeta_A 
    \,, \quad & 
    \phi_m &= \phi_m^\perp + \frac{1}{7}\gamma_*\gamma_m \chi_S + \nabla_m\zeta_S - \frac{1}{7}\gamma_m \gamma_n\nabla^n \zeta_S \,,
    \\ 
    \phib_\mu &= \phib_\mu^\perp - \frac{1}{4}\chib_A\gamma_\mu  + \nabla_\mu\zetab_A - \frac{1}{4}\nabla^\nu \zetab_A\gamma_\nu\gamma_\mu   
    \,, \quad & 
    \phib_m &= \phib_m^\perp - \frac{1}{7}\chib_S\gamma_m\gamma_*  + \nabla_m\zetab_S - \frac{1}{7}\nabla^n \zetab_S\gamma_n\gamma_m  \,,
      \\
        & J = \det_{\frac{1}{2},\frac12}\left[-\nablasl_A^2 + 16 \right]^{-\frac{1}{2}}
        \,, \quad &  & J = \det_{\frac12,\frac{1}{2}}\left[-\nablasl_S^2 - \frac{49}{4} \right]^{-\frac{1}{2}}.
\end{align}  
After some algebra we arrive at
\begin{equation}
\begin{split}
   e^{-1} L^{(2) \star }_{\text{F}}
                    = {} & 
                          \frac{1}{2}i\, \phib^\mu_\perp \left(i\slashed{\nabla}  +\frac 32  \gamma_* \right)\phi^{\perp}_\mu  
                        + \frac{1}{2}i\, \phib^m_\perp \left(i\slashed{\nabla} -\frac 32  \gamma_* \right)\phi^{\perp}_{m}   
                        + \frac{1}{16}i \,\chib_A\cD^A_{\chi}\chi_A 
                        + \frac{1}{14}i \,\chib_S\cD^S_{\chi}\chi_S
                    \\ &
                        +\frac{3}{16}i \,\zetab_A\cD ^2_A \left(i\, \nablasl_A + 2\gamma_* \, i \nablasl_S + 3  \gamma_*\right) \zeta_A  
                        +\frac{3}{98}i \,\zetab_S \cD ^2_S \left(14\, i\nablasl_A + 10 \, \gamma_* \, i\nablasl_S - 21   \gamma_*\right)\zeta_S
                    \\ & 
                        -\frac{3}{8} \,\chib_A\cD^2_A \zeta_A 
                        -\frac{12}{49}\chib_S\cD^2_S \zeta_S\, ,
    \end{split}
    \label{eq:LFermionInteractionXS}
\end{equation}
where we introduced the shorthand notation 
\begin{equation}\label{def:mDA}
\begin{aligned}
        \mD^2_A  &:= -\nablasl_A^2 + 16 \,, \qquad
         &
                \mD^A_{\chi} &:= i\nablasl_A + 2 \gamma_*\,i\nablasl_S - 13 \gamma_*  \,,\\
        \mD^2_S &:= -\nablasl_S^2 - \frac{49}{4} \,, \qquad
         &
                \mD^S_{\chi} &:= i\nablasl_A +\frac{5}{7} \gamma_* \, i\nablasl_S - \frac{11}{2} \gamma_*\,.
\end{aligned}
\end{equation}
The lagrangian 
\eqref{eq:LFermionInteractionXS}
displays non-diagonal terms which mix the \(\chi\) and \(\zeta\) components. We can resolve this interaction by redefining $\chi_A, \chi_S$ as 
    \begin{equation} 
          \chi_A =  \chi_A' +3\frac{1}{\mathcal{D}^A_{\chi}}\mD^2_A\zeta_A
        \,, \qquad
          \chi_S =  \chi_S' -\frac{12}{7} \frac{1}{\mD^S_{\chi}} \mD^2_S\zeta_S\,.
        \label{eq:redefChiSChiprimeS} 
\end{equation}
These redefinitions are algebraic and bring the action in its final form 
\begin{equation}
    \label{eq:LfermionsFinal}
\begin{split}
   e^{-1} L^{(2) \star }_{\text{F}}
                        = {} &
                             \frac{i}{2}\, \phib_\mu^\perp\left(i\nablasl + \frac{3}{2} \gamma_*\right)\phi^{\mu \perp}
                            +\frac{i}{2}\, \phib_m^\perp\left(i\nablasl - \frac{3}{2} \gamma_*\right)\phi^{m \perp}
                        \\ &    
                            +\frac{i}{16}\, \chib'_A\mD^A_{\chi} \chi'_A
                            +\frac{i}{14}\, \chib'_S\mD^S_{\chi} \chi'_S 
                            +\frac{3}{4}\,i\,\overline{\zeta}_A\cD^2_\zeta   \frac {\cD^2_A}{\cD^A_{\chi}}\zeta_A
                            +\frac{3}{7}\,i\,\overline{\zeta}_S\cD^2_\zeta  \frac{\cD^2_S}{\cD^S_{\chi}}\zeta_S \,,
                        \\
    \cD^2_\zeta & =   \left(i\nablasl  - \frac{9}{2} \gamma_*\right)\left(i\nablasl  - \frac{1}{2} \gamma_*\right)\,,
\end{split}
\end{equation}
where we note that the \(\zeta_{A,S}\) fields obtained a remarkably symmetric kinetic term. There is however an important difference between them, relevant for the evaluation of the functional integral.
Inserting \eqref{eq:nablaslEigenvalues} into
   $\mD_S^2$ acting on $\zeta_S$  we find the zero mode
            \begin{equation}\label{eds}
            0 = \mathcal D_S^2 =  -\left(l+\frac{7}{2}\right)^2 + \frac{49}{4} = l(l + 7) \quad \implies  \quad l=0, \quad m_{l=0} = 8\,,
            \end{equation}
   which induces the fermionic counterpart of the AdS modes discussed in section \ref{sec::path-integral-general}.  The associated operator comes from  evaluating $l=0$ in the kinetic operator of $\zeta_S$,
\begin{align*}
    \frac{\mD^2_\zeta  }{\D^S_\chi}  \quad \to\quad     \frac{ \left[i\nablasl_A  - (\frac{9}{2} \pm \frac 72 )\gamma_*\right]  
        \left[i\nablasl_A  - (\frac{1}{2} \pm \frac{7}{2} ) \gamma_*\right]}
        {i\nablasl_A \pm \frac{5}{2} \gamma_* - \frac{11}{2}\gamma_*} \,. 
\end{align*} 
            The AdS-mode contribution is thus (cf.~\eqref{det2})
            \begin{equation}
                \begin{split}
                    Z_{\text{Am}} 
                    = {} & 
                        \det_{\frac{1}{2},\emptyset}\left[i\nablasl_A -3 \gamma_*\right]^{\frac{4}{2}} 
                        \det_{\frac{1}{2},\emptyset}\left[i\nablasl_A -\gamma_*\right]^{-\frac{4}{2}}
                        \det_{\frac{1}{2},\emptyset}\left[i\nablasl_A -4 \gamma_*\right]^{-\frac{4}{2}}
                        \det_{\frac{1}{2},\emptyset}\left[i\nablasl_A +3\gamma_*\right]^{-\frac{4}{2}} \, ,\\
                \end{split}
            \end{equation}
            where the multiplicity is halved w.r.t the one in \eqref{eds} because we have here the $\nablasl_S$ operator not squared and we consequently divided the positive and negative parts of its spectrum.
            
            The final result for the partition function is
            \begin{equation}\label{yep}
                 Z_{\text F} = Z_{\frac{3}{2},\frac{1}{2}} \times Z_{\frac{1}{2},\frac{3}{2}}
                 \,,
            \end{equation}
            where
            \begin{equation}
                Z_{\frac{3}{2},\frac{1}{2}} = \det_{\frac{3}{2},\frac{1}{2}}[\big(i\nablasl +\tfrac{3}{2} \gamma_*\big)g_{\mu\nu}  + 2\, \gamma_\mu \gamma_* \gamma_\nu ]^\frac{1}{2}\, ,\quad  Z_{\frac{1}{2},\frac{3}{2}}= \det_{\frac{1}{2},\frac{3}{2}}[\big(i\nablasl -\tfrac{3}{2} \gamma_*\big)g_{mn} + \gamma_m \gamma_*\gamma_n  ]^\frac{1}{2}\, ,
            \end{equation}
            and, in particular,
            \begin{align}\label{yea}
                Z_{\frac{3}{2},\frac{1}{2}} = & {} 
                            \det\limits_{\frac{3}{2}\perp,\frac{1}{2}}[i\nablasl +\tfrac{3}{2} \gamma_*  ]^\frac{1}{2} \, 
                            \det\limits_{\frac{1}{2},\frac{1}{2}}[i\nablasl_A + 2\gamma_* i\nablasl_S -13 \gamma_* ]^\frac{1}{2}\, 
                            \det\limits_{\frac{1}{2},\frac{1}{2}}[i\nablasl -\tfrac{9}{2}\gamma_*]^\frac{1}{2}\, 
                            \det\limits_{\frac{1}{2},\frac{1}{2}}[i\nablasl - \tfrac{1}{2}\gamma_*]^\frac{1}{2}\, ,\\
                \label{yes}
                Z_{\frac{1}{2},\frac{3}{2}} = & {} 
                    \frac{
                            \det\limits_{\frac{1}{2},\frac{3}{2}\perp}[i\nablasl - \frac{3}{2} \gamma_*  ]^\frac{1}{2} \,
                            \det\limits_{\frac{1}{2},\frac{1}{2}}[i\nablasl_A + \frac{5}{7}\gamma_* i\nablasl_S -\frac{11}{2}\gamma_*   ]^\frac{1}{2}\,
                            \det\limits_{\frac{1}{2},\frac{1}{2}}[i\nablasl -\frac{9}{2}\gamma_*]^\frac{1}{2}\,
                            \det\limits_{\frac{1}{2},\frac{1}{2}}[i\nablasl - \frac{1}{2}\gamma_*]^\frac{1}{2}\,
                            \det\limits_{\frac{1}{2},\emptyset}[i\nablasl_A -3 \gamma_*]^{2} \,
                        }
                        {
                            \det\limits_{\frac{1}{2},\emptyset}[i\nablasl_A -\gamma_*]^{2}\,
                            \det\limits_{\frac{1}{2},\emptyset}[i\nablasl_A -4 \gamma_*]^{2}\,
                            \det\limits_{\frac{1}{2},\emptyset}[i\nablasl_A +3\gamma_*]^{2} \,
                        } \, .
            \end{align}

The KK expansion goes as usual; the eigenvalues and their multiplicities are reported in \ref{appendix:eigenvalues-spinors}.
The evaluation of the divergences proceeds as for the bosons. The main difference is the presence of \(\coth(\pi v)\) in place of \(\tanh (\pi v)\) in the spectral measure \eqref{baf}. 
For general even $d_A$, the $\ads_{d_A}$ zeta function for Majorana fermions is 
\begin{equation}\label{drdGen}
\zeta_{s}(z)  = 
    \frac{  \VOL_{\ads_{d_A}}  }{   \VOL_{S^{d_A-1}} } L_A^{d_A + 2z} 
    \frac{(2s +1) 2^{d_A-2}}{\left(2^{d_A-2} \Gamma\left(\frac{d_A}{2}\right)\right)^2}
	\intl_0^{+\infty}\!\!
 	dv \frac{  v^2 + \left(s + \tfrac12 \right)^2 }{(v^2 +  b_{s} + L_A^{2} \kappa )^z } v \coth (\pi v) \, ,
\end{equation} 
with 
\begin{equation}
    b_{\frac{1}{2}}=\frac{d_A(d_A-1)}{4}\, , \qquad b_{\frac{3}{2}} = \frac{d_A(d_A-1)}{3}\, .
\end{equation}
In the present case we have, using $d_A = 4$,
\begin{align}\label{drd}
\zeta_{\frac12}(z) & = \frac{   L_A^{4 + 2z }   }{   3   }
	\intl_0^{+\infty}\!\!
 	dv \frac{v \,  ( v^2 + 1) }{(v^2 +  3 + L_A^{2} \kappa )^z } \coth (\pi v) \,,
\\ \label{drd2}
%& 
\zeta_{\frac32 \perp} (z)  
 & = \frac{2 \, L_A^{4 + 2z}  }{ 3} \intl_0^{+\infty}\!\!
 	dv \frac{v \, ( v^2 + 4) }{(v^2 + 4 +  L_A^{2}\kappa )^z } \coth(\pi v)\,,
\end{align}
for Majorana spinors and (transverse gamma-traceless) Majorana vector-spinors respectively. Since   \eqref{drd},\eqref{drd2} assume a \emph{quadratic} operator in AdS space \(-\cd^2_A + \kappa\),   those appearing in \eqref{yea},\eqref{yes} have to be squared first.
After that,  the calculation  proceeds in full analogy with the discussion of section~\ref{sec::KKsum}, splitting \(\coth (\pi v) = 1 - 2 (e^{2 \pi v } -1 )^{-1}  \) but otherwise with straightforward extension of the procedure.

All determinants on \((r , \frac 12)\) are finite, cf.~\eqref{doq}. 
The overall nontrivial cancellation comes from the rest, whereby
\begin{equation}\label{dcf}
    \Bigg[ 
        \log\frac{ 
                \det\limits_{\frac{1}{2},\frac{3}{2}\perp} [i\nablasl - \frac{3}{2}  \gamma_*  ]^\frac{1}{2} 
                \det\limits_{\frac{1}{2},\emptyset} [i\nablasl_A -3 \gamma_* ]^{2} 
            }{ 
                \det\limits_{\frac{1}{2},\emptyset} [i\nablasl_A -\gamma_*]^{2} 
                \det\limits_{\frac{1}{2},\emptyset} [i\nablasl_A -4 \gamma_*]^{2} 
                \det\limits_{\frac{1}{2},\emptyset} [i\nablasl_A +3\gamma_*]^{2} 
            } 
            \Bigg]_{\log \Lambda L} = 0 \, .
\end{equation}
Therefore no logarithmic correction arises from the path-integral determinants also for fermions.

\section{Result} \label{sec::results}

In section~\ref{sec:11d-sugra} we have presented all the details of the  evaluation of the one-loop partition function \eqref{pqd}.

We have five types of contributions: 
physical bosonic and fermionic determinants, bosonic and fermionic ghost determinants, zero modes from the quantization.
We have explicitly shown that the bosonic determinants \eqref{jer}  do not give a logarithmic term. The same holds true for the fermionic determinants \eqref{yep}.
The ghost determinants also do not contribute,  as a consequence of \eqref{doq} and \eqref{atm}. 

In full analogy with  \cite{Bhattacharyya:2012ye}, the full result for the logarithmic correction is therefore given by the normalizable zero mode of the 2-form ghost introduced in the gauge-fixing of the 3-form potential as given in \eqref{logL-final}, and we recover
\begin{equation}\label{res}
[\Gamma_{(1)} ]_{\log N}= \frac 14 \log N 
\end{equation}
upon the use of the holographic dictionary.

\section{Discussion} \label{sec:conclusions}

In this paper we have developed a systematic framework for computing the logarithmically divergent part of one-loop partition functions on product spaces \(\ads_{d_A}\times S^{d_S}\), starting from the spectral representation of the AdS determinants and the harmonic decomposition on the sphere. We applied the framework to the full bosonic and fermionic content of 11D supergravity on \(\ads_4 \times S^7\) and reproduced the universal \(\tfrac{1}{4}\log N\) correction to the ABJM free energy predicted by supersymmetric localization. Three technical ingredients play a central role in our analysis.

 (i)  Quantization in eleven dimensions.
We quantize the theory in its eleven-dimensional origin, treating each fluctuation field on \(\ads_4\times S^7\) directly. This approach was used in~\cite{Bhattacharyya:2012ye} for the same setting and has been adopted in related precision-holography setups, e.g.~\cite{Liu:2017vbl}. From the higher-dimensional viewpoint, the odd total dimensionality forces the field-theoretic logarithmic divergences to cancel among the various sectors, so that the entire \(\log L\) correction stems from a single zero-mode contribution, namely the normalizable zero mode of the 2-form ghost introduced in the quantization of the 3-form potential.  

(ii) AdS modes. A recurring theme in our analysis is the proper treatment of zero modes of the sphere Laplacian that retain a non-trivial dependence on the AdS coordinates. Such modes arise systematically from   the transverse/longitudinal decompositions employed to diagonalize the kinetic operators, and they behave as genuine dynamical fields on the AdS factor. They are the analog, in the product-space setting, of the well-understood zero modes on compact spaces, see e.g.~\cite{Fradkin:1983mq,Christensen:1979iy,Tseytlin:2013fca}. Their inclusion is crucial for recovering the expected divergence structure, and we have verified this in a number of independent examples (appendix~\ref{appendix:divergences}). Their   systematic accounting is, in our view, the most subtle conceptual ingredient of our calculation.

(iii) Resummation of the KK tower. The third ingredient is the careful extraction of the large-\(l\) contributions to the spectral \(\zeta\)-function when evaluating the divergent part of the effective action as \(\zeta(0)\). In particular, our prescription correctly accounts for divergences of the appropriate dimensionality that would be missed by the evaluation in which one first takes \(\zeta_{\Delta(l)}(0)\) at each KK level and only afterwards \(\zeta\)-regularizes the sum over \(l\). This provides a generalization of  the low-dimensional strategy of~\cite{Buchbinder:2014nia,Giombi:2023vzu,Beccaria:2023ujc}, and gives a well-defined framework for handling the interplay between AdS spectral integrals and the KK sum.

Although we have tested our prescription in a variety of examples, we have not been able to prove on general grounds why it captures \emph{all} divergent contributions. In principle, an additional logarithmic piece could arise from the infinite sum over \(l\) of the finite parts \(\Gamma_{l,\text{fin}}\); we have verified that no such contribution is generated in all examples we considered, but a general principle explaining this phenomenon is still lacking. We note that, in the complementary situation of odd \(d_A\), this is precisely the mechanism producing the logarithm (see appendix~\ref{app::adsodd}), so the distinction between the two cases deserves a more conceptual understanding.

Supersymmetry played no direct role in our calculation, in the sense that we worked with the explicit field content of the supergravity multiplet and did not exploit cancellations enforced by the superalgebra. Nevertheless, it would be instructive to organize the various determinants into representations of the relevant supergroup. This might be particularly useful in settings where the higher-dimensional answer is non-trivial, for example because the total spacetime dimension is even, or because of less symmetry in the internal manifold. Supersymmetric organization may make cancellations and simplifications among different multiplet components manifest, in the spirit of~\cite{Beccaria:2024gkq}.

The specific holographic application   discussed here corresponds to   ABJM theory with level \(k=1\)  \cite{Aharony:2008ug}. 
A natural next target is 10D type IIA supergravity on \(\ads_4 \times \CP^3\) \cite{Aharony:2008ug} (\(k= \infty\); the spectrum on $\CP^3$ is available~\cite{Ikeda1978SpectraAE}), where the log correction to the ABJM free energy is known from localization~\cite{Marino:2009dp,Bobev:2022eus} but, to our knowledge, has not been reproduced from a one-loop calculation. Arbitrary  \(k\) analysis, which corresponds to 11D sugra on \(\ads_4 \times S^7/\mathbb{Z}_k\), is instead more subtle, as the \(k\) dependence arises from finite terms of the effective action as well (cf. \cite{Beccaria:2023hhi} for related discussions).  It is possible that the approach presented in this paper and tailored to  the log divergence may be refined to also capture  the finite part. It would be interesting to pursue this direction and compare with the previous (unsuccessful) attempt of reproducing the \(k\) dependence of  \cite{Liu:2016dau}.

Beyond the specific ABJM application, our framework opens the way to precision holography in a wider class of setups: any \(\ads \times X\) background with a sufficiently explicit spectrum is amenable to our treatment, and extensions to doubly-warped products \(\ads\times X\times Y\) should also be within reach, modulo a more involved double-sum analysis. 
Speculatively, extension of the techniques presented here might also be used to obtain nontrivial information   in the context of the $\ads_7 \times S^4$  holographic correspondence, where the  (finite) effective  action  is related to the \(a\)-anomaly of the boundary CFT, cf.~\cite{Beccaria:2014qea}.

\subsection*{Acknowledgments}  
We thank Nikolay Bobev and Mattia Cesàro for discussions and correspondence. 
LC is indebted to Matteo Beccaria and Arkady Tseytlin for many exchanges and discussions that led to several of the ideas underpinning  this paper.
We also thank Arkady Tseytlin for comments on the first draft of this paper.
Most of the work of LC was done as a member of the Institute of Theoretical Physics of the Leibniz Universität Hannover (until June 2025).

\newpage

\appendix

\section{Notation, conventions, explicit formulae}
\label{appA}
We work in 11 Euclidean dimensions and indicate spacetime indices with Latin uppercase  letters.  When referring to \(\ads \times S\) factors we use lowercase Greek and Latin indices respectively.

Curvature tensors are defined as
\begin{gather}\label{pac}
[\cd_M,\cd_N]V_P= R_{MNPQ}V^Q
\,,
\qquad
R_{MP} = R\indices{_M_N_P^N}
\,,
\qquad
R = R\indices{_M^M},
\end{gather}
and similarly for the single factors.
In particular, we have the following explicit expressions

\(\ads_{d_A}\):
\begin{equation}\label{paa}
\begin{gathered}
R_{\mu \nu \rho \sigma} = 
- \frac 1 {L_A^2}(g_{ \mu \rho} g_{ \nu \sigma} -
g_{ \mu \sigma} g_{ \nu \rho} )
\,,
\qquad
R_{\mu \nu} = - \frac{ d_A-1 }{L_A^2}   g_{\mu \nu} 
\,,
\qquad
R = - \frac{d_A (d_A-1)}{L_A^2}   
\\
    \VOL_{\ads_{d_A}} = {}   \pi ^{\frac{d_A-1}{2}} \, \Gamma \big[\tfrac{1-d_A}{2}\big] \quad (d_A  \text{ even} )\, ,\qquad\qquad
    \VOL_{\ads_{d_A}} = {}  2 \log R \, 
    			\frac{(-\pi )^{\frac{d_A-1}{2}}}{\Gamma \big[\frac{d_A+1}{2}\big]   }\, ,
    			 \quad (d_A   \text{ odd}). 
\end{gathered}
\end{equation}

$S^{d_S}$:
\begin{equation}\label{pab}
R_{mnrs} = 
 \frac 1 {L_S^2}(g_{ mr } g_{ ns} -
g_{ ms} g_{ nr} )
\,,
\qquad
R_{m n} = \frac{ d_S-1 }{L_S^2}   g_{m n}
\,,
\qquad
R = \frac{d_S (d_S-1)}{L_S^2} 
\,,\qquad
\VOL_{S^{d_S}} = {}   \frac{2 \, \pi ^{\frac{d_S+1}{2}}}{\Gamma \big[ \frac{d_S+1}{2}\big]}\,.
\end{equation}

Our conventions for gamma matrices are such that
\begin{equation}\label{not:gamma}
\begin{gathered}
\{\Gamma_M,\Gamma_N\} = 2g_{MN}
\,,\qquad
\Gamma_M = \left(\gamma_\mu \otimes \mathbbm{1}, \gamma_* \otimes \gamma_m\right),
\qquad
\{\gamma_\mu  ,\gamma_\nu \} = 2 g_{\mu\nu},
\qquad
\{\gamma_m  ,\gamma_n \} = 2 g_{mn}.
\\
\gamma_* \equiv - \frac{i}{4!}\epsilon^{\mu\nu\rho\sigma}\gamma_{\mu\nu\rho\sigma}, 
\qquad
\gamma_*^2  = \mathbbm{1}_A
\,,
\qquad
\Gamma_* \equiv i \Gamma_0\Gamma_1\Gamma_2\Gamma_3 = \gamma_* \otimes \mathbbm{1}_S\, .
\end{gathered}
\end{equation} 
We note the following relations for spin $\frac{1}{2}$ on \(\ads_4 \times S^7\)  
\begin{equation}
    \begin{gathered}
            \relax
            [\nabla_\mu,\nabla_\nu]\psi = -\frac{1}{2 L_A^2} \gamma_{\mu\nu}\psi,\qquad 
            [\nabla_m,\nabla_n]\psi = \frac{1}{2 L_S^2}\gamma_{mn}\psi.
            \\  
              -\nablasl_A^2\psi = \left(-\nabla_A^2 - \frac{3}{L_A^2}\right)\psi, \qquad 
              -\nablasl_S^2\psi = \left(-\nabla_S^2+\frac{21}{2L_S^2}\right)\psi,
    \end{gathered}
\end{equation}
and for spin $\frac{3}{2}$ 
\begin{equation}
    \label{eq:CommutatorDFermion3/2}
    \begin{gathered}
        \relax 
        [\nabla_\mu,\nabla_\nu]\psi_\rho =  -\frac{1}{L_A^2}\left(2g_{\rho [\mu} \psi_{\nu]}  +\frac12 \gamma_{\mu\nu}\psi_\rho\right),\qquad 
        [\nabla_m,\nabla_n]\psi_r ={}  \frac{1}{L_S^2}\left(2g_{r [m} \psi_{n]}  +\frac12 \gamma_{mn}\psi_r\right).
        \\
             -\nablasl_A^2 \psi_\mu^\perp  = \left(-\nabla_A^2 - \frac{4}{L_A^2}\right)\psi_\mu^\perp , \qquad -\nablasl_S^2 \psi_r^\perp = \left(-\nabla_S^2+\frac{23}{2L_S^2} \right)\psi_r^\perp .
    \end{gathered}
\end{equation}
where       
\(\perp\) means transverse and gamma-traceless.

\subsubsection*{Determinants}  
 We denote determinants in  \(D=d_A + d_S\) dimensions as
 \begin{equation}\label{det1}
 \det_r \Delta\,,
 \end{equation}
 where \(r= (s)\) for a spin-\(s\) field (rank-\(s\) totally symmetric field), \(r=p\) for a \(p\)-form, \(r=\frac12\) for a spinor in \(D\) dimensions  and \(r=\frac32\) for a vector-spinor.
 Unless specified, the operator acts on generic unconstrained fields;  \(\perp\) is explicitly added to denote transverse and (gamma-)traceless fields.
 
When breaking the \(D\)-dimensional covariance  to \(\ads_{d_A} \times S^{d_S}\) factors we distinguish fields of type \((r,r')\), indicating representation \(r\) in AdS and \(r'\) in \(S\), and correspondingly determinants become
 \begin{equation}\label{det2}
  \det_{r,r'  } \Delta\,,
 \end{equation}
where again \( r,r' = (s)  ,p, \frac12,\frac32\), now referring to  quantities in \(d_A\) or \(d_S\) dimensions and possibly restricted to \(\perp \) components. 
 A particular case is that of the \(\ads\) modes, discussed in section~\ref{sec::path-integral-general}, arising from the zero mode of the harmonic decomposition on the sphere and retaining only AdS dependence, which are denoted by \(r' = \emptyset\).

\subsubsection*{Spectral measures}  
For
spin-\(s\) transverse-traceless tensor we have the following representation of the zeta function~\cite{Camporesi:1994ga}  
\begin{equation}\label{baa}
\zeta_\Delta (z) =
\frac{\VOL_{\ads_{d_A}}}{\VOL_{ { S^{(d_{\! A}-1)   }}    }   }   
\frac{2^{d_A-2} } \pi   g(s,d_A)
\intl_0 ^{+\infty} \! d v\,
\frac{ \mu_{d_{\!A}}(v)}{( v^2 + \mathrm{b}^2)^z}
\,,
\qquad \mathrm{b}^2 = \left(\frac{d_{A}-1}{2}\right)^2 + s + \kappa =  \text{const.}\,,
\end{equation}
with
\begin{equation}\label{bab}  
g(s,d_A)=
\frac{(2s+d_A-3) \ (s+d_A-4)!}{s!(d_A-3)!} \, ,
\end{equation}
such that  for \(d_A=3\) one has \(g(0,3)=1\), \(g(s\geq 1,3)=2\) and   
\begin{align}\label{bae}
\mu _{d_A}(v ) & = \pi
\frac{v ^2+\left(s + \frac {d_A-3}2\right) ^2 }{ \left(2^{d_A-2}\Gamma\left[\frac {d_A}2\right]\right)^2 } 
\prod_{j=0}^{ \frac{d_A-5}2} (v ^2 + j^2)  \,, \qquad \text{\(d_A\) odd}\,,
\\
\label{baf}
\mu _{d_A}(v ) & = \pi
\frac{v ^2+\left(s + \frac {d_A-3}2\right) ^2 }{ \left(2^{d_A-2}\Gamma\left[\frac {d_A}2\right]\right)^2 } 
v  \tanh(\pi v )
\prod_{j=\frac12}^{ \frac{d_A-5}2} (v ^2 + j^2)  \,, \qquad \text{\(d_A\) even}.
\end{align}
In the products the step is 1 and they are absent for \(d_A=3,4\).

For transverse $p$-forms one has~\cite{Camporesi199457,Camporesi:1991nw,Camporesi:1993mz}
    \begin{equation}
        \label{cqc}
        \zeta_{\Delta_{p\!\perp}}
            =
                \frac{
                        2^{d_A-2}
                    }{
                        \pi
                    }
                \frac{
                        \VOL_{\ads_{d_A}}
                    }{
                        \VOL_{S^{d_A-1}}
                    }
                \,g_p(p,d_A)\,
                \intl_0 ^{+\infty} \! d v\,
                \frac{ 
                        \mu_p(v)
                    }{
                        \left(v^2 + \mathrm b^2\right)^z
                    }\, ,  
                                        \qquad \mathrm{b}^2 = \left(\frac{d_{A}-1}{2}\right)^2 + p + \kappa =  \text{const.}\,,
    \end{equation}
with 
\begin{align} 
    \mu_p(v) = 
        {} & 
            \frac{
                    \pi  
                }{        
                    \left[(\frac{d_A-1}{2} -p)^2+v^2\right] 
                    \left[2^{d_A-2} \Gamma \left(\frac{d_A}{2}\right)\right]^2
                }     
            \prod _{j=0}^{\frac{d_A-1}{2}} \left(j^2+v^2\right)      \, ,
                        &&d_A \text{ odd}\,,
        \\
    \mu_p(v) = 
        {} & 
            \frac{
                    \pi  
                }{        
                    \left[(\frac{d_A-1}{2} -p)^2+v^2\right] 
                    \left[2^{d_A-2} \Gamma \left(\frac{d_A}{2}\right)\right]^2
                }    
            v \tanh(\pi v) \prod _{j=\frac12}^{\frac{d_A-1}{2}} \left(j^2+v^2\right)       \, ,
                        &&d_A \text{ even}\,,
\end{align}
and, for $d_A > 3$, 
\begin{equation}\label{gp}
    g_p(p, d_A)=\frac{(d_A-1)!}{ (d_A-p-1)! \, p!}\,.
\end{equation}

\subsubsection*{Seeley--DeWitt coefficients}
Seeley--DeWitt coefficients \eqref{kag} for the second order operator \(\Delta = -\cd^2 +X \) relevant for this paper are (see e.g.~\cite{Vassilevich:2003xt,Avramidi:2000bm,Gilkey:1975iq})
 \begin{align} \label{b4}
 \begin{aligned}
 	b_4(\Delta)
 =& \tr\Big[  	 
 		\frac 1{12} W_{mn} W^{mn} + \frac 12 X^2 - \frac 16 XR
 		+ \frac 1{180} R_{mnrs} R^{mnrs} - \frac 1 {180} R_{mn} R^{mn } + \frac 1 {72} R^2
  		\Big] ,
 \end{aligned}
 \end{align} 
 and
 \begin{align} \label{app2}
 \begin{aligned}
 	b_6(\Delta)
 =& \tr\Big[  	 
 		- \frac{1}{60} \left( \cd_m W_{mn}  \right)^2
 		+ \frac{1}{90} W_{mn} W_{nr} W_{rm}
 		- \frac{1}{12} X W_{mn} W_{mn}
 		+ \frac{1}{12} X \cd^2 X
 		- \frac{1}{6} X^3  
 \\
 & \qquad
 		+\frac1 {12} R X^2 
 		- \frac{1}{72} R^2 X 
 		 - \frac1{30} X \cd^2 R
 		+ \frac{1}{180} X R_{mn}R_{mn}  
 		- \frac{1}{180} X R_{mnrs}R_{mnrs} 
 \\
 	& \qquad
 					+\frac{1}{72} R W _{mn} W^{mn} 
 - \frac{1}{90} R_{mn} W^{m r}W^{rn} 
 + \frac{1}{180} R_{mnpq} W_{mn} W_{pq}
 	+  \mathfrak E_6 \cdot  \mathbbm 1 
 		\Big] ,
 \end{aligned}
 \end{align} 
 \begin{equation}\label{geom}
 \begin{aligned}
 \mathfrak E_6 
 ={}
 & 
 		\frac{1}{945} R_{mn} R_{pq} R_{mpnq}
 		+ \frac{1}{7\,560} R_{mn} R_{mpqr} R_{npqr}
 		- \frac{4}{2\,835} R_{mn} R_{np} R_{pm}
 \\
 & \qquad
 		+ \frac{17}{45\,360} R\indices{_{mn}^{pq}}  R\indices{_{pq}^{rs}}  R\indices{_{rs}^{mn}} 
 		- \frac{1}{1\,620} R\indices{_m^p_n^q}  R\indices{_p^r_q^s}  R\indices{_r^m_s^n}  
 		+ \frac{1}{840} R_{mn} \cd^2 R^{mn}
 \\
 & \qquad
 		+ \frac{1}{1\,296} R^3
 		+ \frac{1}{1\,080} R R_{mpqr} R_{mpqr}
 		- \frac{1}{1\,080} R R_{mn} R_{mn}
 		+ \frac{1}{336} R   \cd^2 R 
 		\,,
 \end{aligned}
 \end{equation}
  where \(W_{mn}\) is the curvature associated with the internal degrees of freedom.

\subsubsection*{Hodge--de Rham operators}
We use the following Hodge--de Rham operators
\begin{align}
    \label{eq:Delta0HdR1form}
    \Delta_0 C = & {} -\nabla^2 C\, ,\\
        \label{eq:Delta1HdR1form}
        \Delta_1 C_M = & {} -\nabla^2 C_M + R\indices{_M^N} C_N\, ,\\
    \label{eq:Delta2HdR2form}
    \Delta_2 C_{MN} = & {} -\nabla^2 C_{MN} -2  R\indices{^E_{[M}}C_{N]E} -2 R_{MDNE}C^{DE}\, ,\\
    \label{eq:Delta3HdR3form}
    \Delta_3 C_{MNP} = & {} -\nabla^2C_{MNP} - 6 R\indices{_{[M}^Q_N^R}C_{P]QR} + 3 R\indices{_{[M}^R}C_{NP]R}\,.
\end{align}
In general, they satisfy
\begin{equation} 
       \cd^A\Delta_{p+1} C_{AB_1 \cdots B_p} 
        = \Delta_p \cd^A C_{AB_1 \cdots B_p}  \,,
\end{equation}
on arbitrary geometrical backgrounds.

\subsubsection*{Eigenvalues of Laplacians on spheres}

    The spectrum of the Laplacian on  spin-\(s\) tensors, namely a symmetric traceless transverse tensors  of rank $s$, on unit $S^{d_S}$ is~\cite{Camporesi:1994ga}
    \begin{equation}
        \begin{split}
            \lambda_l^{s,d_S} &= (l+s)(l+s+d_S-1) -s  , \qquad\qquad l=0,1,2,3, \ldots\, ,\\
            m_l^{s,d_S} &= g (s,d_S ) \frac{(l+1)(l+2s + d_S-2)(2l+2s + d_S -1)(l+s+d_S-3)!}{(d_S-1)!\,(l+s+1)!}\,,
        \end{split}
        \label{eq:eigenvaluesSphericalHarmonicsPformSymmetric}
    \end{equation}
    with 
    \begin{equation}
        g (s,d_S ) = \frac{(2s+d_S-3)(s+d_S-4)!}{(d_S-3)!\,s!}\,.
    \end{equation}

    The spectrum of the Hodge--de Rham operator $\Delta_p$ on co-exact $p$-forms on a unit sphere $S^{d_S}$ is given in appendix B of~\cite{Copeland:1984qk}. From it, we obtain the spectrum of the standard Laplacian as
    \begin{equation}
        \label{eq:eigenvaluesSphericalHarmonicsforLaplacian}
        \begin{split}
                       &\lambda_{l}^{p,d_S} = (l+p+1)(l+d_S-p) - p\, (d_S - p)\, , 
                       \qquad\qquad l=0,1,2,3,\ldots\, ,
            \\
            & m_l^{p,d_S} =  \frac{(2l + d_S + 1)\,(l+d_S)!}{\,(l+d_S-p)\,(l+p+1)\,(d_S-p-1)! \, p\,! \, l\,!}\,.
        \end{split} 
    \end{equation} 
    
\subsubsection*{Eigenvalues of the Dirac operator on spinors} \label{appendix:eigenvalues-spinors}
        The spectrum  of the Dirac operator $i\nablasl$ acting on spinors  on unit $S^{d_S}$ is~\cite{Camporesi:1995fb}
        \begin{align} 
                \lambda_l^{\frac{1}{2},d_S} = & {} \pm \left(l+\frac{d_S}{2}\right)\,  , \qquad\qquad l=0,1,2,3,\ldots\,,\label{eq:nablaslEigenvalues}\\
                m_l^{\frac{1}{2},d_S} = & {} \frac{2^{\left\lfloor\frac{{d_S}}{2}\right\rfloor}}{({d_S}-1)!}\frac{({d_S}+l-1)!}{l\, !}\, ,\label{eq:nablaslDeg}
        \end{align}
while acting on transverse gamma-traceless vector-spinors  is~\cite{bures1998eigenvalues,Homma:2020xvg} 
        \begin{align} 
                \lambda_l^{\frac{3}{2},d_S} = {} &  \pm \left(l+1+\frac{d_S}{2}\right)\,  , \qquad \qquad\qquad l=0,1,2,3,\ldots\,,\label{eq:nablaslEigenvalues-32}\\
                m_l^{\frac{3}{2},d_S} = {} &  2^{\left\lfloor \frac{d_S}{2}\right\rfloor } \frac{(d_S-2) }{(d_S-1)!}\frac{ (l+1+d_S)!}{(l+2)  \, (l+d_S)\,l\,!}
        \end{align}

\section{Partition functions on \texorpdfstring{$\ads_{d_A} \times S^{d_S}$}{AdS x S}}  \label{app:partf}

In this appendix we provide several explicit examples of functional determinants expressed in terms of their \(\ads \times S\) components as outlined in section~\ref{sec::path-integral-general}.

\subsection{Vector}
\label{app:vect}

Consider the following operator acting on vectors $A^M$
\begin{equation}\label{dsi}
 [\Delta_1 ]_{MN}
	 = - g_{MN} \cd^2  + c g_{MN} + k R_{MN}\,,
\end{equation}
with \(c\) and \(k\)  constants. 
We want to express the operator in terms of the \(\ads_{d_A} \times S^{d_S}\) components \(A_M = (A_\mu, A_m)\), further split in their transverse-traceless fields,
\begin{equation}\label{A_M-decomposition}
    A_\mu =  A_\mu^\perp + \nabla_\mu \sigma\,,  
    \qquad
    A_m =   A_m^\perp + \nabla_m \rho,
    \qquad
    \nabla^\mu A^\perp_\mu = 0 = \nabla^m A^\perp_m\,.
\end{equation}
Expressing the determinant as a path integral over \(A_M\), the decomposition \eqref{A_M-decomposition} brings a Jacobian factor
\begin{align}
    J_{1,0} = & {} \det_{0,0}\left[-\nabla_A^2\right]^{\frac{1}{2}}\, ,  \qquad J_{0,1} = {} \det_{0,0}\left[-\nabla_S^2\right]^{\frac{1}{2}}\, ,
\end{align}
cf. \eqref{pdg}.
Straightforward evaluation using \eqref{paa},\eqref{pab} gives
\begin{equation}\label{A_M-lagrangian-decomposition}
\begin{split}
    A^M\Delta_1[c,k]_{MN} A^N ={}  &   
    		A^\mu_\perp\left[-\cd^2 + c -  {k (d_A-1)}{L_A^{-2}}\right]A_\mu^\perp 
    		- \sigma \left[-\nabla_A^2\right]\left[-\cd^2 + c +(1 - k) {(d_A-1)}{L_A^{-2}}\right]\sigma
    \\ &{}
    		+ A^m_\perp\left[-\cd^2 + c +  {k (d_S-1)}{L_S^{-2}}\right]A_m^\perp 
    		- \rho \left[-\nabla_S^2\right]\left[-\cd^2 + c +(k -1) {(d_S-1)}{L_S^{-2}}\right]\rho\\
\end{split}
\raisetag{-0.3cm}
\end{equation}
Besides, from \eqref{A_M-lagrangian-decomposition} we have an AdS-mode from the action of $-\cd^2_S$ on the scalar $\rho$. The associated contribution  is
\begin{equation}
   \det_{0,\emptyset}\left[-\cd^2_A + c +(k  -1) {(d_S-1)}{L_S^{-2}}\right] \, .
\end{equation}
As a result
\begin{equation}\label{fds}
    \frac{1}{\det\limits_1[\Delta_1 [c,k]]^\frac{1}{2}}= Z_1 = Z_{1,0} \times Z_{0,1}\, ,
\end{equation}
where 
\begin{align}
    Z_{1,0}   = {} &  \frac{1}{\det\limits_{1,0}  [-\cd^2 + c - \tfrac{d_A-1}{L_A^2}  k]^\frac{1}{2}}\, ,\qquad 
    Z_{0,1}   = {}   \frac{1}{\det\limits_{0,1}   [-\cd^2 + c + \tfrac{d_S-1}{L_S^2}  k]^\frac{1}{2}}\, ,   
\end{align}
and, in particular,
\begin{align}
    Z_{1,0} = {} & 
        \frac{1}
        {
            \det\limits_{1\perp,0}[-\cd^2 +  c - \tfrac{d_A-1}{L_A^2}  k  ]^\frac12 \,
	        \det\limits_{0,0}[-\cd^2 +  c - \tfrac{d_A-1}{L_A^2} (k-1)   ] ^\frac12
        }\, ,\\
    Z_{0,1} = {} & 
        \frac{
            \det\limits_{0,\emptyset}     [-\cd_A^2 +  c + \tfrac{d_S-1}{L_S^2} (k-1 ) ]^\frac12
         }  {
            \det\limits_{0,1\perp}[-\cd^2 +  c + \tfrac{d_S-1}{L_S^2}  k  ] ^\frac12 \,
		    \det\limits_{0,0}     [-\cd^2 +  c + \tfrac{d_S-1}{L_S^2} (k-1 ) ]^\frac12
        }\, .
\end{align}

\subsection{Symmetric 2-tensor} \label{appendix:subsec:sym2}

Consider the operator $\Delta_{(2)} = -\nabla^2 + \kappa$  with constant \(\kappa\) acting on a symmetric tensor $h_{MN}$.
Decomposing $h_{MN}$ in \(\ads_{d_A} \times S^{d_S}\) components we get three types of fields: $h_{\mu\nu}, h_{\mu n}$, and $h_{mn}$,
\begin{equation}\label{dsw}
h^{MN}\Delta_{(2)}  h_{MN}
=
h^{\mu \nu }\Delta_{(2)}   h_{\mu \nu }
+
h^{mn}\Delta_{(2)}   h_{mn}
+
2h^{\mu n}\Delta_{(2)} h_{\mu n}\, .
\end{equation}
We split them according to 
\begin{align}
h_{\mu\nu} &= 
    k_{\mu\nu}^\perp + \nabla_{(\mu}k^\perp_{\nu)} +\Big[ \nabla_\mu \nabla_\nu - \frac{1}{d_A}g_{\mu\nu}\nabla^2_A \Big] k + \frac{1}{d_A}g_{\mu\nu}U\, ,
    & (\cd^\mu k^\perp_{\mu\nu} = g^{\mu \nu } k^\perp _{\mu\nu} = \cd ^\mu k _\mu^\perp =0)\,,
\label{eq:decompositionh_munuR}\\ 
h_{mn} &=   
    k_{mn}^\perp + \nabla_{(m}k^\perp_{n)} + \Big[ \nabla_m \nabla_n - \frac{1}{d_S}g_{mn}\nabla^2_S \Big] \varkappa + \frac{1}{d_S}g_{mn}V\, ,
    & (\cd^m k^\perp_{mn} = g^{m n } k^\perp _{mn} = \cd ^m k _m^\perp =0)\,,
\label{eq:decompositionh_mnR}\\
h_{\mu n} &= 
    h_{\mu n}^{\perp} + \nabla_\mu q^\perp_n + \nabla_n q^\perp_\mu + \nabla_n\nabla_\mu \, q\, ,
    & (\cd^m q_m^\perp = \cd^\mu q_\mu^\perp  =0)\,,
\label{eq:decompositionh_munR}
\end{align}
where as usual \( \perp\) means transverse-traceless in all indices.
The associated Jacobians are
\begin{align} 
    J_{(2),0} & { }=  
    \det_{1\perp,0}\left[-\nabla_A^2 + {(d_A-1)}{L_A^{-2}}\right]^{\frac{1}{2}} \ \det_{0,0}\left[-\nabla_A^2\right]^{\frac{1}{2}}   \
    \det_{0,0}\left[-\nabla^2_A + {d_A}{L_A^{-2}}\right]^{\frac{1}{2}} \,, \notag \\
    \label{eq:2sym-jacobians}
    J_{1,1} & { }=   
        \det_{0,1\perp}\left[-\nabla_A^2\right]^\frac{1}{2}   \
        \det_{1\perp,0}\left[-\nabla_S^2\right]^\frac{1}{2}   \
        \det_{0,0}\left[-\nabla_S^2\right]^\frac{1}{2}   \ 
        \det_{0,0}\left[-\nabla_A^2\right]^\frac{1}{2} \ 
        \det_{0,\emptyset}\left[-\nabla_A^2\right]^\frac{1}{2}   \  \,, \\
    J_{0,(2)} & { }=  
        \det_{0,1\perp}\left[-\nabla_S^2 - {(d_S-1)}{L^{-2}_S}\right]^{\frac{1}{2}}    \
        \det_{0,0}\left[-\nabla_S^2\right]^{\frac{1}{2}} \
        \det_{0,0}\left[-\nabla^2_S - {d_S}{L_S^{-2}}\right]^{\frac{1}{2}} \,. \notag 
\end{align}
Direct evaluation of \eqref{dsw} with this decomposition gives 
\begin{align}
 h_{\mu\nu}[-\nabla^2 + \kappa]h^{\mu\nu} ={} &
 	 k^{\mu\nu}_\perp[-\nabla^2 + \kappa] k_{\mu\nu}^\perp 
 	 +  k_\perp^{ \mu}  [-\nabla_A^2 +\tfrac{d_A-1}{L_A^2} ]
 	 				 [-\nabla^2 + \kappa + \tfrac{d_A+1}{L_A^2} ]
 	 				k_\mu^\perp 
		\notag \\ &{}
		+ \tfrac{d_A-1}{d_A} k \, [-\nabla_A^2 ]
					  [-\nabla_A^2 + \tfrac{d_A}{L_A^2} ]
					 		\big[-\nabla^2 + \kappa + \tfrac{2d_A}{L_A^2}\big] k 
		 + U  [-\nabla^2 + \kappa ] U\, ,\\ 
 h_{mn}[-\cd^2 + \kappa]h^{mn} = {}& 
	  			k^{mn}_\perp [-\cd^2 + \kappa] k_{mn}^\perp 
	  			+ k_\perp^{ m}  [-\nabla^2_S -\tfrac{d_S-1}{L^2_S} ]
	  						 [-\cd^2 + \kappa - \tfrac{d_S+1}{L_S^2} ]k_m^\perp 
 \notag \\ &{} 
 			 + \tfrac{d_S-1}{d_S}\varkappa
	 			 				 [-\nabla_S^2 ]
	 			 				 [-\nabla_S^2 - \tfrac{d_S}{L_S^2} ]
	 			 				 [-\cd^2 + \kappa - \tfrac{2d_S}{L^2_S} ]
	 			 	\varkappa
 			 + V \left[-\cd^2 + \kappa\right] V\, ,
 \\
 h_{\mu n}[-\nabla^2 + \kappa]h^{\mu n} ={}&
 				 h^{\mu n}_{\perp}[-\nabla^2 + \kappa]h^{\perp}_{\mu n} 
 				 + q^{n}_\perp  [-\nabla^2 +  \kappa + \tfrac{d_A-1}{L_A^2} ]
 				 			[-\cd_A^2]q_n^\perp 
 				 + q^{ \mu}_\perp  [-\nabla^2 +\kappa - \tfrac{d_S-1}{L_S^2} ]
 				 			[-\nabla_S^2]q_\mu^\perp
 	\notag \\ &{}
 				+ q  [-\cd^2 + \kappa + \tfrac{d_A-1}{L_A^2} - \tfrac{d_S-1}{L_S^2} ]
 				[-\nabla_S^2][-\nabla_A^2]q \, .
\end{align}
When evaluating the path integral on \(h_{mn}\) and \(h_{\mu n}\) we have several sources for zero modes. For example, \(k_m^\perp\) displays a zero mode due to the operator 
\begin{equation}\label{iws}
-\nabla^2_S -\frac{d_S-1}{L_S^2}	   \to  
\lambda_l(1,d_S) L_S^{-2} - (d_S-1)L_S^{-2}
=
l(l+d_S+1)   L_S^{-2}  \, ,
\end{equation} 
for
\(l=0\), with multiplicity  $m_{l=0}(1,d_S) = \frac{1}{2} d_S (d_S+1)$.
The induced AdS operator is   
\( -\cd^2 + \kappa - \frac{d_S+1}{L_S^2} 
\to -\cd^2_A + \kappa -2  \).
The following table shows the results of the same analysis for all the other fields.
\begin{center}
\begin{tabular}{ccccc}
\hline
Field & Operator & Zero mode & Induced AdS operator  & Multiplicity \\
\hline
\(k_m^\perp\) & \(-\nabla^2_S -\frac{d_S-1}{L_S^2}\)  & \(l=0\) & \(-\cd^2_A + \kappa -2 \) & \(\frac{1}{2} d_S (d_S+1)\) \\[0.5em]
\(\varkappa\) & \(-\nabla_S^2 \)  & \(l=0\) & \(-\cd_A^2 + \kappa - 2 \tfrac{d_S}{L_S^2}\) & \(1\)\\ 
\(\varkappa\)& \(-\nabla_S^2-\frac{d_S}{L^2_S} \)& \(l=1\) & \(-\cd_A^2 + \kappa - \tfrac{d_S}{L_S^2}\) & \(d_S+1\)\\[0.5em]
\(q^\perp_\mu\) & \(-\nabla_S^2 \)  & \(l=0\) & \(-\cd^2_A + \kappa - \tfrac{d_S-1}{L_S^2}\)& \(1\)\\[0.5em]  
\(q\) & \(-\nabla_S^2 \)  & \(l=0\) & \(-\cd_A^2 + \kappa - \tfrac{d_S}{L_S^2} +\tfrac{d_A}{L_A^2}\)& \(1\)\\
\hline
\end{tabular}
\end{center}                     
As a result,
\begin{equation}\label{dsr}
\frac{1}{ \det\limits_{(2)} [-\cd^2 +\kappa] ^{\frac12}  }  = Z_{(2)}
= Z_{(2),0} \times 
Z_{0,(2)}\times
Z_{1,1}\, ,
\end{equation}
where 
\begin{align}
    Z_{(2),0}   = {} &  \frac{1}{\det\limits_{(2),0}  [-\cd^2 + \kappa]^\frac{1}{2}}\, ,\quad 
    Z_{0,(2)}   = {}   \frac{1}{\det\limits_{0,(2)}  [-\cd^2 + \kappa]^\frac{1}{2}}\, , \quad 
    Z_{1,1}     = {}   \frac{1}{\det\limits_{1,1}    [-\cd^2 + \kappa]^\frac{1}{2}}\, ,
\end{align}
and, in particular,
    \begin{align}\label{dsr1}
        Z_{(2),0}=
        & \frac{1}{
            \det\limits_{(2)\perp,0}[-\nabla^2+\kappa]^{\frac{1}{2}} 
            \det\limits_{1\perp,0}[-\nabla^2+\kappa + \frac{(d_A+1)}{L_A^2}]^{\frac{1}{2}}\det\limits_{0,0}[-\nabla^2+\kappa - \frac{2d_A}{L^2_A}]^{\frac{1}{2}}\det\limits_{0,0}[-\nabla^2+\kappa]^{\frac{1}{2}}
            }\, ,\\
        \label{dsr2}
        Z_{0,(2)}=
        & \frac{
            \det\limits_{0,\emptyset}[-\nabla^2_A + \kappa -2]^{\frac{d_S(d_S+1)}{4}} \det\limits_{0,\emptyset}[-\nabla^2_A + \kappa -2\frac{d_S}{L_S^2}]^{\frac{1}{2}} 
            \det\limits_{0,\emptyset}[-\nabla^2_A + \kappa -\frac{d_S}{L_S^2}]^{\frac{d_S+1}{2}}}
            {
            \det\limits_{0,(2)\perp}[-\nabla^2+\kappa]^{\frac{1}{2}}
            \det\limits_{0,1\perp}[-\nabla^2+\kappa - \frac{(d_S+1)}{L^2_S}]^{\frac{1}{2}}\det\limits_{0,0}[-\nabla^2+\kappa - \frac{2d_S}{L^2_S}]^{\frac{1}{2}}\det\limits_{0,0}[-\nabla^2+\kappa]^{\frac{1}{2}}
            } \, ,\\
        \label{dsr3}
        Z_{1,1} =& 
        \frac{
            \det\limits_{1\perp,\emptyset}[-\nabla^2_A+\kappa -\frac{d_S-1}{L_S^2}]^\frac{1}{2}
            \det\limits_{0,\emptyset}[-\nabla^2_A+\kappa -\frac{d_S}{L^2_S}+\frac{d_A}{L_A^2}]^\frac{1}{2}}
            {
            \det\limits_{1\perp,1\perp}[-\nabla^2+\kappa]^{\frac{1}{2}}\det\limits_{1\perp,0}[-\nabla^2+\kappa - \frac{d_S-1}{L^2_S}]^{\frac{1}{2}}\det\limits_{0,1\perp}[-\nabla^2+\kappa+ \frac{d_A-1}{L^2_A}]^{\frac{1}{2}}\det\limits_{0,0}[-\nabla^2+\kappa + \frac{d_A-1}{L^2_A} - \frac{d_S-1}{L_S^2}]^{\frac{1}{2}}
            } \,,
    \end{align} 
where we have already simplified the Jacobians in \eqref{eq:2sym-jacobians}.

\subsection{Two-form}
Consider the operator $\Delta_2  = -\nabla^2 + \kappa$ with constant $\kappa$ acting on a two-form $A_{MN}$. Decomposing $A_{MN}$ in $\ads_{d_A} \times S^{d_S}$ components we get three types of fields: $A_{\mu\nu}, A_{\mu n}, A_{nr}$:
\begin{equation} \label{eq:AM}
    A_{MN}\Delta_2 A^{MN} =
        A_{\mu \nu } \Delta_2  A^{\mu\nu} 
        + 
        2 A_{\mu n }\Delta_2  A^{\mu n} 
        + 
         A_{m n} \Delta_2  A^{m n} \, .
\end{equation}
We decompose them according to
\begin{align}
    A_{\mu\nu} = & {} 
        A_{\mu\nu}^\perp + \nabla_{[\mu}a_{\nu]}^\perp, \label{eq:decompositionA_munu}\\ 
    A_{\mu n} = & {} A_{\mu n}^{\perp} + \nabla_\mu \mathfrak a ^\perp_n + \nabla_n \mathfrak a ^\perp_\mu + \nabla_\mu \nabla_n \mathfrak a , \label{eq:decompositionA_mun}\\
    A_{mn} = & {} A_{mn}^\perp + \nabla_{[m}a_{n]}^\perp,\label{eq:decompositionA_mn}
\end{align}
where ${}^\perp$ means transverse with respect to all indices,
   \(0=  \nabla^\mu A^\perp_{\mu\nu} =    \nabla^\mu A^\perp_{\mu n} =    \nabla^n A^\perp_{\mu n}  = \) etc.
The associated Jacobians are
\begin{align}\label{eq:2form-jacobians}
    J_{2,0} =   {} &  \det_{1\perp,0} [-\nabla_A^2 -\tfrac{d_A-1}{L^2_A} ]^{\frac{1}{2}},\notag\\
    J_{1,1} =   {} &   \det_{1\perp,0} [-\nabla_S^2  ]^{\frac{1}{2}}\det_{0,1\perp} [-\nabla_A^2 ]^{\frac{1}{2}}\det_{0,0} [-\nabla^2_S ]^{\frac{1}{2}}\det_{0,0} [-\nabla^2_A ]^{\frac{1}{2}},\\
    J_{0,2} =   {} &  \det_{0,1\perp} [-\nabla_S^2 +\tfrac{d_S-1}{L^2_S} ]^{\frac{1}{2}}\notag\,.
\end{align}
From this decomposition, we get
\begin{align} 
    A_{\mu\nu}[-\cd^2 + \kappa]A^{\mu\nu} 
        = {}  & 
              A^{\mu\nu}_\perp [-\cd^2 + \kappa ]  A_{\mu\nu}^\perp 
                +  \tfrac12 a^{\mu}_\perp [ - \cd_A^2 - \tfrac{d_A-1}{L^2_A}][-\cd^2 + \kappa + \tfrac{(d_A-3)}{L_A^2}]a_\mu^\perp\notag\, , \\ 
    A_{\mu n}[-\cd^2 + \kappa]A^{\mu n} 
        = {} &  
             A^{\mu n}_\perp[-\cd^2 + \kappa ] A^{\perp,\perp}_{\mu n} 
                + \mathfrak a ^{n}_\perp[-\cd^2 + \kappa + \tfrac{d_A-1}{L^2_A}][-\cd^2_A]\mathfrak a _n^\perp  \\
                &+ \mathfrak a ^{\mu}_\perp [-\cd^2 + \kappa - \tfrac{d_S-1}{L^2_S}][-\cd^2_S]\mathfrak a _\mu^\perp\notag 
               + \mathfrak a  [-\cd^2 + \kappa + \tfrac{d_A-1}{L^2_A} - \tfrac{d_S-1}{L^2_S}] [-\cd^2_S][-\cd^2_A]\mathfrak a  \notag\, ,\\ 
    A_{mn}[-\cd^2 + \kappa ] A^{mn} 
        = {} &  
             A^{mn}_\perp  [-\cd^2 + \kappa ] A_{mn}^\perp 
                + \tfrac12 a^{m}_\perp [-\cd^2_S  + \tfrac{d_S-1}{L^2_S}][-\cd^2 + \kappa - \tfrac{(d_S-3)}{L_S^2}]a_m^\perp\, . \notag
\end{align}

When evaluating the path integral on \(A_{\mu r}\) we have two sources for zero modes, summarized in the following table 
\begin{center}
\begin{tabular}{ccccc}
\hline
Field & Operator & Zero mode & Induced AdS operator  & Multiplicity \\
\hline
\(\mathfrak a _{\mu}^\perp\) & \(-\nabla^2_S \)  & \(l=0\) & \(-\cd_A^2 + \kappa - \tfrac{d_S-1}{L^2_S}\) & \(1\) \\[0.5em]
\(\mathfrak a\)& \(-\nabla_S^2 \)  & \(l=0\) & \(-\cd^2_A + \kappa + \tfrac{d_A-1}{L^2_A} - \tfrac{d_S-1}{L^2_S}\) & \(1\)\\ 
\hline
\end{tabular}
\end{center} 
As a result,
\begin{equation}\label{dsr-2form}
    \frac{1}{ \det\limits_{2} [-\cd^2 +\kappa] ^{\frac12}  }  = Z_{2}
    = Z_{2,0} \times 
    Z_{1,1}\times
    Z_{0,2}\, ,
\end{equation}
where 
\begin{align}
    Z_{2,0}   = {} &  \frac{1}{\det\limits_{2,0}  [-\cd^2 + \kappa]^\frac{1}{2}}\, ,\quad 
    Z_{0,2}   = {}    \frac{1}{\det\limits_{0,2}  [-\cd^2 + \kappa]^\frac{1}{2}}\, , \quad 
    Z_{1,1}   = {}    \frac{1}{\det\limits_{1,1}    [-\cd^2 + \kappa]^\frac{1}{2}}\, ,
\end{align}
and, in particular,
\begin{align}\label{dsr-2form1}
    Z_{2,0} = 
        {} & 
        \frac{1}{
            \det\limits_{2\perp,0} [-\nabla^2+\kappa]^\frac{1}{2}
            \det\limits_{1\perp,0} [-\nabla^2+\kappa + \frac{d_A-3}{L_A^2}]^\frac{1}{2}
            }\, ,\\ 
    \label{dsr-2form2}
    Z_{1,1} = 
        {} & 
        \frac{
                \det\limits_{1\perp,\emptyset}[-\nabla_A^2 + \kappa - \frac{d_S-1}{L_S^2}]^{\frac{1}{2}} 
                \det\limits_{0,\emptyset}[-\nabla_A^2 + \kappa + \frac{d_A-1}{L_A^2} - \frac{d_S-1}{L_S^2}]^{\frac{1}{2}}
            }{
                \det\limits_{1\perp,1\perp} [-\nabla^2+\kappa]^\frac{1}{2}
                \det\limits_{1\perp,0} [-\nabla^2+\kappa - \frac{d_S-1}{L_S^2}]^\frac{1}{2}
                \det\limits_{0,1\perp} [-\nabla^2+\kappa + \frac{d_A-1}{L_A^2}]^\frac{1}{2}
                \det\limits_{0,0} [-\nabla^2+\kappa + \frac{d_A-1}{L_A^2} - \frac{d_S-1}{L_S^2}]^\frac{1}{2}
            }\, ,\\ 
    \label{dsr-2form3}
    Z_{0,2} = 
        {} & 
        \frac{1}{
            \det\limits_{0,2\perp} [-\nabla^2+\kappa]^\frac{1}{2}
            \det\limits_{0,1\perp} [-\nabla^2+\kappa - \frac{d_S-3}{L_S^2}]^\frac{1}{2}
        }\,.
    \end{align}
where we have already simplified the Jacobians in \eqref{eq:2form-jacobians}.
%In the case of $d_A=4$ it is convenient to further dualize the components with $2$ $\ads$-indices as
%\begin{align*}
%    A^\perp_{ \mu \nu } = 
%        {} & 
%        \frac{1}{2}   \epsilon_{\mu\nu\rho\sigma} \frac{1}{-\nabla_A^2 - 16}\nabla^\sigma \hat{B}^\rho_{\perp},  
%\end{align*}

\subsection{Three-form}   \label{appendix:Z-3form}
Consider the operator $\Delta_3  = -\nabla^2 + \kappa$ with constant $\kappa$ acting on a three-form $A_{MNP}$. Decomposing $A_{MNP}$ in $\ads_{d_A} \times S^{d_S}$ components we get four types of fields: $A_{\mu\nu\rho}, A_{\mu \nu r}, A_{\mu nr}$ and $ A_{mnr}$:
\begin{equation} \label{eq:AMN}
    A_{MNP}\Delta_3 A^{MNP} =
        A_{\mu \nu \rho} \Delta_3  A^{\mu\nu\rho} 
        + 
        3 A_{\mu \nu n}\Delta_3  A^{\mu \nu n} 
        + 
        3 A_{\mu mn} \Delta_3  A^{\mu mn} 
        + 
        A_{m n p}\Delta_3   A^{m n p}.
\end{equation}
We decompose them according to
\begin{align}
    \label{eq:decompositionA_munurhoR}
    A_{\mu\nu\rho} = 
        {}&  
        A_{\mu\nu\rho}^\perp + 
        \nabla_{[\mu}a_{\nu\rho]}^\perp\, , \\
    \label{eq:decompositionA_munR}
    A_{\mu\nu n} =
        {}&  
        A_{\mu\nu n}^{\perp} + 
        \nabla_{[\mu} \mathfrak a_{\nu] n}^{\perp} + 
        \nabla_n \mathfrak a_{\mu\nu}^\perp + 
        \nabla_{n}\nabla_{[\mu} \mathfrak a^\perp_{\nu]}\, ,\\ 
    \label{eq:decompositionA_mnR}
    A_{\mu mn} = 
        {}&  
        A_{\mu mn}^{\perp} + 
        \nabla_\mu  a_{mn}^\perp + 
        \nabla_{[m}  a_{n]\mu}^{\perp} + 
        \nabla_\mu \nabla_{[m} a^\perp_{n]}\, , \\
    \label{eq:decompositionA_mnpR}
    A_{mnp} = 
        {}&
        A_{mnp}^\perp + \nabla_{[m}\mathfrak a_{np]}^\perp, 
\end{align}
where ${}^\perp$ means transverse with respect to all indices, 
\(0 = \cd^\mu A_{\mu\nu\rho}^\perp = \nabla^\mu A_{\mu\nu r}^\perp =    \nabla^r A_{\mu\nu r}^{\perp} \) etc. 
The associated Jacobians are 
\begin{align} \label{eq:3form-jacobians}
    J_{3,0} =  {} &   
        \det_{2\perp,0} [-\nabla_A^2 - \tfrac{2(d_A-2)}{L_A^2} ]^{\tfrac{1}{2}}\, ,\\
    J_{2,1} =  {} &  
        \det_{1\perp,1\perp} [-\nabla_A^2 -\tfrac{d_A-1}{L^2_A} ]^{\tfrac{1}{2}}
        \det_{2\perp,0} [-\nabla_S^2 ]^{\tfrac{1}{2}}
        \det_{1\perp,0} [-\nabla^2_S ]^{\tfrac{1}{2}}
        \det_{1\perp,0} [-\nabla_A^2 -\tfrac{d_A-1}{L^2_A} ]^{\tfrac{1}{2}}
        \det_{1\perp,\emptyset} [-\nabla_A^2 -\tfrac{d_A-1}{L^2_A} ]^{\tfrac{1}{2}}\, ,\notag\\
    J_{1,2} = {} &  
        \det_{1\perp,1\perp} [-\nabla_S^2 +\tfrac{R_S}{d_S} ]^{\tfrac{1}{2}}
        \det_{0,2\perp} [-\nabla_A^2 ]^{\tfrac{1}{2}}
        \det_{1\perp,0} [-\nabla^2_A ]^{\tfrac{1}{2}}
        \det_{1\perp,0} [-\nabla_S^2 +\tfrac{d_S-1}{L^2_S} ]^{\tfrac{1}{2}}\, ,\notag\\ 
    J_{0,3} = {} &  
        \det_{0,2\perp} [-\nabla_S^2 +\tfrac{2(d_S-2)}{L_S^2} ]^{\tfrac{1}{2}}\, ,\notag
\end{align}
and the various pieces in \eqref{eq:AMN} become
\begin{align} 
    A_{\mu\nu\rho}[-\cd^2 + \kappa]A^{\mu\nu\rho} = {}  
        & A^{\mu\nu\rho}_\perp [-\nabla^2 + \kappa ] A_{\mu\nu\rho}^\perp  
        +  \tfrac13 a_\perp^{\nu \rho} [-\nabla_A^2 - \tfrac{2(d_A-2)}{L^2_A}][-\nabla^2 + \kappa + \tfrac{(d_A-5)}{L^2_A}]a_{\nu\rho}^\perp\, ,\\ 
    A_{\mu\nu n}[-\cd^2 + \kappa]A^{\mu \nu n} = {} 
        & A_\perp^{\mu \nu n}[-\nabla^2 + \kappa ]A^{\perp}_{\mu \nu n}  + \tfrac12 \mathfrak a_{\perp}^{\mu n} [-\nabla^2 +\kappa + \tfrac{(d_A -3)}{L^2_A}][-\nabla_A^2 - \tfrac{d_A-1}{L^2_A}]\mathfrak a_{\mu n}^\perp \notag \\
        &+ \mathfrak a_\perp^{\mu \nu} [-\nabla^2 +\kappa - \tfrac{d_S-1}{L^2_S}][-\nabla_S^2]\mathfrak a_{\mu\nu}^\perp \notag \\
        &+ \tfrac12 \mathfrak a_\perp^{ \nu} [-\nabla^2 + \kappa + \tfrac{(d_A-3)}{L^2_A} - \tfrac{d_S-1}{L^2_S}] [-\nabla_S^2] [-\nabla_A^2 - \tfrac{d_A-1}{L^2_A}]\mathfrak a_\nu^\perp\, , \\ 
    A_{\mu mn}[-\cd^2 + \kappa]A^{\mu mn} = {} 
        & A_\perp^{\mu m n}[-\nabla^2 + \kappa ]A^{\perp}_{\mu m n} + \tfrac12  a_\perp^{\mu n} [-\nabla^2 + \kappa - \tfrac{(d_S-3)}{L^2_S}][-\nabla_S^2 + \tfrac{d_S-1}{L^2_S}] a_{\mu n}^{\perp} \notag \\
        &+  a_\perp^{ m n} [-\nabla^2 + \kappa + \tfrac{d_A-1}{L^2_A}][-\nabla_A^2] a_{m n}^\perp \notag \\
        &+ \tfrac12 a_{\perp}^m [-\nabla^2 + \kappa + \tfrac{d_A-1}{L^2_A} - \tfrac{(d_S-3)}{L^2_S}] [-\nabla_A^2] [-\nabla_S^2 + \tfrac{d_S-1}{L^2_S}]a_m^{\perp} \, ,\\
    A_{mnp}[-\cd^2 + \kappa]A^{m n p} = {}  
        &A^{m n p}_\perp [-\nabla^2 + \kappa ] A_{m n p}^\perp  +\tfrac13 \mathfrak a_\perp^{n p} [-\nabla_S^2 + \tfrac{2(d_S-2)}{L^2_S}][-\nabla^2 + \kappa - \tfrac{(d_S-5)}{L_S^2}]\mathfrak a_{n p}^\perp\, .
\end{align}
When evaluating the path integral on \(A_{\mu\nu r}\) we have two sources for zero modes, summarized in the following table 
\begin{center}
\begin{tabular}{ccccc}
\hline
Field & Operator & Zero mode & Induced AdS operator  & Multiplicity \\
\hline
\(\mathfrak a _{\mu \nu}^\perp\) & \(-\nabla^2_S \)  & \(l=0\) & \(-\cd^2_A + \kappa - \tfrac{d_S-1}{L_S^2} \) & \(1\) \\[0.5em]
\(\mathfrak a_\mu^\perp\)& \(-\nabla_S^2 \)  & \(l=0\) & \(-\cd_A^2 + \kappa + \tfrac{d_A-3}{L_A^2} - \tfrac{d_S-1}{L_S^2}\) & \(1\)\\ 
\hline
\end{tabular}
\end{center} 
As a result,
\begin{equation}\label{dsr-3form}
\frac{1}{ \det\limits_{3} [-\cd^2 +\kappa] ^{\frac12}  }  = Z_{3}
= Z_{3,0} \times 
Z_{2,1}\times
Z_{1,2}\times
Z_{0,3}\, ,
\end{equation}
where 
\begin{align}
    Z_{3,0}   = {} &    \frac{1}{\det\limits_{3,0}  [-\cd^2 + \kappa]^\frac{1}{2}}\, ,\quad 
    Z_{2,1}   = {}      \frac{1}{\det\limits_{2,1}  [-\cd^2 + \kappa]^\frac{1}{2}}\, , \quad 
    Z_{1,2}   = {}      \frac{1}{\det\limits_{1,2}  [-\cd^2 + \kappa]^\frac{1}{2}}\, , \quad 
    Z_{0,3}   = {}      \frac{1}{\det\limits_{0,3}  [-\cd^2 + \kappa]^\frac{1}{2}}\, ,
\end{align}
and, in particular,
\begin{align}\label{dsr-3form1}
    Z_{3,0} = 
        {} & 
        \frac{1}{
            \det\limits_{3\perp,0} [-\nabla^2+\kappa]^\frac{1}{2}
            \det\limits_{2\perp,0} [-\nabla^2+\kappa + \frac{d_A-5}{L_A^2}]^\frac{1}{2}
            }\, ,\\ 
    \label{dsr-3form2}
    Z_{2,1} = 
        {} & 
        \frac{
                \det\limits_{2\perp,\emptyset}[-\nabla_A^2 + \kappa - \frac{d_S-1}{L_S^2}]^{\frac{1}{2}} 
                \det\limits_{1\perp,\emptyset}[-\nabla_A^2 + \kappa + \frac{d_A-3}{L_A^2} - \frac{d_S-1}{L_S^2}]^{\frac{1}{2}}
            }{
                \det\limits_{2\perp,1\perp} [-\nabla^2+\kappa]^\frac{1}{2}\det\limits_{1\perp,1\perp} [-\nabla^2+\kappa + \frac{d_A-3}{L_A^2}]^\frac{1}{2}
                \det\limits_{2\perp,0} [-\nabla^2+\kappa - \frac{d_S-1}{L_S^2}]^\frac{1}{2}
                \det\limits_{1\perp,0} [-\nabla^2+\kappa + \frac{d_A-3}{L_A^2} - \frac{d_S-1}{L_S^2}]^\frac{1}{2}
            }\, ,\\  
    \label{dsr-3form3}
    Z_{1,2} = 
        {} & 
        \frac{1}{
            \det\limits_{1\perp,2\perp} [-\nabla^2+\kappa]^\frac{1}{2}
            \det\limits_{1\perp,1\perp} [-\nabla^2+\kappa - \frac{d_S-3}{L_S^2}]^\frac{1}{2}
            \det\limits_{0,2\perp} [-\nabla^2+\kappa + \frac{d_A-1}{L_A^2}]^\frac{1}{2}
            \det\limits_{0,1\perp} [-\nabla^2+\kappa + \frac{d_A-1}{L_A^2} - \frac{d_S-3}{L_S^2}]^\frac{1}{2}
            }\, ,\\
    \label{dsr-3form4}
    Z_{0,3} = 
        {} & 
        \frac{1}{
            \det\limits_{0,3\perp} [-\nabla^2+\kappa]^\frac{1}{2}
            \det\limits_{0,2\perp} [-\nabla^2+\kappa - \frac{d_S-5}{L_S^2}]^\frac{1}{2}
        }\,.
    \end{align}
where we have already simplified the Jacobians in \eqref{eq:3form-jacobians}.

\section{Explicit evaluation of divergences in \texorpdfstring{\(\ads_{d_A} \times S^{d_S}\)}{AdS x S}} \label{appendix:divergences}

In this appendix we provide several explicit examples of the procedure discussed in section~\ref{sec::KKsum}, as well as some consequences of the results.

  \subsection{Scalar field on \texorpdfstring{\(\ads_4 \times S^2(L_S)\)}{AdS4 x S2(LS)}}
  \label{sec::scalar_ads4s2}
  
  The effective action for a scalar field is
  \begin{equation}\label{cdc}
  \Gamma = \frac12 \log \det_{0}[-\cd^2 + \kappa ] = \frac12 \log \det_{0,0}[-\cd^2_A - \cd_S^2 + \kappa ]    \, .
  \end{equation}
  Evaluate divergence via Seeley--DeWitt coefficient with unit radius \(\ads\) space
  \begin{equation}\label{ccd}
  \begin{aligned}
  \Gamma_{\log\Lambda L } 
  & = - \frac1 {2 (4 \pi)^3 } \log \big(\Lambda L\big)^2 \int\! d^6x \sqrt{g} \,b_6
  \\
  & =  \left(
  -\frac{1}{36} L_S^2 \kappa ^3+\frac{1}{36} \left(1-6 L_S^2\right) \kappa ^2-\frac{\left(29 L_S^4-10 L_S^2+1\right) \kappa }{90
     L_S^2}+\frac{-370 L_S^6+203 L_S^4-42 L_S^2+4}{1890 L_S^4}
    \right)
  \log {\Lambda L }\, ,
  \end{aligned} 
  \end{equation}
  where we used \( \VOL_{\ads_4}  = \frac{4}{3}  \pi^2  \) and \( \VOL_{S^2} = 4 \pi L_S^2\). Notice that since \(L_A=1\), both \(L_S\) and \(\kappa\) are dimensionless.
  
  Alternatively we reduce \eqref{cdc} on spherical harmonics of the sphere \(S^2\)
  \begin{equation}\label{cdp}
  \Gamma = \sum_ {l =0} ^\infty
  \frac {1} 2(2l+1) \log \det[-\cd^2_A + l (l+1) L_S^{-2} + \kappa ]
  \,,
  \qquad
  \Delta(l)=
  -\cd^2_A + l (l+1) L_S^{-2} + \kappa \, ,
  \end{equation}
  where \(2l+1\) is the multiplicity.
  The associated  function \(\zeta_{\Delta(l)}\) is thus given by
  \begin{equation}\label{cda}
  \zeta_{\Delta(l)}(z) = 
  \frac16
  \intl_0^\infty 
  	\! dv \frac{ v  (v^2 + \frac14) \tanh (\pi v )   }{
  	    \left[
  	    	 (l+\frac12)^2 L_S^{-2}
  	    	 -\frac{1}{4  }L_S^{-2}  + \kappa +v^2+\frac{9}{4}
  	      \right]^z }\, ,
  \end{equation}
  so that we want to  evaluate
  \begin{equation}\label{faq}
  \Gamma _{\log\Lambda L }  = 
   -\frac12   \sum_ {l =0} ^\infty
     {(2l+1)} \zeta_{\Delta(l)}(z) \log \big(\Lambda L\big)^2
     \,,\qquad z\to0\,.
  \end{equation}
  Following the discussion in section~\ref{sec::KKsum}, we substitute 
\(  \tanh \pi v = 1 - 2( e^{2 \pi v } + 1 )^{-1}\)
and  we split the contribution of the `1' and of the `\(-2(...)^{-1}\)' in the integrand. 
  
  In \(\zeta_1\), the integral in \(v\) can be done; expanding the result in series of \(l\) we get
  \begin{equation}\label{cdi}
  \zeta_1(z) = \sum_{l=0}^\infty \sum_{n=0}^\infty
  \frac{
  		(-)^n 
  		2^{-2 n-3} 
  		L_S^{2 z-4}
  		  (l-\frac{1}{2} )^{-2 n-2 z+3} 
  		  (L_S^2 \left(4 \kappa +9\right)-1 )^n 
  		 \Gamma (n+z-1) \left(4
      (l-\frac{1}{2} )^2+L_S^2 \left(4 \kappa +z+7\right)-1\right)
    }{
    		3 (z-2) \Gamma (n+1) \Gamma (z)
    }\, ,
  \end{equation}
  and performing the sum in \(l\) by zeta-function regularization we get 
  \(  \sum_l (l-\frac12)^k =\zeta_{\text H}(-k,\frac12) =  - (1- 2^{-k})  \zeta_{\text R}(k)   \) 
  \begin{equation}\label{cdk}
  \begin{aligned}
  \zeta_1(z) =  \sum_{n=0}^\infty &
  \frac{
  		(-)^n
  		 2^{-2n-6} 
  		 L_S^{2 z-4} 
  		  (L_S^2 \left(4 \kappa +9\right)-1 )^n 
  		  \Gamma (n+z-1) 
     }{
     		3 (z-2) 
     		\Gamma (n+1)
     		 \Gamma (z)
     		}
  \\
  &   	\times
     		\left[
     			 (2^{2n+2z}-8 ) 
     		 (L_S^2  (4  \kappa +z+7 )-1 )
     		  \zeta_{\text R} (2 n+2 z-3)+ (2^{2n+2z}-32 ) \zeta_{\text R} (2 n+2 z-5)
     		  \right]\, .
  \end{aligned}
  \end{equation}
  In the limit \(z\to0\) the first 4 terms of the sum over \(n\) contribute, and the result is 
  \begin{equation}\label{cpi}
  \zeta_1(0) =
  	\frac{1}{945 L_S^4}
  	-\frac{1}{72} L_S^2 \kappa ^3
  	+\frac{1}{72} \left(1-6 L_S^2\right) \kappa ^2
  	+\left(-\frac{21 L_S^2}{128}
  	-\frac{1}{180 L_S^2}
  	+\frac{1}{18}\right) \kappa 
  	-\frac{27 L_S^2}{256}
  	-\frac{1}{90 L_S^2}
  +\frac{7}{128}\,.
  \end{equation}

  In \(\zeta_2\), we first expand the denominator of \eqref{cda} and we obtain
  \begin{equation}\label{cpy}
  \zeta_2(z) = 
   \intl_0^\infty dv
   \frac{v \left(4 v^2+1\right)}{ e^{2 \pi  v}+1} 
  \sum_{l=0}^\infty \sum_{n=0}^\infty
  	 \left(l-\frac{1}{2}\right)^{1-2 n-2 z}
  \frac{ 
  	(-)^{n+1}  L_S^{2 (n+z)}
  	\Gamma (n+z)
    }{
    		6  \Gamma (n+1) \Gamma (z)
    }
   \left(\kappa +v^2+\frac{9}{4}-\frac{1}{4 L_S^2}\right)^n\, ,
  \end{equation}
  and summing on \(l\) as above we get,
  \begin{equation}\label{cpt}
  \zeta_2(z) = 
   \intl_0^\infty dv
   \frac{v \left(4 v^2+1\right)}{ e^{2 \pi  v}+1} 
   \sum_{n=0}^\infty 
  \frac{ 
  	(-)^{n}  L_S^{2 (n+z)} (1-2^{2n+2z-1 } )
  	\zeta(2n+2z-1)
  	\Gamma (n+z)
    }{
    		6  \Gamma (n+1) \Gamma (z)
    }
   \left(\kappa +v^2+\frac{9}{4}-\frac{1}{4 L_S^2}\right)^n\, .
  \end{equation}
  In the \(z\to0\) limit only the terms \(n=0,1\) survive; the integral in \(v\) can then be performed and the result is
  \begin{equation}\label{cpr}
  \begin{aligned}
  \zeta_2(0) &=
  \frac{1}{144}
  \intl_0^\infty dv
  \frac{ 
  		v 
  		 (4 v^2+1 ) 
  		( 12 L_S^2 v^2 + 12 L_S^2 \kappa +27 L_S^2-4)
  	}{
  		e^{2 \pi v} +1
  } 
  = \frac{17 L_S^2 \kappa }{5760}+\frac{367 L_S^2}{48384}-\frac{17}{17280}\, .
  \end{aligned}
  \end{equation}

  The result for \( -(\zeta_1(0) + \zeta_2(0))\log \Lambda L \) is equal to \eqref{ccd}. The complicated dependence on \(\kappa \) and \(L_S\) is fully recovered.

\subsection{Vector on \texorpdfstring{\(\ads_4\times S^2\)}{AdS4 x S2 }   }
 
Consider  a vector \(A_M\) on \(\ads_4 (L_A) \times S^2(L_S)\) with partition function
\begin{align}\label{boa}
Z &  {}  = e^{-\Gamma} 
=  \int \! \DD A e^{- \int \! A \Delta_1[c,k] A} 
= \Big[{\det}_1 \, \Delta_1[c,k] \Big]^{-\frac12}\, ,
\\
& A \Delta_1[c,k] A
= A^M \Delta_1[c,k]_{MN} A^N
\,, \qquad
\Delta_1[c,k]_{MN}
	 = - g_{MN} \cd^2  + c\,  g_{MN} + k\,  R_{MN}\, .
\end{align}
We do expect a log divergence from 6D perspective with the Seeley--DeWitt coefficient
\begin{equation}\label{bob}
\begin{aligned}
B_6 = 
\frac{c^2 k  }{2}  L_A^2  L_S^2 
-\frac{ c^2 k   }{12}   L_A^4 
-\frac{ c^3       }{12} L_A^4 L_S^2
-\frac{ c^2  }{2}     L_A^2  L_S^2
+\frac{ c^2}{12}    L_A^4
-\frac{c k^2}{12   L_S^2}   L_A^4 
+\frac{c k}{18 L_S^2}    L_A^4 
-\frac{2  c k   }{3}   L_A^2 
-\frac{ c}{180 L_S^2}   L_A^4
\\
+\frac{ c}{3}    L_A^2
-\frac{k^3}{36 L_S^4}    L_A^4 
+\frac{ k^2}{36 L_S^4}   L_A^4
-\frac{  k^2}{6 L_S^2}   L_A^2 
+\frac{3 k^3 }{2 L_A^2}    L_S^2
-\frac{3 k^2}{L_A^2}    L_S^2
+\frac{k}{60 L_S^4}    L_A^4 
+\frac{8 k}{45 L_S^2}    L_A^2 
+\frac{43 k }{30 L_A^2}    L_S^2
\\
-\frac{13 }{1260L_S^4}   L_A^4
-\frac{1}{90 L_S^2}   L_A^2
-\frac{59 }{315 L_A^2}   L_S^2
-\frac{3}{2} c k^2 L_S^2
+2 c k L_S^2
-\frac{4 c }{5}   L_S^2
+\frac{k^2}{2}
-\frac{89 k}{90}
+\frac{4}{15}\,.
\end{aligned}
\end{equation}

The operator divides 
\begin{equation}\label{boc}
A^M \Delta_1[c,k]_{MN} A^N
 = A^\mu [- \cd^2 + c - 3 k L_A^{-2}] A_\mu
 + A^m [- \cd^2 + c +  k L_S^{-2}] A_m\,,
\end{equation}
and splitting transverse + longitudinal we get 
\begin{equation}\label{bog}
\det_1 \Delta_1[c,k] =
\frac{	\det\limits_{1\perp,0} [-\cd^2 + c - \tfrac{3}{L_A^2} k  ]    \ 
	\det\limits_{0,1\perp} [-\cd^2 + c + \tfrac{1}{L_S^2}  k  ]   \
	\det\limits_{0,0} [-\cd^2 + c - \tfrac{3}{L_A^2} (k-1)   ]   \
	\det\limits_{0,0}      [-\cd^2 + c + \tfrac{1}{L_S^2} (k-1 ) ]
 }{
 	\det\limits_{0,\emptyset} [-\cd^2 +  c + \tfrac{1}{L_S^2} (k-1 ) ]
	} \,.
\end{equation} 

Each determinant can be expressed in terms of \(\ads_4\) determinants by expanding in \(S^2\) harmonics.   We have a logarithmic divergence in each product term.
Let us distinguish the contributions according to the field type,
\begin{align}\label{bqd}
\Gamma 
	&  
	= \Gamma_{0,0}  
			+   \Gamma_{0,1\perp}   
			+   \Gamma_{1\perp,0}
						+   \Gamma_{\text{Am}}
	=  - \frac12 \zeta_1'(0) - \frac 12 \zeta_1(0) \log \big(\Lambda L\big)^2 \, .
\end{align}
The AdS-mode contribution is immediate,
\begin{equation}\label{bqo}
\begin{gathered}
(\Gamma_{\text{Am}} )_{\log\Lambda L } 
=  \frac 12 \left[\log 	\det_{0,\emptyset} [-\cd^2 +  c + \tfrac{1}{L_S^2} (k-1 ) ]\right]_{\log\Lambda L } 
= - B_{4; \text{Am}   } \log \Lambda\, ,
\\
B_{4 ; \text{Am}} =
\frac 1 {360}\left[
\frac{(c L_S^2+k-1 )^2}{L_S^4} 15L_ A^4  
+\frac{  (c L_S^2+k-1 )}{L_S^2}60 L_A^2\,
+58\right]\,.
\end{gathered}
\end{equation}
The other determinants are evaluated as described in section~\ref{sec::KKsum} and we get 
\begin{equation}\label{bof}
\begin{gathered}
(\Gamma_{0,0})_{\log\Lambda L }  = -\frac { \log \big(\Lambda L\big)^2}2\Big(
\frac{1}{8} L_A^2 c^2 k L_S^2
-\frac{1}{24} L_A^4 c^2 k
-\frac{1}{36} L_A^4 c^3 L_S^2
-\frac{7}{24} L_A^2 c^2 L_S^2
+\frac{5 L_A^4 c^2}{72}
-\frac{L_A^4 c k^2}{24L_S^2}
   +\frac{L_A^4 c k}{9 L_S^2}
\\
   -\frac{1}{4} L_A^2 c k
   -\frac{29 L_A^4 c}{360 L_S^2}
   +\frac{13 L_A^2 c}{36}
   -\frac{L_A^4 k^3}{72 L_S^4}
   +\frac{L_A^4 k^2}{18   L_S^4}
   -\frac{L_A^2 k^2}{12 L_S^2}
   +\frac{3 k^3 L_S^2}{8 L_A^2}
   -\frac{15 k^2 L_S^2}{8 L_A^2}
   -\frac{3 L_A^4 k}{40 L_S^4}
      \\
   +\frac{43 L_A^2 k}{180 L_S^2}
   +\frac{373 k L_S^2}{120  L_ A^2} 
   +\frac{67 L_A^4}{1890 L_S^4}
   -\frac{8 L_A^2}{45 L_S^2}
   -\frac{13639 L_S^2}{7560 L_A^2}
   -\frac{3}{8} c k^2 L_S^2
   +\frac{5}{4} c k L_S^2
   -\frac{431 c   L_S^2}{360}
   +\frac{k^2}{8}-\frac{26 k}{45}
   +\frac{121}{216}
    \Big)\,,
\end{gathered}
\end{equation}
\begin{equation}\label{bqe}
\begin{gathered}
(\Gamma_{0,1\perp})_{\log\Lambda L } 
= -\frac {\log \big(\Lambda L\big)^2} 2 \Big( 
-\frac{1}{24} L_A^4 c^2 k
-\frac{1}{72} L_A^4 c^3 L_S^2
-\frac{1}{12} L_A^2 c^2 L_S^2
+\frac{L_A^4 c^2}{72}
-\frac{L_A^4 c k^2}{24 L_S^2}
+\frac{L_A^4 c k}{36   L_S^2}
   -\frac{1}{6} L_A^2 c k
   +\frac{L_A^4 c}{120 L_S^2}
   +\frac{L_A^2 c}{18}
   \\
   -\frac{L_A^4 k^3}{72 L_S^4}
   +\frac{L_A^4 k^2}{72 L_S^4}
   -\frac{L_A^2 k^2}{12 L_S^2}
   +\frac{L_A^4   k}{120 L_S^4}
   +\frac{L_A^2 k}{18 L_S^2}
   -\frac{11 L_A^4}{1512 L_S^4}
   +\frac{L_A^2}{60 L_S^2}
   -\frac{37 L_S^2}{378 L_A^2}
   -\frac{29 c L_S^2}{180}
   -\frac{29   k}{180}
   +\frac{29}{540}
  \Big)\,,
\end{gathered}
\end{equation}
and
\begin{equation}\label{bqq}
\begin{gathered}
(\Gamma_{1\perp,0})_{\log\Lambda L } 
= -\frac {\log \big(\Lambda L\big)^2} 2 \Big( 
\frac{3}{8} L_A^2 c^2 k L_S^2
-\frac{1}{24} L_A^4 c^3 L_S^2
-\frac{1}{8} L_A^2 c^2 L_S^2
+\frac{L_A^4 c^2}{24}
-\frac{1}{4} L_A^2 c k
-\frac{L_A^4 c}{60 L_S^2}
+\frac{L_A^2   c}{12}
   +\frac{9 k^3 L_S^2}{8 L_A^2}
   -\frac{9 k^2 L_S^2}{8 L_A^2}
   \\
   +\frac{L_A^2 k}{20 L_S^2}
   -\frac{67 k L_S^2}{40 L_A^2}
   +\frac{L_A^4}{315 L_S^4}
   -\frac{L_A^2}{60   L_S^2}
   +\frac{4321 L_S^2}{2520 L_A^2}
   -\frac{9}{8} c k^2 L_S^2
   +\frac{3}{4} c k L_S^2
   +\frac{67 c L_S^2}{120}
   +\frac{3 k^2}{8}
   -\frac{k}{4}
   -\frac{67}{360}
  \Big)\,.
\end{gathered}
\end{equation}

Despite the remarkable complexity of these expressions, we do recover the divergence from    \(B_6\)   in \eqref{bob},
\begin{equation}\label{bqi}
(\Gamma_{1\perp,0})_{\log\Lambda L } 
+ (\Gamma_{0,1\perp})_{\log\Lambda L } 
+ (\Gamma_{0,0})_{\log\Lambda L } 
+ (\Gamma_{\text{Am}} )_{\log\Lambda L } 
= -     \frac 12  B_6 \log \big(\Lambda L\big)^2\, ,
\end{equation}
We emphasize that to get such agreement it is necessary to include the contribution of the AdS-mode \eqref{bqo}.

\subsection{Divergences in  \texorpdfstring{$\ads_{\mathrm{even} } (L_A) \times S^{\mathrm{odd} } (L_S)$}{AdS even x S odd}} 
\label{app::AdSevenSodd}
As the total dimension \(D = d_A + d_S\) is odd,  we expect no net log divergence. However, the single determinants in the partition functions in \eqref{fds}, \eqref{dsr}, \eqref{dsr-3form} display non-vanishing divergences and it is the combination of the different terms that conspires to cancel the divergent piece.

From several examples we noticed regularities in the cancellations. In particular, each of the different factors appearing in the partition functions above is finite by itself.
More specifically, we find explicitly that all determinant on fields in the representation \((\mathcal R,0)\) are finite. This also holds for fermions \((\mathcal R',\tfrac12)\),
\begin{equation}\label{doq}
\big[\log \det_{\mathcal R,0} [-\nabla^2+\kappa ] \,\big]_{\log\Lambda L } =0 \, ,
\quad 
 \forall \, \mathcal R\,;
\qquad
 \big[\log \det_{\mathcal R',\frac12} [i\slashed{\cd}_A + \alpha\,\gamma_* \, i\slashed{\cd}_S - \kappa  \gamma_* ]\, \big]_{\log\Lambda L }  =0\, ,
\quad \forall \, \mathcal  R'
\,.
\end{equation} 
We collect in the following the combination of non-finite contributions, which are finite once put together. 
\\[0.5em]
\textbf{Bosonic Combinations:}
\begin{subequations}\label{eq:finite_determinants}
\begin{align}
%    \intertext{\textbf{Bosonic Combinations:}}
    \bigg[ \log \frac{ \det\limits_{1\perp,\emptyset} [-\nabla^2_A+\kappa -\frac{d_S-1}{L_S^2} ]^{\frac{1}{2}} }{ \det\limits_{1\perp,1\perp} [-\nabla^2+\kappa ]^{\frac{1}{2}} } \bigg]_{\log\Lambda L }  &= 0 \,, \label{eq:det_bos_1} \\
    \bigg[ \log \frac{ \det\limits_{0,\emptyset} [-\nabla^2_A+\kappa +\frac{d_A-1}{L^2_A}-\frac{d_S-1}{L_S^2} ]^{\frac{1}{2}} }{ \det\limits_{0,1\perp} [-\nabla^2+\kappa+ \frac{d_A-1}{L^2_A} ]^{\frac{1}{2}} } \bigg]_{\log\Lambda L }  &= 0 \,, \label{eq:det_bos_2} \\
    \bigg[ \log\frac{ \det\limits_{0,\emptyset} [-\nabla^2_A + \kappa -2 ]^{\frac{d_S(d_S+1)}{4}} \det\limits_{0,\emptyset} [-\nabla^2_A + \kappa -\frac{d_S}{L_S^2} ]^{\frac{d_S+1}{2}} }{ \det\limits_{0,(2)\perp} [-\nabla^2+\kappa ]^{\frac{1}{2}} } \bigg]_{\log\Lambda L }  &= 0 \,, \label{eq:det_bos_3} \\
    \bigg[ \log \frac{ \det\limits_{0,\emptyset} [-\nabla^2_A + \kappa -2\frac{d_S}{L_S^2} ]^{\frac{1}{2}} }{ \det\limits_{0,1\perp} [-\nabla^2+\kappa - \frac{d_S+1}{L^2_S} ]^{\frac{1}{2}} } \bigg]_{\log\Lambda L }  &= 0 \,, \label{eq:det_bos_4} \\
    \bigg[ \log \det\limits_{0,2\perp} [-\nabla^2+\kappa ]^{\frac{1}{2}} \det\limits_{0,1\perp} [-\nabla^2+\kappa - \tfrac{d_S-3}{L_S^2} ]^{\frac{1}{2}} \bigg]_{\log\Lambda L }  &= 0 \,, \label{eq:det_bos_6} \\
    \bigg[ \log \det\limits_{1\perp,2\perp} [-\nabla^2+\kappa]^\frac{1}{2} \det\limits_{1\perp,1\perp} [-\nabla^2+\kappa - \tfrac{d_S-3}{L_S^2}]^\frac{1}{2}\bigg]_{\log\Lambda L }  & = 0 \, , \label{eq_det_bos_7}\\
    \bigg[ \log \det\limits_{0,3\perp}  [-\nabla^2+\kappa ]^{\frac{1}{2}} \det\limits_{0,2\perp} [-\nabla^2+\kappa - \tfrac{d_S-5}{L_S^2} ]^{\frac{1}{2}} \bigg]_{\log\Lambda L }  &= 0 \,, \label{eq:det_bos_8} \\
    \bigg[ \log \det\limits_{0,3\perp} [-\nabla^2+\kappa + \alpha Q]^{\frac{1}{2}} \det\limits_{0,2\perp} [-\nabla^2+\kappa - \tfrac{2}{L_S^2}]^{\frac{1}{2}}\bigg]_{\log\Lambda L }  &= 0   \qquad (d_S = 7)  \,, \label{eq:det_bos_10} 
    \intertext{with \(Q A_{mnr}= \frac 16 \varepsilon_{mnrabcs}\cd^s A^{abc}\) and \(\alpha\) arbitrary.  \vspace{-1em} }
     \intertext{\textbf{Fermionic Combination:}}
    \bigg[ \log\frac{ \det\limits_{\frac{1}{2},\frac{3}{2}\perp} [i\nablasl - \frac{3}{2}  \gamma_*  ]^{\frac{1}{2}} \det\limits_{\frac{1}{2},\emptyset} [i\nablasl_A -3 \gamma_* ]^{2} }{ \det\limits_{\frac{1}{2},\emptyset} [i\nablasl_A -\gamma_* ]^{2} \det\limits_{\frac{1}{2},\emptyset} [i\nablasl_A -4\gamma_* ]^{2} \det\limits_{\frac{1}{2},\emptyset} [i\nablasl_A +3\gamma_* ]^{2} } \bigg]_{\log\Lambda L }  &= 0 \,. \label{eq:det_ferm_1}
\end{align}
\end{subequations}

As a first result let us reconsider the partition functions in appendix~\ref{app:partf} and apply the cancellations just observed to the factors of the partition function. 
What we have effectively shown is that all partition functions of the form \(Z_{r, r'}\) are finite, namely all   determinants  on unconstrained fields have no logarithmic divergence,
\begin{equation}\label{atm}
\log \det_{r, r'} [-\cd^2 + \kappa] : \quad
\zeta(0 ) = 0\,.
\end{equation}
Note that this applies only to determinants where the fields do not obey differential constraints; when expressing the determinants in terms of their irreducible representations, individual terms may exhibit divergences, but they are precisely matched by other contributions. In this cancellation, the effect of the AdS modes is crucial.

An immediate consequence is that partition functions of quadratic fields in \(d_A\) even, \(d_S\) odd spaces are finite.
This reproduces the expected cancellation of the overall divergences from a \(D\)-dimensional perspective.

\subsection{Divergences in  \texorpdfstring{$\ads_{4} (L_A)\times S^{7 } (L_S)$}{AdS4 (LA) x S7 (LS)}} 
% ----------------------------------------------------------------- DIVERGENCE TABLE -------------------------------------------------------------------------------------
 
Some explicit results for divergences used in the main text are
\begin{equation}\label{eq:log-div-results}
    \renewcommand{\arraystretch}{1.5} 
    \begin{array}{cc}
        \toprule
        \multicolumn{1}{c}{\mathbf{Field} } & \multicolumn{1}{c}{
        \zeta(0)  \text{ of }  -\nabla^2 + \kappa \text{ on }  \ads_4 (L_A) \times S^7 (L_S) } \\
        \midrule
        \multicolumn{2}{c}{\mathbf{AdS-Modes}} \\
        \midrule
        (0,\emptyset)\, \quad \, & \displaystyle \tfrac{29}{180} + \tfrac{1}{6} L_A^2 \kappa  + \tfrac{1}{24} L_A^4 \kappa ^2 \\
        (1{\perp},\emptyset)\, \quad \, & \displaystyle - \tfrac{67}{120} + \tfrac{1}{4} L_A^2 \kappa  + \tfrac{1}{8} L_A^4 \kappa ^2 \\
        (\tfrac12,\emptyset)\, \quad \, & \displaystyle \tfrac{101}{180} + \tfrac{2}{3} L_A^2 \kappa  + \tfrac{1}{6} L_A^4 \kappa ^2 \\
        (\tfrac32,\emptyset)\, \quad \, & \displaystyle \tfrac{71}{45} + \tfrac{8}{3} L_A^2 \kappa  + \tfrac{2}{3} L_A^4 \kappa ^2 \\ 
        \midrule
        \multicolumn{2}{c}{\mathbf{Bosonic \  contributions}} \\
        \midrule
        (0,1{\perp})\, \quad \,  & \displaystyle \tfrac{29}{180} + \tfrac{3 L_A^4}{2 L_S^4}  -\tfrac{L_A^2}{L_S^2} + \tfrac{1}{6} L_A^2 \kappa  - \tfrac{L_A^4 \kappa }{2 L_S^2} + \tfrac{1}{24} L_A^4 \kappa ^2 \\
        (1{\perp},1{\perp})\, \quad \,  & \displaystyle - \tfrac{67}{120} + \tfrac{9 L_A^4}{2 L_S^4} - \tfrac{3 L_A^2}{2 L_S^2} + \tfrac{1}{4} L_A^2 \kappa   - \tfrac{3 L_A^4 \kappa }{2 L_S^2} + \tfrac{1}{8} L_A^4 \kappa ^2 \\
        (0,(2){\perp})\, \quad \,  & \displaystyle \tfrac{29}{5} + \tfrac{21 L_A^4}{L_S^4} - \tfrac{56 L_A^2}{3 L_S^2} + 6 L_A^2 \kappa  - \tfrac{28 L_A^4 \kappa }{3 L_S^2} + \tfrac{3}{2} L_A^4 \kappa ^2 \\
        (0,2{\perp})\, \quad \,  & \displaystyle - \tfrac{29}{180} - \tfrac{25 L_A^4}{6 L_S^4} + \tfrac{5 L_A^2}{3 L_S^2} - \tfrac{1}{6} L_A^2 \kappa  + \tfrac{5 L_A^4 \kappa }{6 L_S^2} - \tfrac{1}{24} L_A^4 \kappa ^2 \\
        (1,2{\perp})\, \quad \,  & \displaystyle \tfrac{67}{120} - \tfrac{25 L_A^4}{2 L_S^4} + \tfrac{5 L_A^2}{2 L_S^2} - \tfrac{1}{4} L_A^2 \kappa  + \tfrac{5 L_A^4 \kappa }{2 L_S^2} - \tfrac{1}{8} L_A^4 \kappa ^2 \\
        (0,3{\perp})\, \quad \,  & \displaystyle - \tfrac{116}{15} - \tfrac{288 L_A^4}{L_S^4} + \tfrac{96 L_A^2}{L_S^2} -8 L_A^2 \kappa  + \tfrac{48 L_A^4 \kappa }{L_S^2} -2 L_A^4 \kappa ^2 \\ 
        \midrule
        \multicolumn{2}{c}{\mathbf{Fermionic \ contributions}} \\ 
        \midrule
        (\tfrac12,\tfrac32{\perp})\, \quad \,  & \displaystyle \tfrac{202}{45} + \tfrac{147 L_A^4}{4 L_S^4}  - \tfrac{28 L_A^2}{L_S^2} + \tfrac{16}{3} L_A^2 \kappa   - \tfrac{14 L_A^4 \kappa }{L_S^2} + \tfrac{4}{3} L_A^4 \kappa ^2 \\
        \bottomrule
    \end{array}
\end{equation}
 
We also have  
\begin{equation}\label{alla}
\begin{split}
   \log \det_{0,0}&\left[-\nabla^2 + \kappa +  \tfrac{A}{L_S} \sqrt{-\nabla_S^2 + \tfrac{9}{L_S^2}  }\, \right] :\\
   \zeta(0) &=  - \tfrac{29}{360} A 
        - \tfrac{2059}{77760} A^3 
        - \tfrac{29}{16200} A^5 
        - \tfrac{29}{907200} A^7
        - \tfrac{55901 A L_A^4}{684288 L_S^4} 
        - \tfrac{A^3 L_A^4}{16 L_S^4} 
        - \tfrac{23 A^5 L_A^4}{2400 L_S^4} 
        - \tfrac{71 A^7 L_A^4}{129600 L_S^4} 
        - \tfrac{29 A^9 L_A^4}{2177280 L_S^4} 
    \\
        &\quad
                - \tfrac{A^{11} L_A^4}{8553600 L_S^4} 
                + \tfrac{1480159 A L_A^2}{10886400 L_S^2} + \tfrac{197 A^3 L_A^2}{2880 L_S^2} 
        + \tfrac{317 A^5 L_A^2}{43200 L_S^2} 
        + \tfrac{7 A^7 L_A^2}{25920 L_S^2} 
        + \tfrac{A^9 L_A^2}{311040 L_S^2} 
        + \tfrac{5809 A L_S^2}{272160 L_A^2} 
    \\  
        &\quad
                + \tfrac{1147 A^3 L_S^2}{272160 L_A^2} 
                + \tfrac{37 A^5 L_S^2}{272160 L_A^2} 
                - \tfrac{1639 A L_S^4}{680400 L_A^4} 
        - \tfrac{149 A^3 L_S^4}{680400 L_A^4} 
        + \tfrac{179 A L_S^6}{2494800 L_A^6} 
    \\
        &\quad
        +\Big(- \tfrac{1}{12} A L_A^2 - \tfrac{71}{2592} A^3 L_A^2 - \tfrac{1}{540} A^5 L_A^2 - \tfrac{1}{30240} A^7 L_A^2 + \tfrac{1480159 A L_A^4}    {21772800 L_S^2} + \tfrac{197 A^3 L_A^4}{5760 L_S^2} + \tfrac{317 A^5 L_A^4}{86400 L_S^2} + \tfrac{7 A^7 L_A^4}{51840 L_S^2} 
    \\  
        &\quad
        + \tfrac{A^9 L_A^4}{622080 L_S^2} + \tfrac{4553}{129600} A L_S^2 + \tfrac{899}{129600} A^3 L_S^2 + \tfrac{29}{129600} A^5 L_S^2 - \tfrac{407 A L_S^4}{68040 L_A^2} - \tfrac{37 A^3 L_S^4}{68040 L_A^2} + \tfrac{149 A L_S^6}{453600 L_A^4}\Big) \kappa  
    \\
        &\quad
        + \Big(- \tfrac{1}{48} A L_A^4 - \tfrac{71}{10368} A^3 L_A^4 - \tfrac{1}{2160} A^5 L_A^4 - \tfrac{1}{120960} A^7 L_A^4 + \tfrac{157}{8640} A    L_A^2 L_S^2 + \tfrac{31}{8640} A^3 L_A^2 L_S^2 + 
    \\
        &\qquad
        \tfrac{1}{8640} A^5 L_A^2 L_S^2 - \tfrac{319}{64800} A L_S^4 - \tfrac{29}{64800} A^3 L_S^4 + \tfrac{37 A L_S^6}{90720 L_A^2}\Big) \kappa ^2 
    \\  
        &\quad
        + \Big(\tfrac{157}{51840} A L_A^4 L_S^2 
        + \tfrac{31}{51840} A^3 L_A^4 L_S^2 
        + \tfrac{1}{51840} A^5 L_A^4 L_S^2 - \tfrac{11}{6480} A L_A^2 L_S^4 - \tfrac{1}{6480} A^3 L_A^2 L_S^4 + \tfrac{29}{129600} A L_S^6\Big) \kappa ^3 
    \\
        &\quad
        +\left(- \tfrac{11}{51840} A L_A^4 L_S^4 - \tfrac{1}{51840} A^3 L_A^4 L_S^4 + \tfrac{1}{17280} A L_A^2 L_S^6\right) \kappa ^4 
    \\  
        &\quad
        + \tfrac{1}{172800} A L_A^4 L_S^6 \kappa ^5\, ,
\end{split}
\end{equation}
which is odd in $A$.
Similarly,
\begin{equation}\label{eq:logDetsqrt}
    \begin{split}
        \log  \det_{1\perp, 1\perp}  & \Big[-\nabla^2 + 6   \pm 6 \sqrt{-\nabla_S^2 + 10  } \, \Big]  :
        \\
        \zeta(0)  & {}= 
            - \frac{67}{120} + \frac32    L_A^2 + \frac92   L_A^4 + \frac{45 L_A^4}{2L_S^4} - \frac{3 L_A^2}{2L_S^2} - \frac{9    L_A^4}{L_S^2} 
            \pm ( \ldots)
            \,,
    \end{split}
\end{equation} 
where the complicated terms in the \(\pm\) brackets cancel each other in the sum.
Also,
\begin{equation}
\begin{aligned}
    \label{eq:det_1/2_3/2linear}
    \log    \det_{\frac{1}{2}\perp,\frac{3}{2}\perp}  &
    			 [-\cd^2 + \tfrac{3}{L_S}\nablasl_S + \kappa  ]: \\
     \zeta(0)& { }=  
            \frac{101}{45}
            -\frac{14 L_A^2}{L_S^2}
            +\frac{447 L_A^4}{8 L_S^4}
            -\left(\frac{7  L_A^4}{L_S^2}+\frac{8 L_A^2}{3}\right)\kappa
            +\frac{2  L_A^4}{3}\kappa ^2
            \,.
\end{aligned}
\end{equation}

\subsection{Scalar on \texorpdfstring{\(\ads_{4} \times S^{2}\)}{AdS4 x S2} with fourth-order operator} \label{subsec:zRegScalar}
In order to provide a nontrivial check of the discussion in section~\ref{sec::op4},
we  compute the log divergence to the logarithm of the determinant of the following quartic operator  
\begin{equation}\label{delta4-ads4xs2}
    \Delta= \left[-\nabla^2 + \kappa +  \frac{A}{L_S} \sqrt{-\nabla_S^2 + \tfrac{1}{4 \, L_S^2}}\right]\left[-\nabla^2 + \kappa - \frac{A}{L_S} \sqrt{-\nabla_S^2 + \tfrac{1}{4 \, L_S^2}}\right]
\end{equation}
for the explicit example of $\ads_4 \times S^2$, where we can also compute determinants with the heat kernel and compare the two results. 
For the operator \eqref{delta4-ads4xs2},  the square root   simplifies once we plug in the eigenvalues of the scalar harmonics on $S^2$, $\lambda_l^{0,2} =  l(l+1)$.  

Proceeding as  discussed in  section~\ref{sec::op4}, both   factors in \eqref{delta4-ads4xs2} give the same contribution; 
their sum is 
\begin{equation}\label{eq:mnp}
    \begin{split}
        \log \det_{0,0} & \Delta  : 
        \\
           \zeta(0 ) & = {} 
            \tfrac{29}{270} 
            -\tfrac{37 L_S^2}{189 L_A^2}
            -\tfrac{L_A^2}{45 L_S^2}
            +\tfrac{29 A^2}{180}
            +\tfrac{2 L_A^4}{945 L_S^4}
            +\tfrac{A^6 L_A^4}{720 L_S^4}
            +\tfrac{A^4 L_A^4}{288 L_S^4}
            -\tfrac{A^4 L_A^2}{36 L_S^2}
            +\tfrac{A^2 L_A^4}{720 L_S^4}
            -\tfrac{A^2 L_A^2}{24 L_S^2}
        \\
           &\quad
           +\kappa  \left(-\tfrac{A^4 L_A^4}{72 L_S^2}-\tfrac{A^2 L_A^4}{48 L_S^2}+\tfrac{A^2 L_A^2}{6}-\tfrac{L_A^4}{90 L_S^2}+\tfrac{L_A^2}{9}-\tfrac{29 L_S^2}{90}\right)
        \\
           &\quad
           +\kappa ^2 \left(\tfrac{A^2 L_A^4}{24}+\tfrac{L_A^4}{36}-\tfrac{L_A^2 L_S^2}{6}\right) 
           -\tfrac{1}{36} \kappa ^3 L_A^4 L_S^2\, .
    \end{split}
\end{equation}

The operator \eqref{delta4-ads4xs2} belongs to the more general family
\begin{equation}\label{del4}
    \begin{split}
        \hat \Delta  
        =& \nabla^4 + 2\kappa (-\nabla_A^2)  + \left(2\kappa - \tfrac{A^2}{L_S^2}\right)(-\nabla_S^2) - \tfrac{A^2}{L_S^2} C + \kappa^2\,,
    \end{split}
\end{equation}
whose divergence can  also be evaluated via the SdW coefficient $b_6$ of 
 ~\cite{Casarin:2021fgd,Casarin:2019aqw,Casarin:2023ifl} (keeping in mind the comment around \eqref{edn}).  
   We get 
\begin{equation}
    \begin{split}
        B_6   [\hat \Delta  ] = \displaystyle
            & {}
                 \tfrac{1}{12} L_A^4 L_S^2 
                    \Big(
                         \tfrac{A^6}{60 L_S^6}+\tfrac{A^4 C}{6 L_S^4}
                        -\tfrac{A^4}{3 L_A^2 L_S^4}
                        -\tfrac{2 A^2 C}{L_A^2 L_S^2}
                        +\tfrac{A^2 C}{3 L_S^4}
                        +\tfrac{29 A^2}{15 L_A^4 L_S^2}
                   \\ &\quad\quad\qquad
                        -\tfrac{A^2}{15 L_S^6}
                        -\tfrac{148}{63 L_A^6}
                        +\tfrac{58}{45 L_A^4 L_S^2}
                        -\tfrac{4}{15 L_A^2 L_S^4}
                        +\tfrac{8}{315 L_S^6}
                     \Big) 
            \\
                & +\tfrac{1}{12} \kappa  L_A^4 L_S^2 
                    \left(
                        -\tfrac{A^4}{6 L_S^4}
                        -\tfrac{A^2 C}{L_S^2}
                        +\tfrac{2 A^2}{L_A^2 L_S^2}
                        -\tfrac{58}{15 L_A^4}
                        +\tfrac{4}{3 L_A^2 L_S^2}
                        -\tfrac{2}{15 L_S^4}
                    \right)
            \\ 
                &+\tfrac{1}{12} \kappa ^2 L_A^4 L_S^2 \left(\tfrac{A^2}{2 L_S^2}-\tfrac{2}{L_A^2}+\tfrac{1}{3 L_S^2}\right) 
                -\tfrac{1}{36} \kappa ^3 L_A^4 L_S^2\ ,
    \end{split}
\end{equation}
which for $C = \frac{1}{4 \, L_S^2}$ coincides with \eqref{eq:mnp}. This highlights the consistency of our regularization scheme. 

\section{Odd-dimensional \texorpdfstring{$\ads$}{AdS}} \label{app::adsodd} 
\subsection{General discussion}
The discussion is analogous to \(\zeta_1(z)\) in \eqref{dsf}, with the crucial difference that the integrand has a polynomial of \emph{even} powers of \(v\),
\begin{equation}\label{dsf2}
\zeta_\Delta (z)  = 
	  	    \log R \, L_A^{d_A+2z}  \sum_{l\geq 0} m_{l}^{s,d_S}
	 	\intl_0 ^{+\infty} \! d v  \
	 	{( v ^2 + b_l^2)^{-z}}
	 \sum_{j=0}^{ \frac{d_A-2 }2}  q_{j} v ^{ 2j } \,,
\end{equation} 
  \(q_{j}\) being \(l\)-dependent coefficients.
The integral in \(v\) can be directly done term by term, and  using \eqref{fpd} we have 
\begin{equation}\label{fdn-odd0}
\zeta_\Delta(z) = 	\log R \, (L_A)^{{d_A}+2z}      \sum_{l\geq 0} m_{l}^{s,{d_S}}
\sum_{j=0}^{ \frac{{d_A}-2 }2}  q_{j} \,
	 	b_l^{2 (j+\frac12  - z)} 
	 	    \frac{\Gamma  ( j+\frac12 ) \Gamma  (z -j -\frac12)}{2 \Gamma (z)}
	 	    \,.
\end{equation}   
Since we are on an odd-dimensional $\ads$ space, there will be no contribution to the divergence from $\zeta_\Delta(0)$. The only possible divergence comes from a divergence in the   sum over $l$ arising from $\zeta_{\Delta}'(0)$. Expanding the overall Gamma factor near $z \to 0$, we have $\Gamma(z) = \frac{1}{z} + \mathcal{O}(1)$, which implies
\begin{align}
    \frac{\Gamma[z - j - \frac12]}{\Gamma[z]} =  \Gamma \left[-j-\tfrac{1}{2}\right] \, z +\mathcal O (z^2 )\,.
\end{align}
Plugging this into the expression for $\zeta_\Delta(z)$ allows us to isolate the derivative at $z=0$. Using the reflection identity  
\(    \Gamma  [j+\tfrac{1}{2}] \ \Gamma [-j-\tfrac{1}{2}] = {2\pi} (-)^{j+1}  /{(2j+1)}\) 
we obtain
\begin{equation}\label{fdn-odd1}
    \zeta_\Delta'(0) 
        =   -\log R \, (L_A)^{{d_A}}  \sum_{l\geq 0} m_{l}^{s,{d_S}}
            \sum_{j=0}^{ \frac{{d_A}-2 }2} (-)^{\,j}  q_{j} \,
            b_l^{2 (j+\frac12)} \frac{ \pi }{2 j+1}\, .
\end{equation}     
The (finite) sum over \(j\) is a combination of polynomials of \(l\) raised to a half-integer power. Since we want to understand its UV behavior, we  expand it for large \(l\),
\begin{equation}\label{fdn-odd2}
    \zeta_\Delta'(0) 
        =   -\log R \, (L_A)^{{d_A}}  \sum_{l\geq 0} 
        		\left( \ldots 	+ a_2 l^2 
        		        		+ a_1 l 
   		       					+ a_0  
        						+ a_{-1} \frac 1 l  
        						+   a_{-2 } \frac 1 {l^2} 
        						+ \ldots
        		\right)\,.
\end{equation}    
Negative powers \(l^{-n}\), \(n\geq 2\), are convergent; non-negative powers of \(l\) can be regularized via Riemann zeta function \( \zeta_{\text R}\). However, 
the asymptotic expansion may   contain a $1/l$ term, which is not \( \zeta_{\text R}\)-regularizable.  In this case, summing up to a hard momentum cutoff $l_{\text{max}}$ yields a logarithmically divergent contribution, since the sum reduces to the harmonic numbers, $\sum_{l=1}^{n} l^{-1} = H_n \sim \log n$. By identifying the dimensionless cutoff eigenvalue index with the UV scale, $ l_{\text{max}} =   \Lambda L $, this $1/l$ tail   reproduces the logarithmic UV divergence expected from the even-dimensional Seeley--DeWitt coefficients of the full space, despite the lack of pure odd-$\ads_{d_A}$ divergences. The exact relation between \(l_{\text{max}}\) and \(\Lambda L\) is inferred from examples below.

\subsection{Scalar field on \texorpdfstring{\(\ads_3 \times S^2(L_S)\)}{AdS3 x S2}}\label{sect::scalar_ads3xs}
Consider a scalar field on $\ads_3 \times S^2$. The total spacetime dimension is \(D=5\), consequently, the Seeley--DeWitt coefficient $b_5$ vanishes identically, and we expect no logarithmic divergence.

To verify this via the zeta-function regularization, we expand the operator $\Delta = -\nabla^2 + \kappa$ in spherical harmonics on $S^2$, 
\begin{equation}\label{dac}
\Delta
=
-\cd^2  + \kappa 
= - \cd^2_A - \cd^2_S + \kappa 
\quad \to \quad    - \cd^2_A + l(l+1) L_S^{-2} + \kappa 
\equiv \Delta(l)\, .
\end{equation}
Inserting this in the  representation of $\zeta_\Delta$ for $\ads_3$, we get:
\begin{equation}\label{caf}
\zeta'_{ \Delta(l)}(0) = 
\frac1{3 \pi }
{\left( l (l+1) L_S^{-2}+\kappa +1\right)^{3/2}} \, .
\end{equation}
The total contribution requires summing over the $S^2$ multiplicities $m_l = 2l+1$,
\begin{equation}\label{cae}
\log \det \Delta =
\sum_{l = 0}^\infty
(2l+1) \log \det  \Delta(l)
=
-\sum_{l = 0}^\infty
(2l+1)
\zeta'_{ \Delta(l)}(0) \, ;
\end{equation}
expanding the summand asymptotically for large $l$, we obtain
\begin{equation}\label{djn}
(2l+1)
\zeta'_{ \Delta(l)}(0) 
=
\text{positive powers of \(l\)}
+ a_0 + a_2 \frac{1}{l^2} 
+ \text{more negative powers of \(l\)}\, .
\end{equation}
We see that the $l^{-1}$ term is absent, hence   the sum does not produce a $\log l_{\text{max}}$ divergence upon regularization, consistent with the higher-dimensional expectation.

\subsection{Scalar field on \texorpdfstring{\(\ads_3 \times S^3 (L_S)\)}{AdS3 x S3}}\label{sec::scalar_ads3s3}
Consider the operator $\Delta = -\nabla^2 + \kappa$ acting on a scalar field on $\ads_3 \times S^3(L_S)$. From the six-dimensional perspective, the effective action contains a logarithmic divergence determined by the Seeley--DeWitt coefficient \(b_6\),
\begin{equation}\label{dub}
\Gamma_{\log\Lambda L }  = 
\frac 12 
( \log \det \Delta)_{\log\Lambda L } 
=
 -\frac{\log \Lambda L}{(4 \pi )^ 3} \VOL_{\ads_3} \VOL_{S^3(L_S)}  b_6(\Delta) 
= 
-\log R \log \Lambda   L
\frac{ \left(L_S^2 \kappa +L_S^2-1\right)^3}{96 L_S^3}.
\end{equation}
Expanding the operator in harmonics we have 
\begin{equation}\label{duc}
\Delta =  - \cd^2_A - \cd_S^2 + \kappa 
\quad \longrightarrow\quad  - \cd_A^2 
	+  \lambda_l   L_S^{-2}
	+ \kappa 
	= \Delta(l)
\,,
\qquad
\lambda_l = l (l+2)
\,,
\quad
m_l = (l+1)^2.
\end{equation}
Thus we consider 
\begin{equation}\label{dud}
\zeta_{\ads_3 \times S^3} (z)
=  \sum_{ l = 0 }^\infty  (l+1)^2 
 	\zeta_{\Delta(l) } (z),
\end{equation}
and, since in \(\ads_3\) there is no logarithmic divergence, we have
\begin{equation}\label{due}
\Gamma =  -\zeta_{\ads_3 \times S^3} (0)
=  -\frac 12  \sum_{ l = 0 }^\infty  (l+1)^2 
 	\zeta'_{\Delta(l) } (0) \,,
\end{equation}
where each term in the summand is finite but divergent in \(l\).

Using the $\ads_3$ spectral function, the derivative evaluates to
\begin{equation}
    m_l \, \zeta_{\Delta(l)}' (0)  = -\frac{1}{3} \log R \left(l (l+2) L_S^{-2}+\kappa +1\right)^{3/2}(l+1)^2\, .
\end{equation}
Expanding this expression asymptotically for large $l$ we have
\begin{equation}
     m_l \, \zeta_{\Delta(l)}' (0) =
        \text{non-negative powers of \(l\)}
        +
        \frac 1 l	
        \frac{ (L_S^2  (\kappa +1 )-1 )^3}{48 L_S^3} \log R
        + \mathcal O(l^{-2})\, .
\end{equation}
We see that we get a non-vanishing contribution to the coefficient of $l^{-1}$. 
As a result for the divergence,
\begin{align}\label{dug}
\Gamma_{\log\Lambda L }  =  -\zeta_{\ads_3 \times S^3} (0)
= -\frac{ (L_S^2  (\kappa +1 )-1 )^3}{96 L_S^3} \log R \log l_{\text {max}}\, .
\end{align}
Identifying the dimensionless cutoff $l_{\text{max}} = \Lambda L $ reproduces the six-dimensional logarithmic divergence \eqref{dub}.

\subsection{Vector field on \texorpdfstring{\(\ads_3 \times S^3 (L_S)\)}{AdS3 x S3}}\label{sec::vectads3s3}
The partition function for a vector field $A_M$ on $\ads_3 \times S^3(L_S)$ is   
\begin{align}\label{ara}
Z &  {}  = e^{-\Gamma} 
=  \int \! \DD A e^{- \int \! A \Delta_1[c,k] A} 
= {\det}_1 \, \Delta_1[c,k] ^{-\frac12}\, ,
\\
& A \Delta_1[c,k] A
= A^M \Delta_1[c,k]_{MN} A^N
\,, \qquad
\Delta_1[c,k]_{MN}
	 = - g_{MN} \cd^2  + c g_{MN} + k R_{MN}\, .
\end{align}
The six-dimensional Seeley--DeWitt coefficient $b_6(\Delta_1)$ determines the divergence 
\begin{equation}\label{edw}
\begin{aligned}
B_6 =  
&
\frac{\log R}{48} L_S^3 \left(9 c^2 k-3 c^3-9 c^2-18 c k^2+18 c k-6 c+12 k^3-18 k^2+3 k+1\right)
\\
& 
+\frac{\log R}{16} L_S \left(-3 c^2 k+3 c^2-12 c k+6 c+6 k^2-9 k+2\right)
\\
& 
-\log R\left( \frac{6 c k^2-6 c k+2 c+6 k^2-9 k+2}{16 L_S} 
+ \frac{12 k^3-18 k^2+3 k+1}{{48 L_S^3}}\right)  .
\end{aligned}
\end{equation}

From the discussion in \ref{app:vect} leading to \eqref{fds}, we have
\begin{align}
Z=
        \frac{
            \det\limits_{0,\emptyset}      [ -\cd^2 + c + \frac{2}{L_S^2} (k-1 ) ]^\frac12
        }
        {
            \det\limits_{0,1\perp} [-\cd^2 + c + \frac{2}{L_S^2}  k  ] ^\frac12 \,
		    \det\limits_{0,0}      [-\cd^2 + c + \frac{2}{L_S^2} (k-1 ) ]^\frac12\,
		    \det\limits_{1\perp,0} [-\cd^2 + c - 2 k  ]^\frac12 \,
		    	        \det\limits_{0,0} [-\cd^2 + c - 2  (k-1)   ] ^\frac12
        }\, .
\end{align}
First, the \(\ads\) mode term in this case does not give any divergent contribution, since it is a standard determinant on an odd-dimensional space.
We proceed to expand the determinants in terms of \(S^3\) harmonics. The scalar contributions are of the form
\begin{equation}\label{arh}
\begin{gathered} 
 \log \det_{0,0} [  -\cd_A^2 - \cd_S^2  + \kappa]
	= \sum_{ l = 0 }^\infty (l+1)^2
		 \log
		 			\det_{0,0} [  -\cd_A^2  + l (l+2) L_S^{-2}  + \kappa] 
\,, 
\end{gathered}
\end{equation}
where the determinants in the sum are on a scalar operator of \(\ads_3\).
For the contribution \(1{\perp},0\), the \(S^3\) part is also a scalar, so that
\begin{equation}\label{ari}
\begin{gathered}
 \log \det_{1\perp,0} [  -  \cd_A^2 - \cd_S^2  + \kappa' ]
	= \sum_{ l = 0 }^\infty    m_l^{(0,3)}
		 		\log	\det_{1\perp,0} [  -  \cd_A^2  +( \lambda_l^{(0,3)} L_S^{-2}  + \kappa')]
		 		\,,
\end{gathered}
\end{equation}
where the determinant in the sum is over transverse \(1\)-forms.
Finally, we have  \(0,1{\perp}\) sum, which is a transverse 1-form on \(S^3\), and which gives
\begin{equation}\label{arj}
\log \det_{0,1{\perp}} [  -   \cd_A^2 -   \cd_S^2  + \kappa'' ]
	= \sum_{ l = 0 }^\infty2(l+1)(l+3)
		 \log	\det_{0,1{\perp}} [  -   \cd_A^2  +  (( l + 2 )^2 -2)L_S^{-2}  + \kappa'' ] 
\,,  
\end{equation}
here the determinants in the sum are on an \(\ads_3\) scalar.
Evaluating the $\zeta'(0)$ derivatives for the scalar components, isolating the $l^{-1}$ asymptotic terms, and summing up to $l_{\text{max}}$ yields
\begin{align}\label{ark}
\zeta_{0,0}' (0) & = 
		\frac{ (L_S^2  (\kappa +1 )-1 )^3}{48 L_S^3} \log R \log l_{\text {max}}
		\,, \qquad
		\kappa  = \Big\{  c + 2 (k-1) L_S^{-2} \,, \  c -2 (k-1)    \Big\}  \, ,
			\\
\zeta_{0,1\perp}' (0) & =
 	\frac{ 
 					(c L_S^2+2 k+L_S^2-2 )^2
 				  (c L_S^2+2 k+L_S^2+4 )
			 }{   24 L_S^3 }
 	\log R \log l_\text{max}\, .
\end{align}
The vector contributions are similarly given by  
\begin{gather}\label{arl}
\zeta_{1\perp,0} (z)
 = - \frac{2 \log R}{\pi }  
\intl_0^\infty \! dv
		\frac{v^2-1}{  [  c-2 k+ l (l+2) L_S^{-2} +v^2+2  ] ^{z}} 
=
\frac{\log R \ \Gamma  (z-\frac{3}{2} ) }{2 \sqrt{\pi } \Gamma  (z)} 
\frac{ c-2 k+2 z-1+  l (l+2) L_S^{-2}}{
	 [c-2 k+ 2 +l (l+2) L_S^{-2}]^{z-\frac{1}{2}}  
} \, ,\raisetag{-0.4cm}
\\
\zeta_{1\perp,0}' (0)
=
\sum_l ^{l_{\text {max}}}
\left[
 			  m_l^{(0,3)}  \, 
  			 \zeta_{1\perp (l)} (z)
   \right]_{\mathcal O ( l^{-1} ) }
 = 
\frac{  
	(c L_S^2-2 k L_S^2-4 L_S^2-1 ) 
	 (c L_S^2-2 k L_S^2+2 L_S^2-1 )^2
			}{
	24 L_S^3
	}
\log R \log l_\text{max}\, .\raisetag{-0.1cm}
\end{gather}
Combining the four contributions,
the result exactly matches the Seeley--DeWitt 6d calculation \eqref{edw}
\begin{equation}\label{arn}
\Gamma_{\log\Lambda L }  = -\zeta_\Delta(0) = -B_6 \log\Lambda  L 
\,, \qquad
l_{\text{max} } = \Lambda L \,.
\end{equation} 
 
\newpage 

\bibliographystyle{JHEP-v2.9}
  
\bibliography{biblio} 
 
\end{document}